\documentclass[twocolumn]{aastex63}

\accepted{for Publication in ApJS}

\usepackage{natbib,aas_macros,amsmath}
\usepackage{multirow,color}
\usepackage{natbib}

\newcommand{\carcsec}{$\!\!\arcsec$}
\newcommand{\m}[1]{\mathrm{#1}}

\newcommand{\redc}[1]{\textcolor{black}{#1}}
\newcommand{\redcc}[1]{\textcolor{black}{#1}}

\usepackage{booktabs}

\begin{document}
\shortauthors{Harikane et al.}

\shorttitle{
JWST Galaxies at $z\sim 9-17$
}

\title{
A Comprehensive Study on Galaxies at $z\sim9-\redc{16}$ Found in the Early JWST Data:\\
UV Luminosity Functions and Cosmic Star-Formation History at the Pre-Reionization Epoch
}

\email{hari@icrr.u-tokyo.ac.jp}
\author[0000-0002-6047-430X]{Yuichi Harikane}
\affiliation{Institute for Cosmic Ray Research, The University of Tokyo, 5-1-5 Kashiwanoha, Kashiwa, Chiba 277-8582, Japan}

\author[0000-0002-1049-6658]{Masami Ouchi}
\affiliation{National Astronomical Observatory of Japan, 2-21-1 Osawa, Mitaka, Tokyo 181-8588, Japan}
\affiliation{Institute for Cosmic Ray Research, The University of Tokyo, 5-1-5 Kashiwanoha, Kashiwa, Chiba 277-8582, Japan}
\affiliation{Kavli Institute for the Physics and Mathematics of the Universe (WPI), University of Tokyo, Kashiwa, Chiba 277-8583, Japan}

\author[0000-0003-3484-399X]{Masamune Oguri}
\affiliation{Center for Frontier Science, Chiba University, 1-33 Yayoi-cho, Inage-ku, Chiba 263-8522, Japan}
\affiliation{Department of Physics, Graduate School of Science, Chiba University, 1-33 Yayoi-Cho, Inage-Ku, Chiba 263-8522, Japan}

\author[0000-0001-9011-7605]{Yoshiaki Ono}
\affiliation{Institute for Cosmic Ray Research, The University of Tokyo, 5-1-5 Kashiwanoha, Kashiwa, Chiba 277-8582, Japan}

\author[0000-0003-2965-5070]{Kimihiko Nakajima}
\affiliation{National Astronomical Observatory of Japan, 2-21-1 Osawa, Mitaka, Tokyo 181-8588, Japan}

\author[0000-0001-7730-8634]{Yuki Isobe}
\affiliation{Institute for Cosmic Ray Research, The University of Tokyo, 5-1-5 Kashiwanoha, Kashiwa, Chiba 277-8582, Japan}
\affiliation{Department of Physics, Graduate School of Science, The University of Tokyo, 7-3-1 Hongo, Bunkyo, Tokyo 113-0033, Japan}

\author{Hiroya Umeda}
\affiliation{Institute for Cosmic Ray Research, The University of Tokyo, 5-1-5 Kashiwanoha, Kashiwa, Chiba 277-8582, Japan}
\affiliation{Department of Physics, Graduate School of Science, The University of Tokyo, 7-3-1 Hongo, Bunkyo, Tokyo 113-0033, Japan}

\author[0000-0003-4985-0201]{Ken Mawatari}
\affiliation{National Astronomical Observatory of Japan, 2-21-1 Osawa, Mitaka, Tokyo 181-8588, Japan}

\author[0000-0003-3817-8739]{Yechi Zhang}
\affiliation{Institute for Cosmic Ray Research, The University of Tokyo, 5-1-5 Kashiwanoha, Kashiwa, Chiba 277-8582, Japan}
\affiliation{Department of Astronomy, Graduate School of Science, The University of Tokyo, 7-3-1 Hongo, Bunkyo, Tokyo 113-0033, Japan}

\begin{abstract}
We conduct a comprehensive study on dropout galaxy candidates at $z\sim9-\redc{16}$ using the first 90 arcmin$^2$ JWST/NIRCam images taken by the early release observations (ERO) and early release science (ERS) programs. With the JWST simulation images, we find that a number of foreground interlopers are selected with a weak photo-$z$ determination ($\Delta\chi^2>4$). We thus carefully apply a secure photo-$z$ selection criterion ($\Delta\chi^2>9$) and conventional color criteria with confirmations of the ERO NIRSpec spectroscopic redshifts, and obtain a total of \redc{23} dropout galaxies at $z\sim9-\redc{16}$, including two candidates at \redc{$z_\mathrm{phot}=16.25_{-0.46}^{+0.24}$ and $16.41_{-0.55}^{+0.66}$}. We perform thorough comparisons of dropout galaxies found in our work with recent JWST studies, and conclude that our galaxy sample is reliable enough for statistical analyses. We derive the UV luminosity functions at $z\sim9-\redc{16}$, and confirm that our UV luminosity functions at $z\sim 9$ and $12$ agree with those determined by other HST and JWST studies. The cosmic star-formation rate density decreases from $z\sim9$ to $12$, and perhaps to $16$, but the densities at $z\sim12-16$ are higher than the constant star formation efficiency model. Interestingly, there are six bright galaxy candidates at \redc{$z\sim10-\redc{16}$} with $M_\mathrm{UV}<-19.5$ mag and $M_*\sim10^{8-9} M_\odot$. Because a majority ($\sim80$\%) of these galaxies show no signatures of AGNs in their morphologies, the high cosmic star-formation rate densities and the existence of these UV-luminous galaxies are explained by no suppression of star-formation by the UV background radiation at the pre-reionization epoch and/or an efficient UV radiation production by a top-heavy IMF with Population III-like star formation.
\end{abstract}

\keywords{%
galaxies: formation ---
galaxies: evolution ---
galaxies: high-redshift 
}

\section{Introduction}\label{ss_intro}

%

%
One of the most important goals in astronomy today is to
understand galaxy formation 
from their birth stage to current stage \citep{2016ARA&A..54..761S,2018PhR...780....1D,2020ARA&A..58..617O,2021arXiv211013160R}.
To accomplish the goal, observations for
present galaxies to first galaxies
are key to revealing the entire process of
galaxy formation, while observations of 
early high redshift galaxies, especially first galaxies,
are missing \citep[e.g.,][]{2011ApJ...740...13Z,2022MNRAS.513.5134N}.

Over the past 2--3 decades, large telescopes have driven
observational studies of galaxy formation 
with millions of galaxies at a redshift up to $z\sim 10$
since the start of deep imaging observations represented by the
legendary Hubble Deep Field project with Hubble Space Telescope (HST; \citealt{1996AJ....112.1335W}). 
%
%
%
To date, deep-field imaging observations have reached
detection limits of $\simeq 30$ mag in the wavelength range
of $0.4-1.6\ \mu$m with HST/ACS and WFC3 instruments in the Hubble Ultra Deep Field (HUDF; \citealt{2006AJ....132.1729B}, see \citealt{2021AJ....162...47B} and references therein) with the moderately deep ultra-violet (UV) extension, UVUDF 
\citep[$0.2-0.4\ \mu$m;][]{2011ApJS..193...27W,2013AJ....146..159T}. Albeit with shallower detection limits of $\sim 26-29$ mag, HST GOODS, COSMOS, and CANDELS, and the associated parallel-field programs have covered a total area of square degrees in the blank fields \citep{2004ApJ...600L..93G,2007ApJS..172...38S,2011ApJS..197...35G,2011ApJS..197...36K}.
Complementary ground-based observations of Subaru Hyper Suprime-Cam survey have been completed optical imaging covering $\sim1000$ deg$^2$ with $\sim 26$ mag depth (\citealt{2022PASJ...74..247A}; see also \citealt{2022ApJS..259...20H}), while the ground-based near-infrared ($1-2\ \mu$m) and Spitzer Space Telescope imaging ($3-8\ \mu$m) are limited to a total of few deg$^2$ with the similar or moderate depths of $\sim 25-26$ mag.
Beyond deep imaging in blank fields, the HST programs, Hubble Frontier Fields (HFF) and Reionization Lensing Cluster Survey (RELICS), target 6 and 41 massive clusters, respectively, with depths of $\sim 26-29$ mag to study faint high redshift galaxies behind the clusters, exploiting gravitational lensing magnification 
\citep{2017ApJ...837...97L,2019ApJ...884...85C}.
%
These deep imaging data provide more than 4 million photometrically-selected dropout galaxies at $z\sim 4-10$ \citep{2021AJ....162...47B,2022ApJS..259...20H} and up to $z\sim 13$ \citep{2022ApJ...929....1H}. Albeit with much small high redshift galaxy samples, spectroscopic observations confirm galaxies up to $z=9.1$ with ALMA \citep{2018Natur.557..392H} and $z=11.0$ by HST/WFC3 grism and Keck/MOSFIRE spectroscopy
\citep{2016ApJ...819..129O,2021NatAs...5..256J}.
Star formation in even higher redshift ($z\gtrsim14$) is discussed based on Balmer break galaxy candidates at $z\sim6$ \citep{2020ApJ...889..137M}.

%
With the galaxy samples photometrically selected in the rest-frame UV wavelengths, a number of studies have derived rest-frame UV luminosity functions
reaching up to $z\sim 10-13$. The UV luminosity functions show the redshift evolution
from $z\sim 3$ to $10$ 
with a decrease of the normalization $\phi^*$
and an increase of the faint-end slope $\alpha$, and no evolution of characteristic luminosity $L^*$ 
on the basis of the Schechter function parameterization
\citep{2015ApJ...803...34B,2021AJ....162...47B,2015ApJ...810...71F,2018ApJ...854...73I}.
At $z\sim 4$ and above, there are claims that
the bright-end of the UV luminosity function is explained with the double power-law function, but not with the Schechter function, due to the excessive number of
bright galaxies 
\citep{2014MNRAS.440.2810B,2020MNRAS.493.2059B,2018PASJ...70S..10O,2018ApJ...863...63S,2022ApJS..259...20H}.
Such bright galaxy population includes the galaxies with
spectroscopic redshifts at $z\sim 10$ 
\citep{2016ApJ...819..129O,2021NatAs...5..256J}
and perhaps galaxy candidates at $z\sim 13$ 
\citep{2022ApJ...929....1H},
while it is not clearly concluded with the sufficient statistical accuracy and the spectroscopic confirmations
\citep[e.g.,][]{2019ApJ...883...99S,2020ApJ...891..146R,2020MNRAS.493.2059B}.
%
%

Over the cosmological volumes, 
the redshift evolution of cosmic star formation rate (SFR) density
is revealed with the UV luminosity function measurements, 
and shows a monotonic decrease from $z\sim 3$ to $z\sim 10$ 
with a small contribution of dusty starbursts at $z\gtrsim 6$
\citep[e.g.,][]{2022arXiv220511526B,2022arXiv220714733B}.
The UV luminosity function measurements provide the physical picture of galaxy formation over 
the redshift range of $z\sim 0-10$, tying galaxies and dark-matter halos via abundance matching techniques 
(e.g. \citealt{2013ApJ...770...57B,2019MNRAS.488.3143B,2013MNRAS.428.3121M,2018MNRAS.477.1822M,2015ApJ...814...95F}).
There is an increasing trend of stellar-to-halo mass ratio towards high redshift for a given halo mass
\citep[e.g.,][]{2013ApJ...770...57B,2019MNRAS.488.3143B,2016ApJ...821..123H,2018PASJ...70S..11H}, 
which is consistent with the original idea of the galaxy-formation downsizing picture \citep{1988ApJ...332L..29C}.
The galaxy-dark matter halo connection probed by the clustering analysis indicates that the star formation efficiency, defined by the ratio of the SFR to the dark matter accretion rate, $SFR/\dot{M}_\m{h}$, is almost constant across the redshift of $z\sim2-7$ given the dark matter halo mass \citep{2018PASJ...70S..11H,2022ApJS..259...20H}, and the constant star formation efficiency model can reproduce the evolutional trend of the cosmic SFR density (e.g., \citealt{2010ApJ...718.1001B}, \citealt{2015ApJ...813...21M}, \citealt{2018PASJ...70S..11H,2022ApJS..259...20H}, \citealt{2018ApJ...868...92T}).

The UV luminosity function measurements, especially at the faint end, are clue to understanding galaxy formation 
\citep{2016MNRAS.463.1968Y}
as well as cosmic reionization \citep{2021arXiv211013160R}, 
where abundant faint star-forming galaxies are thought to be sources of cosmic reionization. 
The faint-end ($\gtrsim -15$ mag) UV luminosity function 
at $z\sim 6-10$ is probed with galaxies 
behind massive clusters such with the HFF data
via gravitational lensing magnification
\citep{2015ApJ...814...69A,2018MNRAS.479.5184A,2015ApJ...799...12I,2018ApJ...854...73I,2017ApJ...835..113L,2016ApJ...820...98L,2018ApJ...855..105O}, while the faint-end slopes and luminosity function turnovers are poorly constrained, due to the limited statistics and lensing magnification systematics
\citep{2017ApJ...843..129B,2022arXiv220511526B,2018ApJ...855....4K,2022MNRAS.514.1148Y}.
%

%
Here the James Webb Space Telescope (JWST) was launched in the end of 2021, and just started its operation in the early 2022. The first data sets of JWST were released on July 12, 2022, taken by the early release observations (ERO) whose targets include a massive cluster SMACS J0723.3-7327 (SMACS J0723, $z=0.39$)
and Stephan's Quintet. 
The ERO imaging data taken with NIRCam \citep{2005SPIE.5904....1R}
are deep enough to detect high redshift galaxies 
with the depths of $\sim 30$ mag, and multi-band data covering
$\gtrsim 2\mu$m wavelengths allow us to detect galaxies at 
the previously unreachable redshift range up to $z\sim 20$.
Rest-frame optical emission at $z\gtrsim 10$ is 
redshifted to the mid infrared bands and can be covered with 
the Mid-Infrared Instrument (MIRI; \citealt{2015PASP..127..612B}).
The ERO spectroscopic data of NIRSpec \citep{2022A&A...661A..80J}
taken in the multi-object spectroscopy mode 
confirmed galaxies up to $z=8.5$
with rest-frame optical lines in the $2-5\ \mu$m wavelengths.
The slit-less spectroscopy of NIRISS \citep{2012SPIE.8442E..2RD}
supplement spectroscopic redshift determinations 
in the wavelength range of $\sim 1-2\ \mu$m.
All of these data sets are revolutionizing the galaxy formation
studies. The JWST observatory subsequently releases 
the director's discretionary early release science (ERS) data
that include NIRCam, NIRSpec, and NIRISS data taken by
the ERS programs of 
Cosmic Evolution Early Release Science 
(CEERS; \citealt{2017jwst.prop.1345F,2022arXiv221105792F}) and 
GLASS James Webb Space Telescope Early Release Science 
(GLASS; \citealt{2022arXiv220607978T}).
Further releases will deliver data of Cycle 1 observations
that include Public Release IMaging for Extragalactic Research (PRIMER; \citealt{2021jwst.prop.1837D}), UNCOVER \citep{2021jwst.prop.2561L}, and COSMOS-Webb \citep{2021jwst.prop.1727K} once the observations complete. 
Programs of guaranteed time observations (GTO), such as
JWST Advanced Deep Extragalactic Survey (JADES; \citealt{2020IAUS..352..342B})
will be also completed in the early years.

This is a great development of observational astronomy,
presenting the unprecedentedly deep and 
high quality data covering the infrared band ($>2\mu$m).
In fact, after the releases of the ERO and ERS data sets, 
we find the explosive progresses of galaxy formation studies.
The mass models of the ERO target cluster, SMACS J0723, are
improved with the NIRCam imaging and NIRSpec spectroscopic data
\citep{2022arXiv220707101M,2022arXiv220707102P,2022arXiv220707567C}.
High redshift galaxies are searched in the ERO SMACS J0723 and ERS CEERS and GLASS fields, and are identified at $z\sim 9-20$
 \citep{2022arXiv220709434N,2022arXiv220709436C,2022arXiv220515388L,2022arXiv220711217A,2022arXiv220711671M,2022arXiv220712338A,2022arXiv220712356D,2022arXiv220712474F,2022arXiv220711558Y}.
The morphological properties are investigated
with the NIRCam images of the ERO SMACS J0723 and the CEERS observations via the comparisons of HST images for galaxies at $z\sim 3-6$ \citep{2022arXiv220709428F} and 
the rest-frame optical and near-infrared bands
for galaxies at $z\sim 1-2$ \citep{2022arXiv220710655S}, respectively.
The infrared photometric properties of galaxies at $z\sim 1-2$ are studied with the NIRCam and MIRI images of the ERO SMACS J0723 observations in conjunction with the ALMA archival data \citep{2022arXiv220708234C}.
The ERO NIRSpec observations in SMACS J0723 
provide high-quality spectra that allow to
identify 10 galaxies at $z=1.2-8.5$, three of which reside at $z=7.7-8.5$ \citep{2022arXiv220708778C}, 
and to characterize the inter-stellar medium of 
the galaxies \citep{2022arXiv220710034S,2022arXiv220712375C}.
NIRISS spectroscopic data complements the NIRSpec observations and provide spectroscopic sample of $z\sim1-8$ galaxies \citep{2022arXiv220711387R,2022arXiv220713113W,2022arXiv220713459B,2022arXiv220713625M}.
More JWST results for galaxy formation are being
actively reported.

In this paper, we present a comprehensive study on high redshift galaxies using the first JWST/NIRCam datasets taken by the ERO and ERS programs.
The deep infrared imaging data taken with NIRCam allow us to search for galaxies at $z\gtrsim9$, and to constrain the UV luminosity function and the cosmic SFR density in the universe 600 Myrs after the Big Bang.
We will also perform thorough comparisons of galaxies found in our work and recent JWST studies.

This paper is organized as follows. Section \ref{ss_data} presents
the JWST observational data sets used in this study.
Section \ref{ss_catalog_selection} explains our sample selection and 
galaxy photometry catalog. In Section \ref{ss_model},
we describe the mass model for the lensing cluster.
We show our main results of UV luminosity functions and cosmic SFR densities
in Section \ref{ss_LF}, and discuss the physical properties of early galaxies in Section \ref{ss_dis}.
Section \ref{ss_summary} summarizes our findings.
Throughout this paper, we use the Planck cosmological parameter sets of the TT, TE, EE+lowP+lensing+BAO result \citep{2020A&A...641A...6P}:
$\Omega_\m{m}=0.3111$, $\Omega_\Lambda=0.6899$, $\Omega_\m{b}=0.0489$, $h=0.6766$, and $\sigma_8=0.8102$.
All magnitudes are in the AB system \citep{1983ApJ...266..713O}.

\begin{deluxetable*}{ccccccccccc}
\setlength{\tabcolsep}{0.1cm}
\tablecaption{Limiting Magnitudes of the JWST Data}
\tablehead{
\colhead{} & \colhead{Area} & & \multicolumn{8}{c}{$5\sigma$ Limiting Magnitude} \\
\cmidrule{2-2}  \cmidrule{4-11} 
\colhead{Field} & \colhead{($\m{arcmin^2}$)} & & \colhead{$F090W$} & \colhead{$F115W$} & \colhead{$F150W$} & \colhead{$F200W$} & \colhead{$F277W$} & \colhead{$F356W$} & \colhead{$F410M$}& \colhead{$F444W$} \\
\colhead{(1)}& \colhead{(2)}& & \colhead{(3)}& \colhead{(4)} &  \colhead{(5)}& \colhead{(6)}& \colhead{(7)}& \colhead{(8)}& \colhead{(9)}& \colhead{(10)}}

\startdata
SMACS J0723 & 11.0 && 29.4 & \nodata & 29.4 & 29.6 & 29.8 & 29.9 & \nodata & 29.6 \\
GLASS & 6.8 && 29.5 & 29.6 & 29.4 & 29.6 & 29.6 & 29.9 & \nodata & 29.6 \\
CEERS1 & 8.4 && \nodata & 29.3 & 29.1 & 29.3 & 29.5 & 29.7 & 28.9 & 29.1 \\
CEERS2 & 8.5 && \nodata & 29.3 & 29.0 & 29.7 & 29.5 & 29.6 & 28.9 & 29.4 \\
CEERS3 & 8.4 && \nodata & 29.4 & 29.2 & 29.4 & 29.6 & 29.7 & 29.0 & 29.2 \\
CEERS6 & 8.4 && \nodata & 29.4 & 29.1 & 29.3 & 29.5 & 29.7 & 29.0 & 29.0 \\
Stephan's Quintet & 37.2 && 27.7 & \nodata & 27.9 & 28.1 & 28.8 & 28.9 & \nodata & 28.6 \\
\hline
PSF FWHM && & 0.\carcsec06 & 0.\carcsec07 & 0.\carcsec07 & 0.\carcsec08 & 0.\carcsec13 & 0.\carcsec14 & 0.\carcsec16 & 0.\carcsec16 
\enddata

\tablecomments{Columns: (1) Field. (2) Effective area in $\m{arcmin^2}$. (3)-(12) Typical limiting magnitudes which correspond to $5 \sigma$ variations in the sky flux measured with a circular aperture of $0.\carcsec2$-diameter in the deepest region.}
\label{tab_jwst_limitmag}
\end{deluxetable*}

\section{Observational Dataset}\label{ss_data}

\subsection{JWST/NIRCam Data}

\begin{figure}
\centering
\begin{minipage}{0.48\hsize}
\begin{center}
\includegraphics[width=0.99\hsize, bb=6 8 352 349]{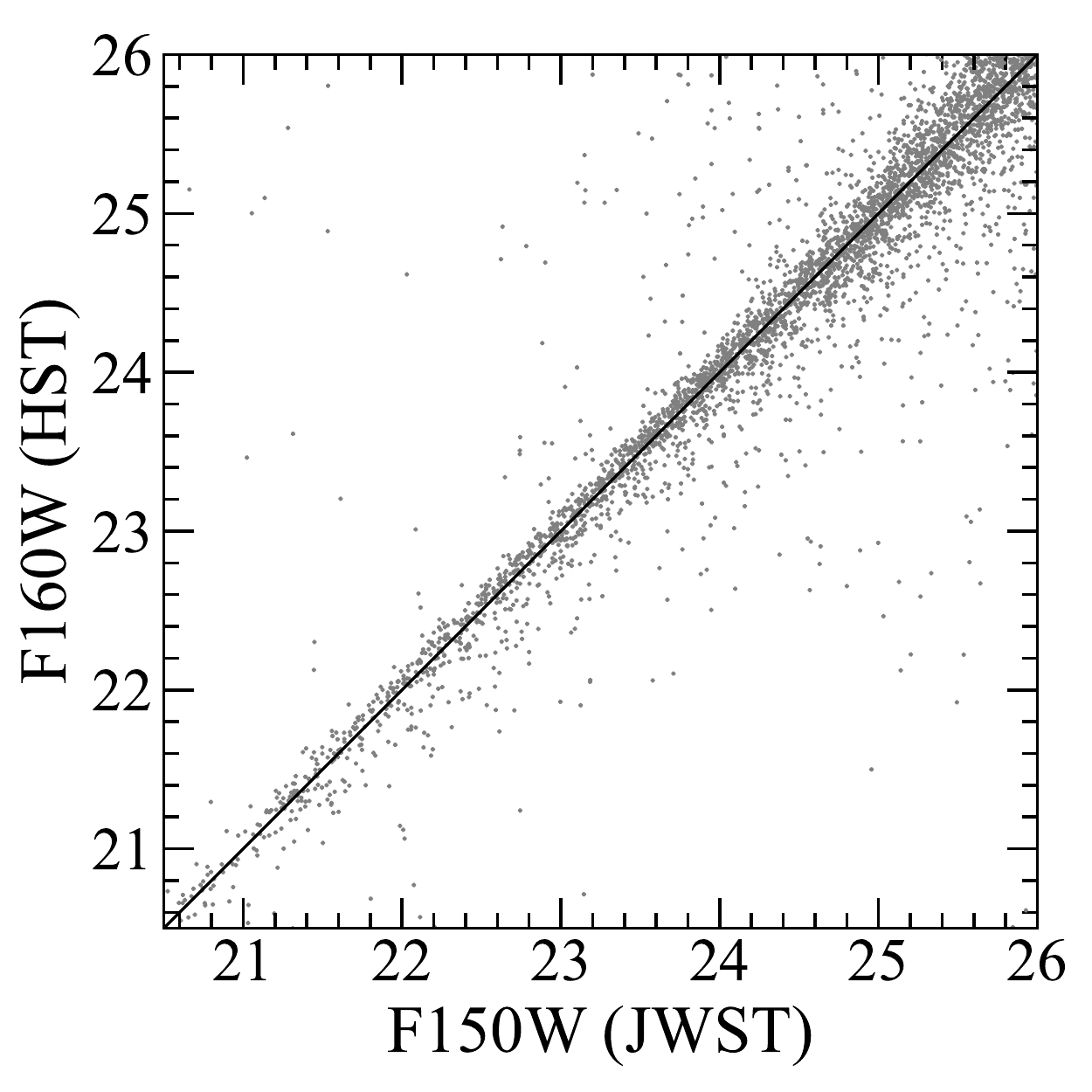}
\end{center}
\end{minipage}
\begin{minipage}{0.48\hsize}
\begin{center}
\includegraphics[width=0.99\hsize, bb=6 8 352 349]{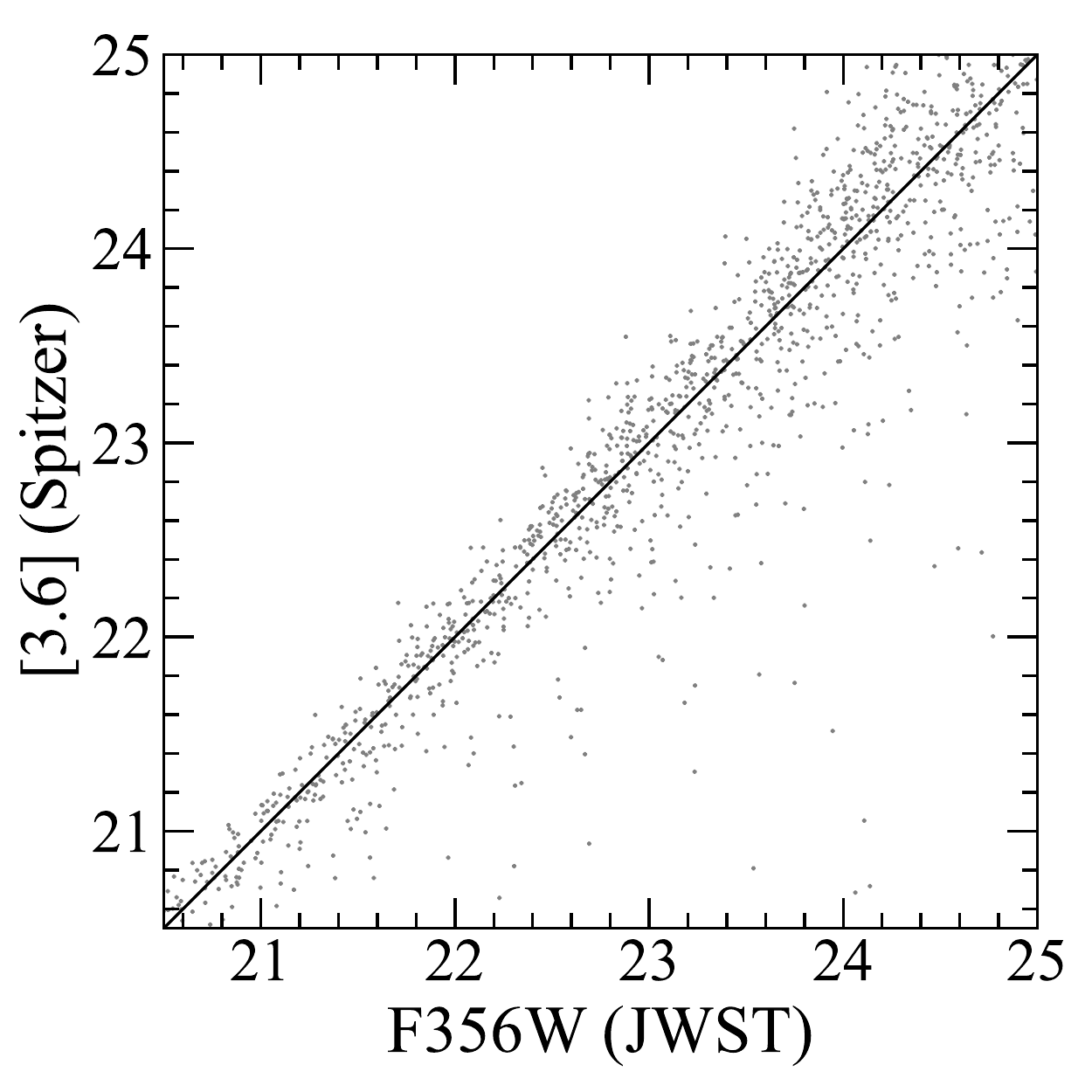}
\end{center}
\end{minipage}
\caption{Comparison of magnitudes.
Magnitudes measured in the JWST $F150W$-band (left) and $F356W$-band (right) are compared with those in the HST $F160W$-band and Spitzer [3.6]-band, respectively.
The measured magnitudes agree well with those in the HST and Spitzer images within $\sim10\%$, indicating that the flux is reasonably calibrated.
Note that we include a $10\%$ error floor on all measured fluxes to account for possible systematic uncertainties.
}
\label{fig_mag_mag}
\end{figure}

We use four JWST NIRCam datasets obtained in the ERO and ERS programs, ERO SMACS J0723, ERO Stephan's Quintet, ERS CEERS, and ERS GLASS (Table \ref{tab_jwst_limitmag}).
The total area is $\sim90\ \m{arcmin^2}$.
\redc{We retrieved law data (\_uncal.fits) from the MAST archive and reduced the data using the JWST pipeline version 1.6.3 development version (1.6.3.dev34+g6889f49, almost the same as 1.7.0).
We use the Calibration Reference Data System (CRDS) context file of {\tt jwst\_0995.pmap} released in October, whose calibration values were derived using calibration observations of three different standard stars placed in all of the 10 NIRCam detectors.
These new flux calibrations were verified using imaging of the globular cluster M92 \citep{2022RNAAS...6..191B}.
In addition to the standard reduction, we added some processes to obtain better reduced images as follows.}
Before the Stage 2 calibration, we subtracted stray light features called ``wisps" by using a script provided by the NIRCam team\footnote{https://jwst-docs.stsci.edu/jwst-near-infrared-camera/nircam-features-and-caveats/nircam-claws-and-wisps}, and removed striping by using a script provided in the CEERS team  \citep{2022arXiv221102495B}\footnote{https://ceers.github.io/releases.html\#sdr1}.
We ran the {\tt SkyMatch} step individually on each frame of Stage 2 calibrated data before Stage 3 calibration, following a suggestion by the CEERS team \citep{2022arXiv221102495B}.
The images were pixel-aligned with a pixel scale of $0.\carcsec 015/$pixel, except for ones in the Stephan's Quintet field with a scale of $0.\carcsec 03/$pixel to reduce image sizes. 
Because the pipeline-processed images still showed a gradient of the sky background, we further subtracted the sky background using {\sc SExtractor} \citep{1996A&AS..117..393B}.
The intracluster light around the cluster center of SMACS J0723 is also removed in this process, \redc{although more sophisticated processes are sometimes employed \citep[e.g.,][]{2017ApJ...835..113L,2021ApJS..256...27P}.}
\redc{Note that our galaxy samples are not severely affected by systematics due to the intracluster light removal, because as shown in Section \ref{ss_sample}, the only candidate selected in the SMACS J0723 field, SM-z12-1, is located in the parallel field.
The results of luminosity functions do not change beyond the errors if we remove the datapoint estimated with the SMACS J0723 data.}
Finally, we corrected for an astrometric offset between each detector and band using iraf tasks {\tt geomap} and {\tt geotran}.
To check the reliability of the flux calibration, we compare our measured magnitudes in the JWST images with those in the HST and Spitzer images.
As shown in Figure \ref{fig_mag_mag}, the measured fluxes are almost consistent with those in the HST and Spitzer images, indicating that the flux is reasonably calibrated.

The limiting magnitudes were measured in $0.\carcsec1$, $0.\carcsec2$, and $0.\carcsec3$-diameter circular apertures by randomly placing apertures in sky areas using Python packages {\sc Astropy/photutils}.
Sky areas were defined as pixels without objects detected by {\sc SExtractor}.
We measured the limiting magnitudes in bins of the weight values to take into account inhomogeneity of the depth. 
Using the weight map, we masked some regions around the edge of the detectors whose exposure time is short.
We also measured a full-widths at half-maximum (FWHM) of the point spread function (PSF) in each image by selecting stellar objects in the magnitude-FWHM diagram.
The measured limiting magnitudes in a $0.\carcsec2$-diameter circular aperture, effective areas, and typical FWHMs of the PSFs are presented in Table \ref{tab_jwst_limitmag}.
\redc{Here the effective area is defined as an area that is observed with all available bands before the foreground removal.
The effect of the foreground will be taken into account in the completeness estimate (Section \ref{ss_completeness}).}
In the following sections we detail our observational dataset in each field.

\subsubsection{ERO: SMACS J0723}

A massive galaxy cluster at $z=0.39$, SMACS J0723, was deeply observed with NIRCam, NIRSpec, MIRISS, and MIRI in the ERO (ERO-2736). 
The NIRCam images were taken in the six bands of $F090W$, $F150W$, $F200W$, $F277W$, $F356W$, and $F444W$, covering $11.0\ \m{arcmin^2}$.
The exposure time in each filter is $\sim7500$ seconds, and the $5\sigma$ limiting magnitude in the $F356W$ band is 29.9 mag.

\subsubsection{ERO: Stephan's Quintet}

Stephan's Quintet, a group of five local galaxies, was observed with NIRCam and MIRI in the ERO (ERO-2732).
The NIRCam images were taken in the six bands of $F090W$, $F150W$, $F200W$, $F277W$, $F356W$, and $F444W$, covering $42\ \m{arcmin^2}$. 
We have masked central regions of the field that is affected by the five local galaxies in Stephan's Quintet, resulting in an effective area of $37.2\ \m{arcmin^2}$, corresponding to $\sim4$ NIRCam pointings.
The exposure time in each filter is roughly $\sim1200$ seconds, and the $5\sigma$ limiting magnitude in the $F356W$ band is 28.9 mag.

\subsubsection{ERS: CEERS}

A part of the HST/CANDELS Extended Groth Strip (EGS) field is observed with JWST in the Cosmic Evolution Early Release Science (CEERS) survey (ERS-1345; \citealt{2017jwst.prop.1345F}, \citealt{2022arXiv221105792F}). 
We use four pointing datasets of NIRCam obtained in June 2022 with the seven bands of $F115W$, $F150W$, $F200W$, $F277W$, $F356W$, $F410M$, and $F444W$, covering a total of $33\ \m{arcmin^2}$.
The exposure time in each filter is $\sim2800-6200$ seconds, and the $5\sigma$ limiting magnitude in the $F356W$ band is 29.7 mag.
Since the exposure times are not uniform across the four NIRCam pointings, we separately analyze the four pointing data. 

\subsubsection{ERS: GLASS}

A massive galaxy cluster, Abell 2744, was observed with JWST in the ERS program of Through the Looking GLASS (ERS-1324; \citealt{2017jwst.prop.1324T,2022arXiv220607978T}).
Deep NIRCam images were taken in June 2022 in a parallel mode of NIRISS observations targeting the center of the cluster, in seven bands $F090W$, $F115W$, $F150W$, $F200W$, $F277W$, $F356W$, and $F444W$, covering $6.8\ \m{arcmin^2}$.
The exposure time in each filter is $\sim5600-23000$ seconds, and the $5\sigma$ limiting magnitude in the $F356W$ band is 29.9 mag.
The lensing magnification is negligible in the field of the NIRCam observations that is $\sim5$ arcmin away from the cluster center of Abell 2744.

\subsection{JWST/NIRSpec}\label{ss_obs_spec}

We use publicly available data from the ERO NIRSpec observations targeting the field of the SMACS J0723 cluster (ERO-2736).
The NIRSpec observations consist of two pointings with the same Multi Shutter Array (MSA) configuration.
NIRSpec observations were carried out by using the disperser-filter combinations of G235M/F170LP and G395M/F290LP, which cover the wavelength range from 1.66 $\mu$m to 5.16 $\mu$m with a spectral resolution of $R\sim1000$.
The total exposure time of the two individual pointings is 8,840 seconds for each grating.
The NIRSpec observations have taken spectra for a total of 35 objects, three of which, s04590 ($z_\m{spec}=8.495$), s06355 ($z_\m{spec}=7.664$), and s10612 ($z_\m{spec}=7.659$), are securely identified at $z>7$ whose dropouts can be covered with the NIRCam $F090W$ band.
See K. Nakajima et al. in prep. for details of the reduction and analysis of the NIRSpec data.

\subsection{HST/ACS and WFC3}

HST multi-band images are available in the fields of SMACS J0723 and CEERS (EGS).
We downloaded HST ACS and WFC3 images in the SMACS J0723 and CEERS fields from the websites of RELICS (\citealt{2019ApJ...884...85C})\footnote{https://archive.stsci.edu/prepds/relics/\#dataaccess} and CEERS\footnote{https://ceers.github.io/releases.html\#hdr1}, respectively.
We found a small offset (\redc{$\sim0.\carcsec 2$}) of the WCS between the JWST and HST images.
In this paper we use coordinates of the JWST images.
  
\section{Photometric Catalog and Sample Selection}\label{ss_catalog_selection}

\subsection{Photometric Catalog}\label{ss_catalog}

We construct multi-band source catalogs from the JWST data to select the $F115W$, $F150W$, and $F200W$-dropout galaxies.
We use {\sc SWarp} \citep{2002ASPC..281..228B} to produce our detection image that is a weighted mean image of the bands redder than the Lyman break in each dropout selection \redc{(i.e., $F150W$, $F200W$, $F277W$, $F356W$, $F410M$, and $F444W$ for the $F115W$-dropout selection, $F200W$, $F277W$, $F356W$, $F410M$, and $F444W$ for the $F150W$-dropout selection, and $F277W$, $F356W$, $F410M$, and $F444W$ for the $F200W$-dropout selection)}.
To measure object colors, we match the image PSFs to the F444W-band images whose typical FWHM of the PSF is $\simeq 0.\carcsec16$, the largest of the NIRCam multi-band images.

\begin{figure*}
\centering
\begin{minipage}{0.32\hsize}
\begin{center}
\includegraphics[width=0.99\hsize, bb=16 5 343 356]{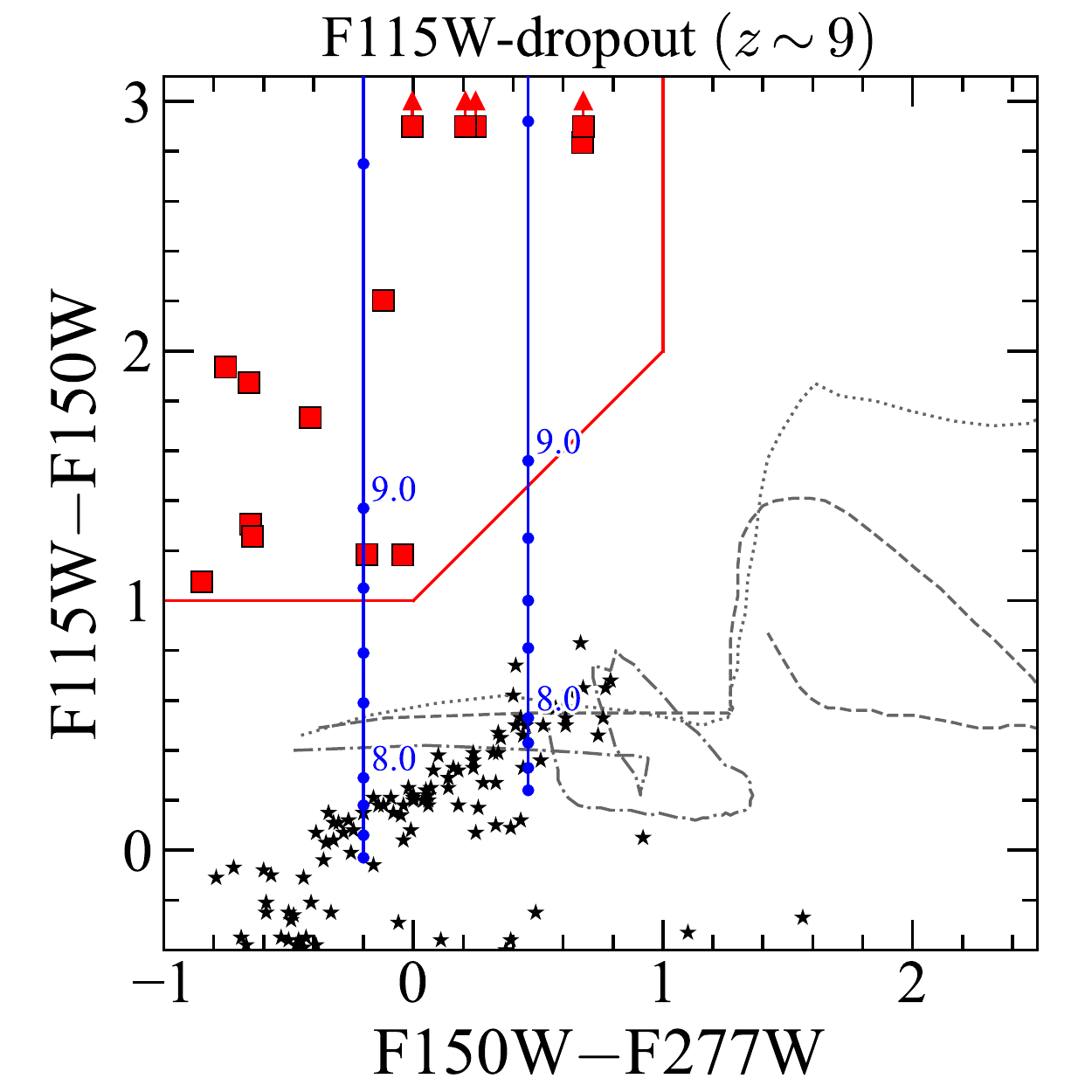}
\end{center}
\end{minipage}
\begin{minipage}{0.32\hsize}
\begin{center}
\includegraphics[width=0.99\hsize, bb=16 5 343 356]{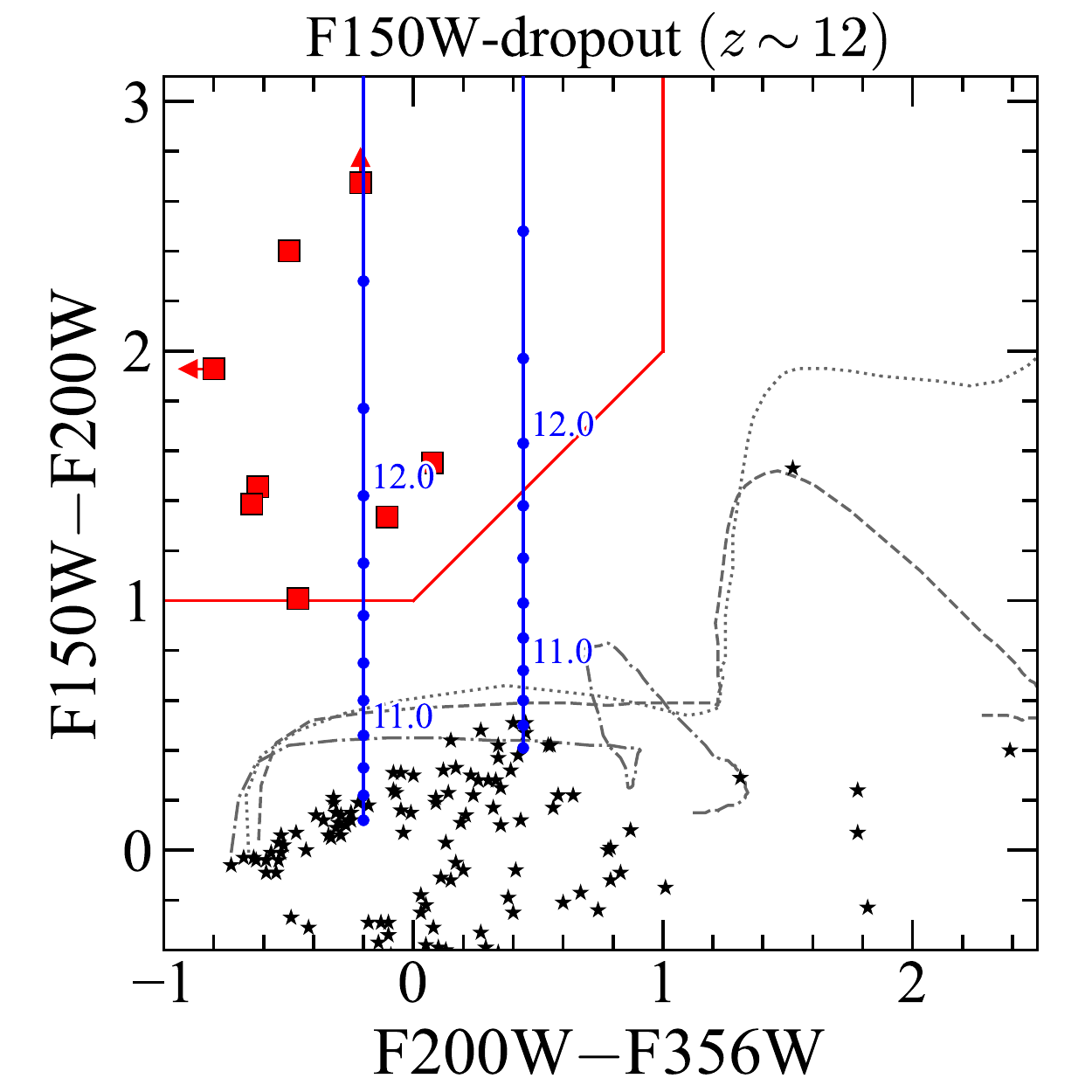}
\end{center}
\end{minipage}
\begin{minipage}{0.32\hsize}
\begin{center}
\includegraphics[width=0.99\hsize, bb=16 5 343 356]{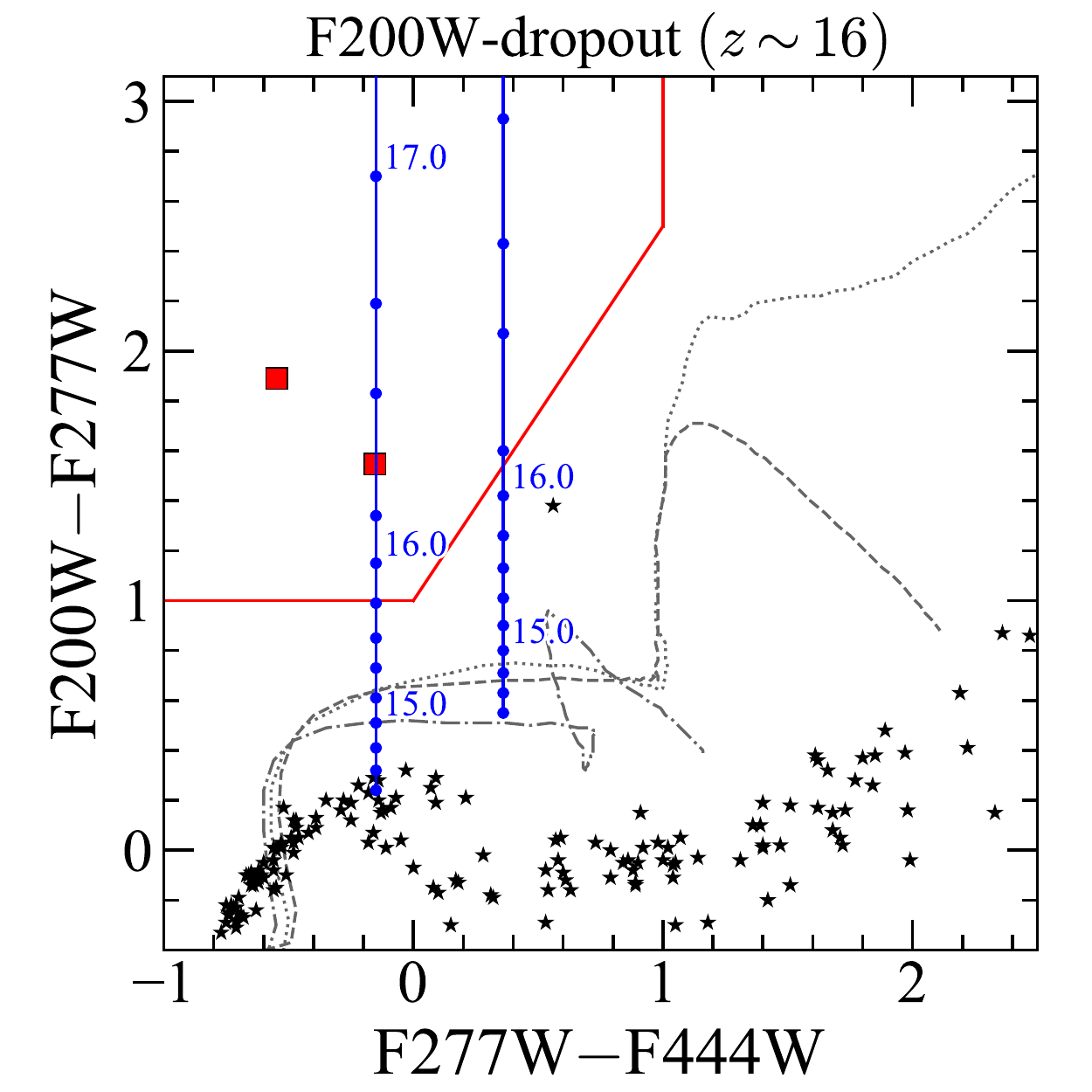}
\end{center}
\end{minipage}
\caption{Two-color diagrams of $F115W-F150W$ vs. $F150W-F277W$ (left), $F150W-F200W$ vs. $F200W-F356W$ (center), and $F200W-F277W$ vs. $F277W-F444W$ (right) corresponding to the color selections for $F115W$-dropouts at $z\sim 9$, $F150W$-dropouts at $z\sim 12$, and $F200W$-dropouts at $z\sim 16$, respectively. The red squares represent our dropout galaxy candidates that meet the color selection criteria indicated with the red lines. The blue lines denote colors of the dropout galaxy models with UV spectral slopes of $\beta_\m{UV}=-2.3$ and $-1.3$ whose redshifts are indicated with the numbers and the blue circles with an interval of $\Delta z=0.2$.  
The black dotted, dashed, dot-dashed lines show colors of typical elliptical, Sbc, and irregular galaxies \citep{1980ApJS...43..393C} redshifted from $z=0$ to $7$. The star marks present expected colors of Galactic dwarf stars \citep{2006ApJ...651..502P,2011ApJS..197...19K}.
These colors of dwarf stars are estimated by interpolating available flux measurements obtained by ground based telescopes ($J$, $H$, and $K$-bands) and the Spitzer telescope ([3.6] and [4.5]).
}
\label{fig_2color}
\end{figure*}

We perform source detection and photometry with {\sc SExtractor} \citep[version 2.5.0;][]{1996A&AS..117..393B}.
We found that the photometry with {\sc SExtactor} {\tt MAG\_AUTO} performs better as a total magnitude than that with photutils {\tt isophotal\_flux}, which is one of the outputs of the JWST calibration pipeline, using CEERS simulated images with a mock galaxy catalog created with the Santa Cruz Semi-Analytic Model \citep{2021MNRAS.502.4858S,2022arXiv220613521Y}.\footnote{https://ceers.github.io/releases.html\#sdr1}
We run {\sc SExtractor} in dual-image mode for each image with its detection image, having the parameter set as follows: ${\tt DETECT\_MINAREA}=5$, ${\tt DETECT\_THRESH}=3.0$, ${\tt ANALYSIS\_THRESH}=3.0$, ${\tt DEBLEND\_NTHRESH}=32$, and ${\tt DEBLEND\_MINCOUNT}=0.005$.
The total number of the objects detected is $\sim$250,000.
We measure the object colors with the {\tt MAG\_APER} magnitudes defined in a $0.\carcsec3$-diameter circular aperture in the PSF-matched images.
Source detections are evaluated with $0.\carcsec1$ and/or $0.\carcsec2$-diameter circular apertures in the original (not PSF-matched) images.
The total magnitudes are estimated from the $0.\carcsec3$-diameter aperture magnitudes with the aperture correction.
The value of the aperture correction is defined as the difference between the {\tt MAG\_AUTO} magnitude and the $0.\carcsec3$-diameter aperture magnitude in an image of the weighted mean of the PSF-matched images whose wavelengths are longer than the Lyman break \redc{(i.e., $F150W$, $F200W$, $F277W$, $F356W$, $F410M$, and $F444W$ for the $F115W$-dropout selection, $F200W$, $F277W$, $F356W$, $F410M$, and $F444W$ for the $F150W$-dropout selection, and $F277W$, $F356W$, $F410M$, and $F444W$ for the $F200W$-dropout selection)}.
Furthermore, we correct for a small offset (\redc{$\sim0.1$ mag}) between the measurement of {\tt MAG\_AUTO} and the true total magnitude due to the wing of the PSF not captured with {\tt MAG\_AUTO} (see Sections 2.2 and 2.5.1 in \citealt{2022ApJ...928...52F}).
We measure this offset by inserting mock galaxy images randomly in the real images, and measure the magnitudes using {\sc SExtractor} in a similar manner as the completeness simulation described later in Section \ref{ss_completeness}.
To account for systematic uncertainties of the flux measurements (e.g., zero-point correction), we include 10\% error floor on all measured fluxes.
Finally we correct for the galactic extinction using \citet{1998ApJ...500..525S} and \citet{2011ApJ...737..103S} and make final photometric catalogs.

\subsection{Dropout Selection}\label{ss_selection}

From the photometric catalogs constructed in Section \ref{ss_catalog}, we construct $z\sim9-\redc{16}$ dropout galaxy catalogs based on the Lyman break color selection technique \citep[e.g.,][]{1996ApJ...462L..17S,2002ARA&A..40..579G}.
As shown in Figure \ref{fig_2color}, galaxy candidates at $z\sim9$, $12$, and $\redc{16}$ can be selected by the $F115W$, $F150W$, and $F200W$-dropout selections,
respectively.

First, to identify secure sources, we select source whose signal-to-noise ratios (S/Ns) in a $0.\carcsec2$-diameter circular aperture are higher than 5 in the detection images.
We also require sources to be detected at $>3.5\sigma$ levels in at least two bands redder than the Lyman break.
We then select dropout galaxy candidates by using their broadband spectral energy distribution (SED) colors. 
We adopt the following color criteria: \\

$F115W$-dropout ($z\sim9$):
\begin{gather}
(F115W-F150W > 1.0) \land \label{eq_f115w_1}\\
(F150W-F277W < 1.0) \land \label{eq_f115w_2} \\
(F115W-F150W > (F150W-F277W) +1.0) \label{eq_f115w_3}
\end{gather}

$F150W$-dropout ($z\sim12$):
\begin{gather}
(F150W-F200W > 1.0) \land \label{eq_f150w_1}\\
(F200W-F356W < 1.0) \land \label{eq_f150w_2} \\
(F150W-F200W > (F200W-F356W) +1.0) \label{eq_f150w_3}
\end{gather}

$F200W$-dropout ($z\sim\redc{16}$):
\begin{gather}
(F200W-F277W > 1.0) \land \label{eq_f200w_1}\\
(F277W-F444W < 1.0) \land \label{eq_f200w_2} \\
(F200W-F277W > 1.5(F277W-F444W) +1.0)  \label{eq_f200w_3}
\end{gather}

We select sources with prominent breaks with the criteria of Equations (\ref{eq_f115w_1}), (\ref{eq_f150w_1}), and (\ref{eq_f200w_1}), and measure the slope of the continuum and remove red interlopers with Equations of (\ref{eq_f115w_2}-\ref{eq_f115w_3}), (\ref{eq_f150w_2}-\ref{eq_f150w_3}), and (\ref{eq_f200w_2}-\ref{eq_f200w_3}).
To measure the slope of the continuum, we use the bands that have a large wavelength difference as much as possible, while not biting the Balmer break in the redder band.
To remove foreground interlopers, we exclude sources with continuum detections at $>2\sigma$ levels in the $0.\carcsec 1$- or $0.\carcsec 2$-diameter apertures in bands bluer than the Lyman break, i.e., the $F090W$ band for the $F115W$-dropouts, $F090W$ and $F115W$ bands for the $F150W$-dropouts, and $F090W$, $F115W$, and $F150W$ bands for the F200W-dropouts.
To select reliable candidates, we restrict our dropout selections in fields where the bluer band than the Lyman break is available; the $F115W$-dropout selection is only performed in the GLASS field.
We also apply a criterion of {\sc SExtractor} stellarity parameter, {\tt CLASS\_STAR}, of $<0.9$, to remove stellar contaminants.
Finally, we visually inspect images of the selected sources to remove spurious sources or sources affected by nearby bright objects and diffraction spikes of bright stars.
\redc{We removed about half of the selected objects in this process.}
\redc{We also visually inspect HST images of the selected sources in the SMACS J0723 and CEERS fields to check whether the source is consistent with being a high redshift galaxy, although the HST images are typically $\sim1-2$ mag shallower than the JWST images in these fields.
}

\subsection{SED Fitting}\label{ss_SEDfit}


To further remove low redshift interlopers, we perform galaxy SED fitting 
with the flexible Bayesian inference code \textsc{prospector}
\citep{2021ApJS..254...22J}, and derive the photometric redshift.
Model spectra are derived from Flexible Stellar Population Synthesis 
\citep[FSPS;][]{2009ApJ...699..486C,2010ApJ...712..833C}
package
with the modules for Experiments in Stellar Astrophysics Isochrones and Stellar Tracks (MIST; \citealt{2016ApJ...823..102C}). The boost of ionizing flux production of massive stars
are included in the MIST isochrones \citep{2017ApJ...838..159C}.
Here we assume the stellar initial mass function (IMF) 
determined by \citet{2003PASP..115..763C}, the \citet{2000ApJ...533..682C} dust extinction law, and the intergalactic medium (IGM) attenuation model by \citet{1995ApJ...441...18M}.
\redc{Note that the choice of the IGM attenuation model does not affect our galaxy selection at $z\sim9-16$ because the flux bluer than the Ly$\alpha$ break is almost absorbed by the highly neutral IGM at these redshifts regardless of the choice of the IGM attenuation model.}
The Ly$\alpha$ emission line is also masked considering the high IGM neutral fraction at these redshifts.
We adopt a flexible star formation history with five bins that are spaced equally in logarithmic times between 0 Myr and a
lookback time that corresponds to $z=30$, where the SFR within each bin is constant.
We change the redshift, optical depth in the $V$-band, star formation history, and total stellar mass as free parameters, while fix the metallicity to $Z=0.2\ Z_\sun$.
We assume a continuity prior for the star formation history, and flat priors for other parameters in the range $0<z<20$, $0<\tau_\m{V}<2$, and $6<\m{log}(M_*/M_\sun)<12$.
We search for the best-fit model to the observed photometry with the MCMC method by using {\sc emcee} \citep{2013PASP..125..306F}.

\begin{figure*}
\centering
\begin{center}
\textbf{\huge Interlopers in Simulations}\par\medskip
\includegraphics[width=0.99\hsize, bb=12 1 707 216]{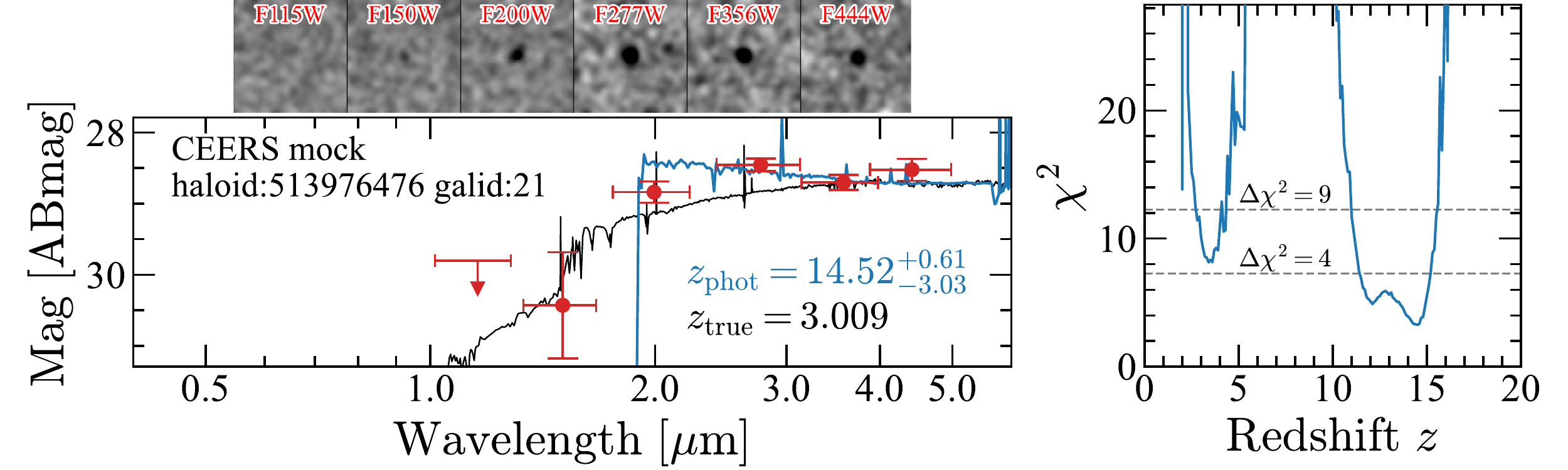}
\includegraphics[width=0.99\hsize, bb=12 1 707 216]{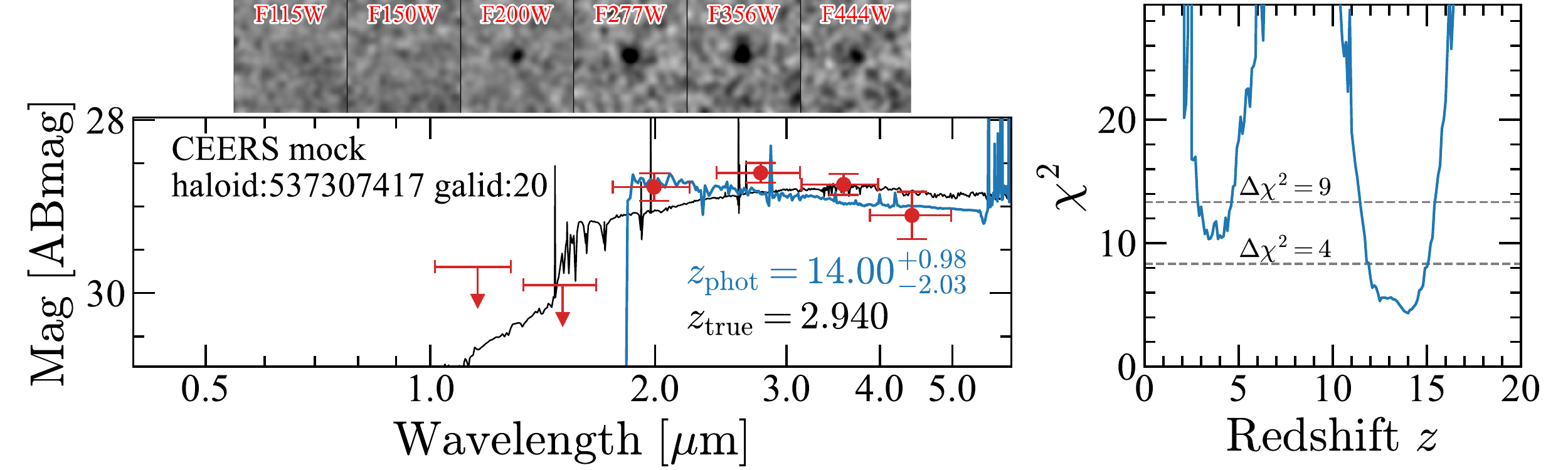}
\includegraphics[width=0.99\hsize, bb=12 1 707 216]{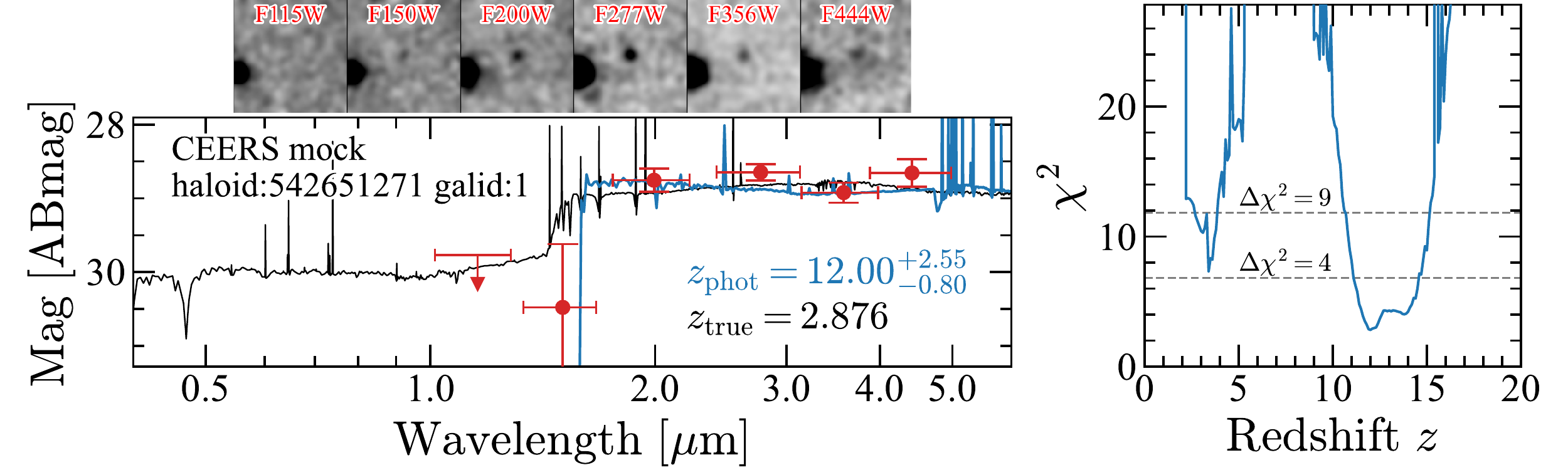}
\end{center}
\caption{Examples of mock galaxies whose true redshifts are $z_\m{true}\sim3$ but selected as $F150W$-dropout galaxies at $z_\m{phot}\sim12-15$ identified in the simulated NIRCam images.
For each object, the top left panel shows the $1.\carcsec5\times1.\carcsec5$ snapshots in NIRCam bands with a 3-pixel smoothing whose band names are indicated with the red labels.
The bottom left panel presents the SED of the object.
The red symbols with error bars are measured magnitudes or $2\sigma$ upper limits, and the blue curve shows the best-fit model.
The true and estimated redshifts with $2\sigma$ errors are indicated with the black and blue texts, respectively.
The $\chi^2$ value is shown in the right panel as a function of redshift.
\redcc{The black curve is the true SEDs at $z\sim3$}, while the blue curve denotes $\chi^2$ values of our SED fitting for our photometric redshift determination. 
These objects meet the weak photometric redshift criterion of $\Delta\chi^2>4$, but do not meet our strict criterion of $\Delta\chi^2>9$, where $\Delta\chi^2$ is the $\chi^2$ difference between the best high redshift solution and a lower redshift solution, $\Delta\chi^2=\chi^2(z_\m{low})-\chi^2(z_\m{high})$.
See texts for details.
}
\label{fig_ceers}
\end{figure*}

Based on the results of the SED fitting, we select objects whose high-redshift solution is more likely than the low-redshift ones, by using $\Delta\chi^2$, which is defined as a difference between the $\chi^2$ values of the best high-redshift solution and the lower-redshift solution, $\Delta\chi^2=\chi^2(z_\m{low})-\chi^2(z_\m{high})$.
Previous studies use a criterion of $\Delta\chi^2>4.0$, corresponding to a $2\sigma$ level \citep[e.g.,][]{2020MNRAS.493.2059B,2022ApJ...929....1H,2022arXiv220712356D,2022arXiv220712474F}.
However, given the small number of the available JWST bands bluer than the Lyman break and the expected small number density of $z>9$ galaxies, it is possible that this criterion is not sufficient to remove low redshift interlopers.
To determine the threshold value for $\Delta\chi^2$, we use the CEERS simulated NIRCam images.
In the CEERS simulated images, mock galaxies at $z=0-10$ in \citet{2019MNRAS.483.2983Y,2022arXiv220613521Y} are inserted using the JWST data simulator {\sc Mirage} \citep{bryan_hilbert_2019_3519262}.
We measure fluxes of mock galaxies in each band in the same manner as our real dropout galaxy selection (Section \ref{ss_catalog}), and conduct the SED fitting using {\sc prospector}.
As shown in Figure \ref{fig_ceers}, at least eight sources at $z_\m{true}\sim3-4$ in the simulations have the best photometric redshifts of $z_\m{phot}\sim12-15$ and $\Delta\chi^2=4-9$, indicating that the criterion of $\Delta\chi^2>4$ is not sufficient to remove low redshift interlopers. 
Thus in this study, we instead adopt a strict screening criterion of $\Delta\chi^2>9.0$, which can remove these interlopers.
\redc{The inclusion of this strict criterion does not introduce a bias with respect to a color of the UV continuum for bright galaxies, because the strength of the break is the most important factor to determine the $\Delta\chi^2$ value.
For faint galaxies, about 40\% of them at $\sim29-30$ mag will be missed due to the inclusion of this criterion, and this effect is taken into account in the completeness estimate in Section \ref{ss_completeness}.}

\subsection{Comparisons with Spectroscopic Redshifts}\label{ss_specz}

\begin{figure*}
\centering
\begin{center}
\includegraphics[width=0.99\hsize, bb=8 8 716 209]{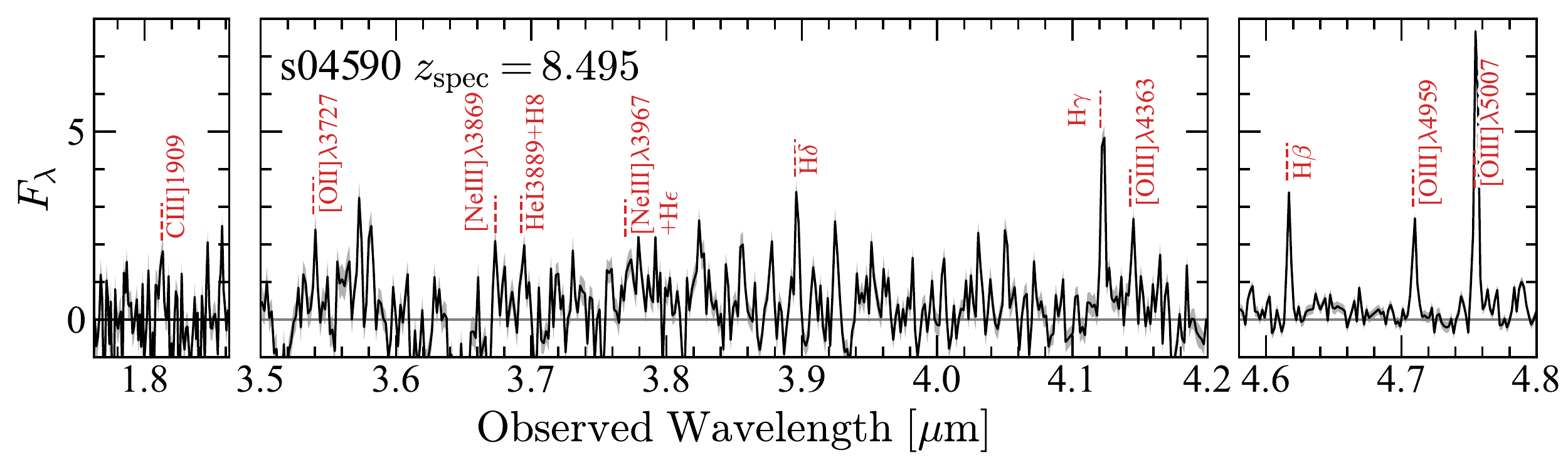}
\includegraphics[width=0.99\hsize, bb=12 1 707 216]{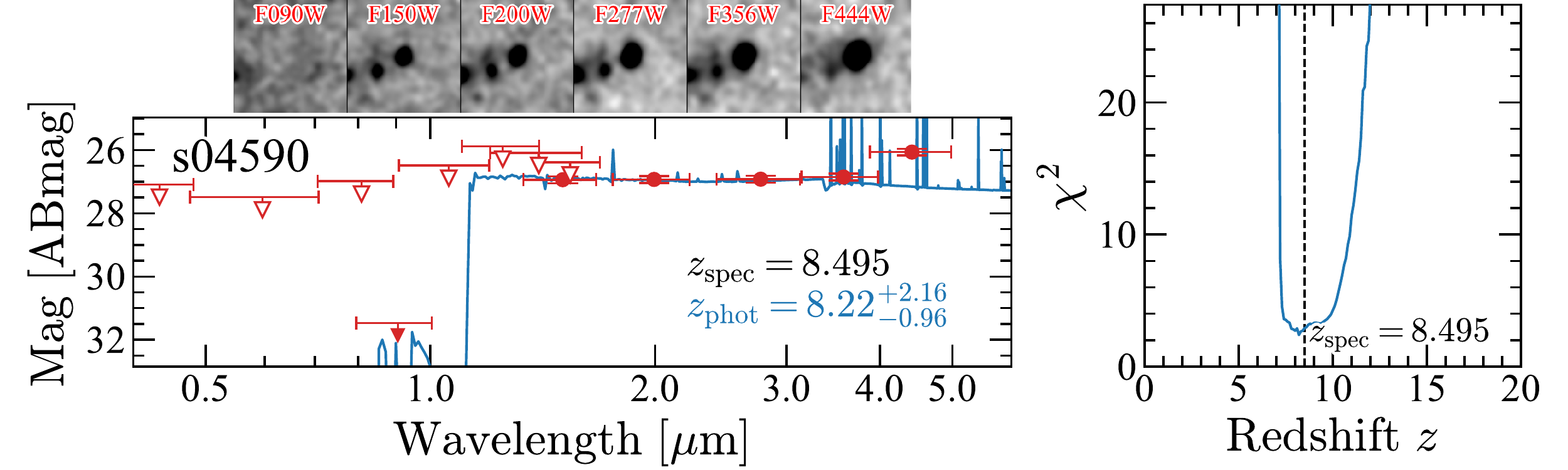}
\end{center}
\caption{
Top: NIRSpec spectrum of s04590 at a spectroscopic redshift of $z=8.495$. The spectroscopic redshift is confirmed with the H$\beta$, H$\gamma$, H$\delta$, {\sc [Oiii]}$\lambda\lambda$5007,4959, {\sc [Oiii]}$\lambda$4363, {\sc [Oii]}$\lambda$3727, [Ne{\sc iii}]$\lambda$3967, [Ne{\sc iii}]$\lambda$3869, and the tentative {\sc Ciii]}$\lambda$1909 lines. 
The flux is arbitrary.
Bottom left: Optical to near-infrared SED of s04590. The red circles and arrows indicate the measured magnitudes and $2\sigma$ limits, respectively. The filled (open) red symbols denote the measurements and the limits obtained with JWST/NIRCam (HST/ACS and WFC3). The blue curve and the blue redshift label represent the best-fit model SED and the photometric redshift with $2\sigma$ errors derived by our photometric redshift technique, which is compared with the spectroscopic redshift indicated with the black label.
The images on this panel show $1.\carcsec5\times1.\carcsec5$ cutout images of s04590 in the NIRCam bands with a 3-pixel smoothing whose band names are indicated with the red labels.
Bottom right: $\chi^2$ as a function of redshift. The blue curve denotes $\chi^2$ values of our SED fitting for our photometric redshift determination. The vertical dashed line
indicates the spectroscopic redshift that agrees well with the photometric redshift.
%
}
\label{fig_specz_1}
\end{figure*}

\begin{figure*}
\centering
\begin{center}
\includegraphics[width=0.99\hsize, bb=8 8 716 209]{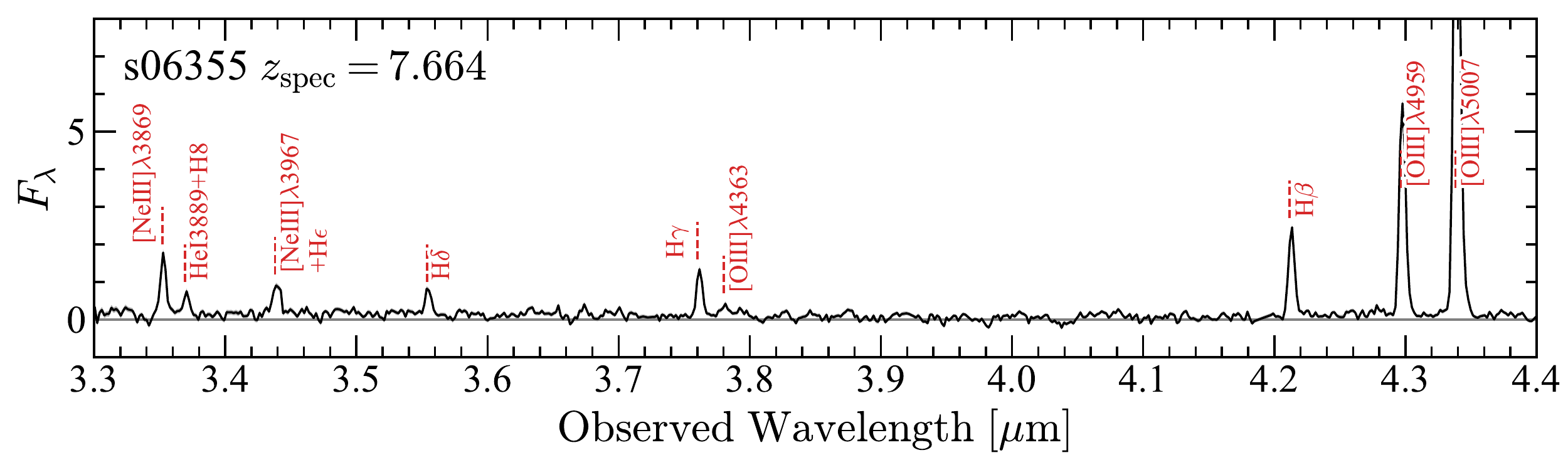}
\includegraphics[width=0.99\hsize, bb=12 -5 707 216]{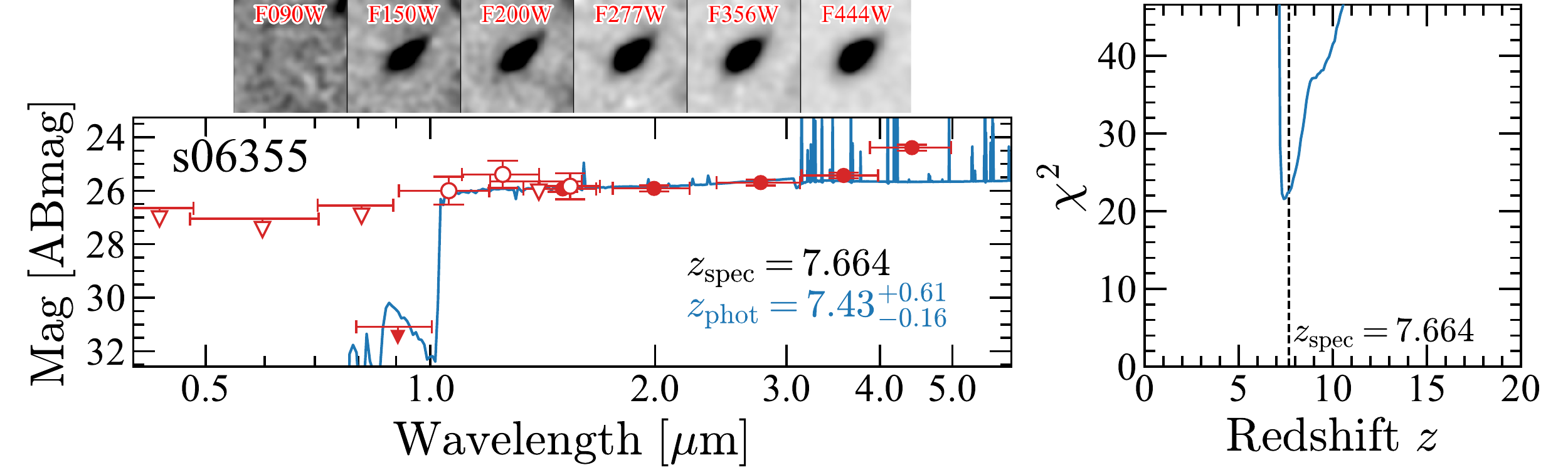}
\includegraphics[width=0.99\hsize, bb=8 8 716 209]{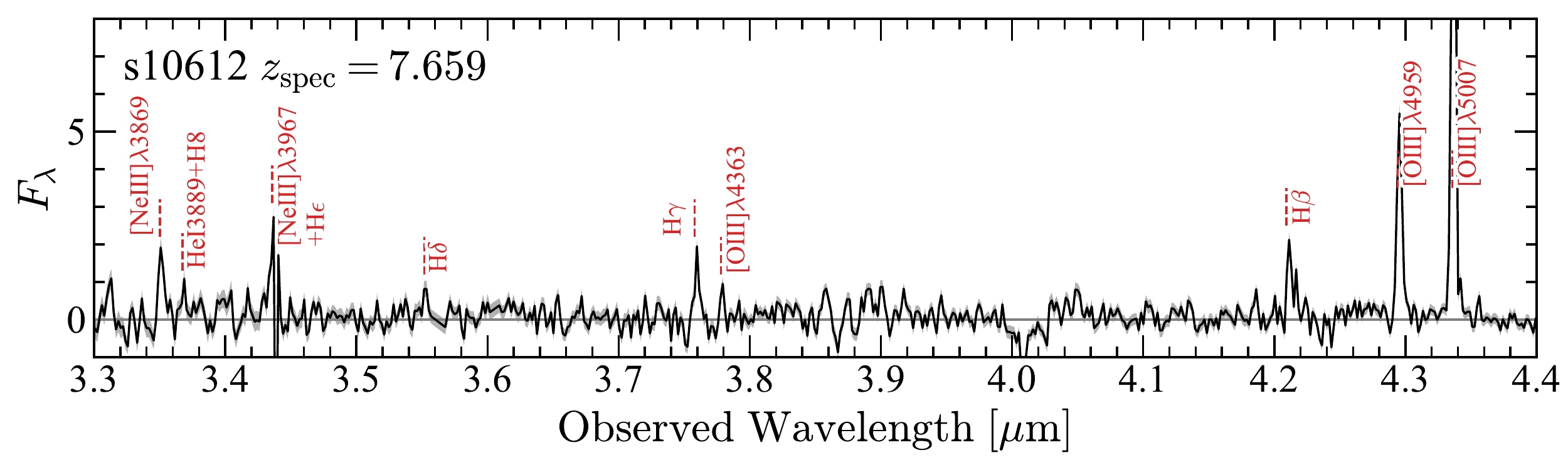}
\includegraphics[width=0.99\hsize, bb=12 1 707 216]{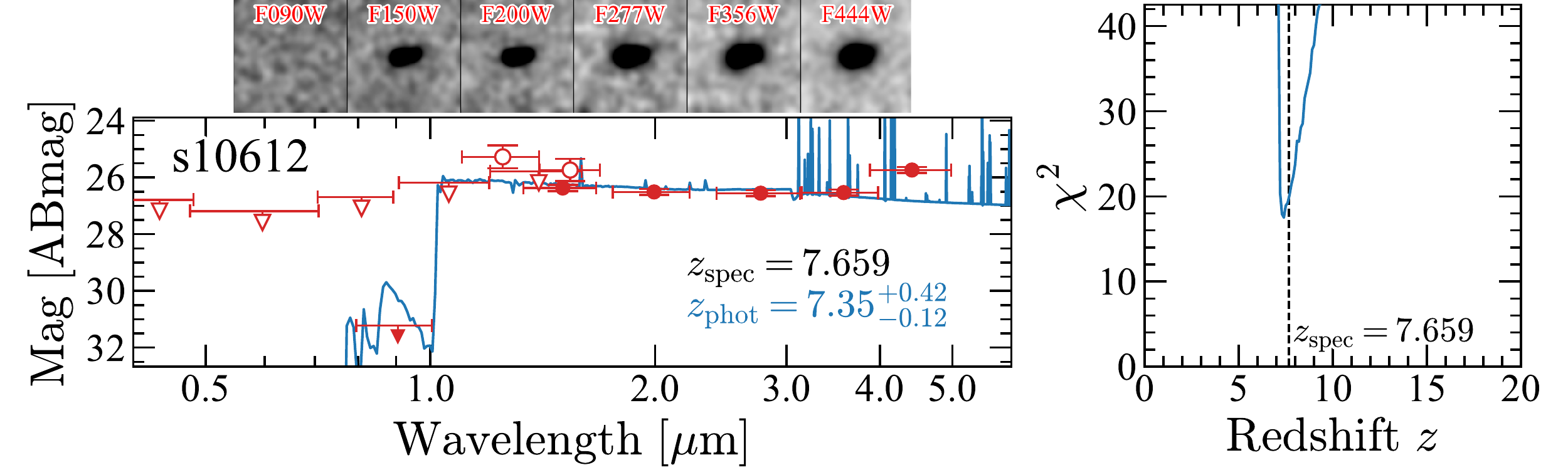}
\end{center}
\caption{Same as Figure \ref{fig_specz_1}, but for s06355 at $z_\m{spec}=7.664$ and s10612 at $z_\m{spec}=7.659$.
Our estimates of the photometric redshifts agree well with the spectroscopic redshifts.
}
\label{fig_specz_2}
\end{figure*}

To test the reliability of our galaxy selections and SED fitting, we compare our photometric redshift estimates with the spectroscopic results obtained in the NIRSpec observations (Section \ref{ss_obs_spec}).
Since there are currently no $z>9$ source spectroscopically confirmed with NIRSpec in the filelds used in this study, we focus on the three galaxies at $z>7$, s04590 ($z_\m{spec}=8.495$), s06355 ($z_\m{spec}=7.664$), and s10612 ($z_\m{spec}=7.659$).
We measure fluxes of the three galaxies in the NIRCam images and estimate photometric redshifts using \textsc{prospector}, in the same manner as our dropout galaxies.
Figures \ref{fig_specz_1} and \ref{fig_specz_2} present results of the SED fitting and comparison with the spectroscopic redshifts.
We found that the estimated photometric redshifts agree well with the spectroscopic redshifts within $\sim2\sigma$ uncertainties, indicating that our SED fitting works well to estimate the redshift from the NIRCam photometry.

\begin{figure*}
\centering
\begin{center}
\includegraphics[width=0.9\hsize, bb=8 7 432 286]{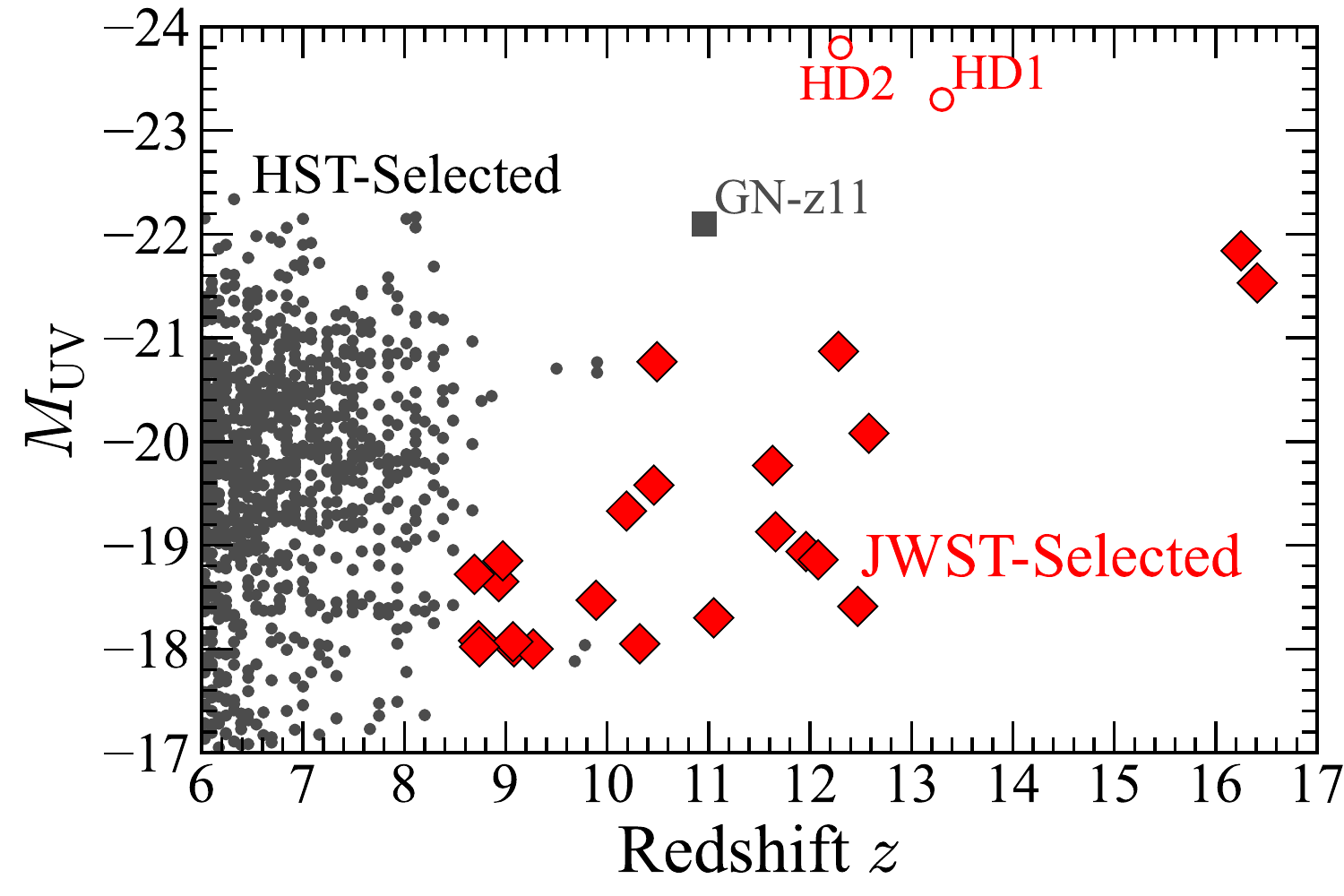}
\end{center}
\caption{Absolute UV magnitude as a function of redshift for galaxies at $6<z<16$. The red diamonds represent our dropout galaxy candidates selected with the JWST images. The red open circles show HD1 and HD2 previously found by the combination of the images taken with Sptizer and ground-based telescopes \citep{2022ApJ...929....1H}. The gray square and circles denote GN-z11 \citep{2016ApJ...819..129O,2021NatAs...5..256J} and dropout galaxies selected with deep HST images \citep{2015ApJ...803...34B}.
}
\label{fig_z_Muv}
\end{figure*}

\subsection{Final Sample}\label{ss_sample}

\begin{figure*}
\centering
\begin{center}
\includegraphics[width=0.9\hsize, bb=12 11 707 288]{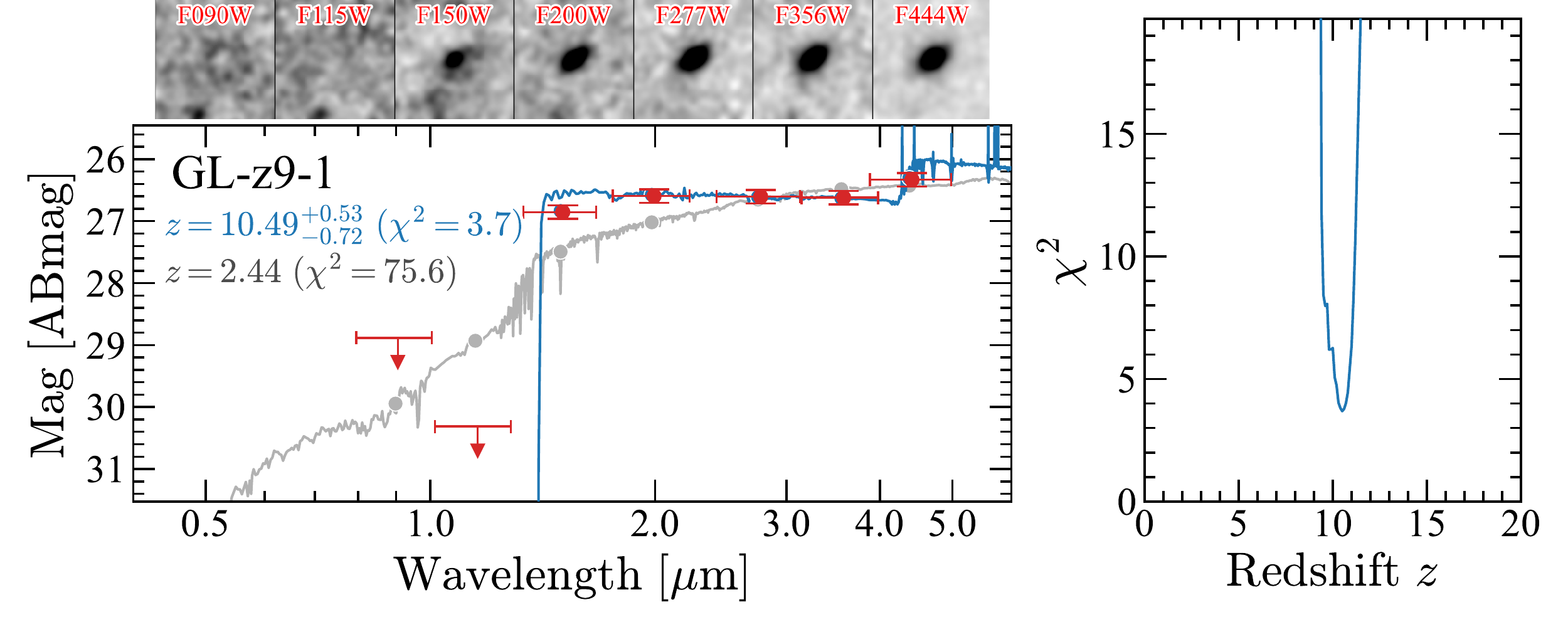}
\end{center}
\begin{center}
\includegraphics[width=0.9\hsize, bb=12 11 707 288]{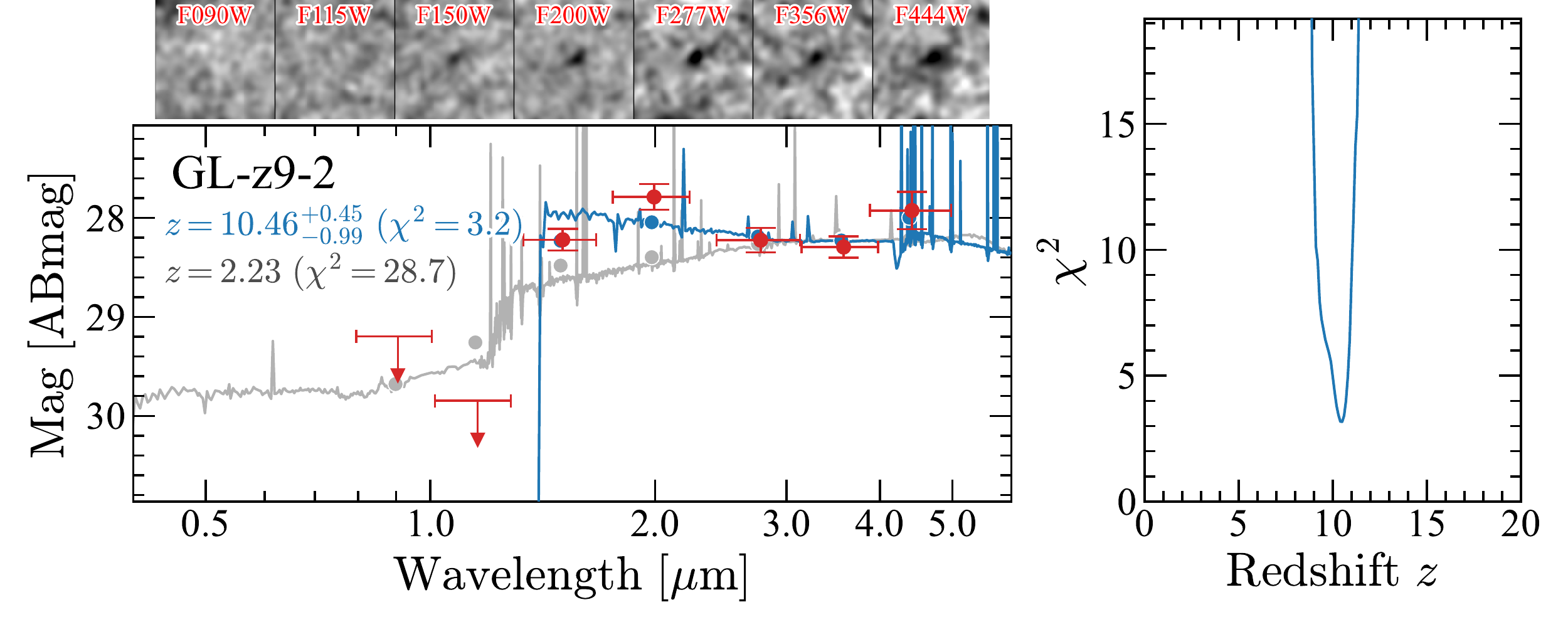}
\end{center}
\begin{center}
\includegraphics[width=0.9\hsize, bb=12 11 707 288]{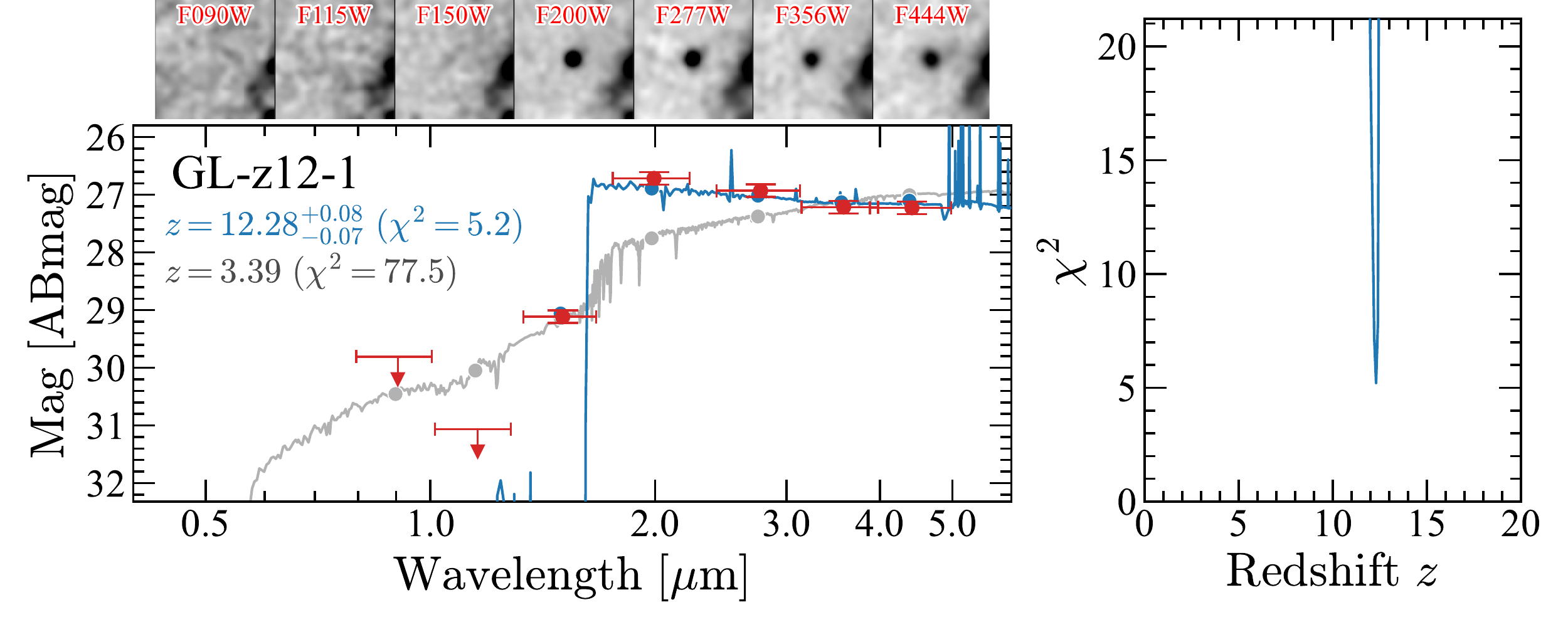}
\end{center}
\caption{
(Top:) The left panel presents the optical to near-infrared SEDs of the $z\sim 9$ dropout galaxy, GL-z9-1. The red circles and arrows show the measured magnitudes and $2\sigma$ upper limits, respectively. The blue curve denotes the best-fit model SED whose redshift and $2\sigma$ errors are presented in the upper left with the blue labels. 
The gray curve is a significantly worse fit of a low redshift solution.
The images on this panel are $1.\carcsec5\times1.\carcsec5$ cutout images of GL-z9-1 in the NIRCam bands with a 3-pixel smoothing whose band names are indicated with the red labels.
The right panel shows $\chi^2$ values of the SED fitting as a function of redshift.
(Middle:) Same as the top panels, but for another $z\sim 9$ dropout galaxy candidate, GL-z9-2.
(Bottom:) Same as the top panels, but for a $z\sim 12$ dropout galaxy candidate, GL-z12-1.
}
\label{fig_sed_z9}
\end{figure*}

\begin{figure*}
\centering
\begin{center}
\includegraphics[width=0.9\hsize, bb=12 11 707 288]{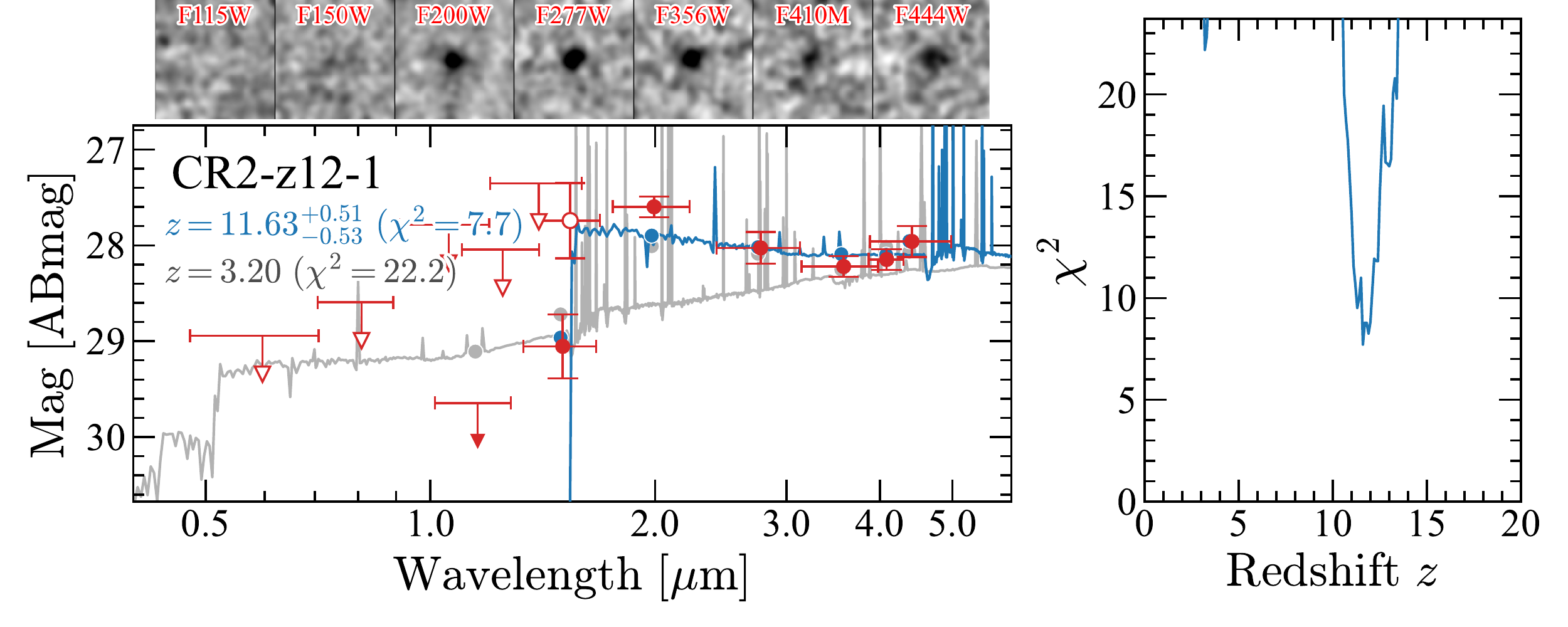}
\end{center}
\begin{center}
\includegraphics[width=0.9\hsize, bb=12 11 707 288]{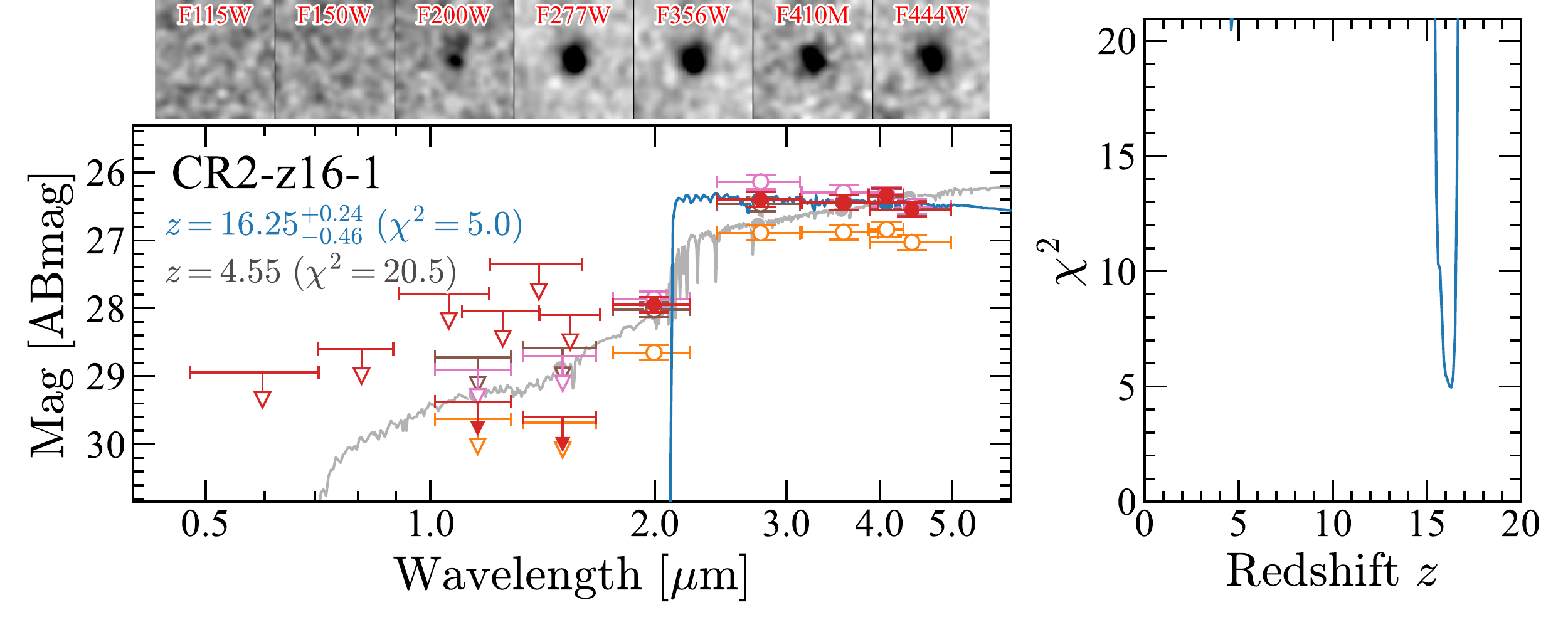}
\end{center}
\begin{center}
\includegraphics[width=0.9\hsize, bb=12 11 707 288]{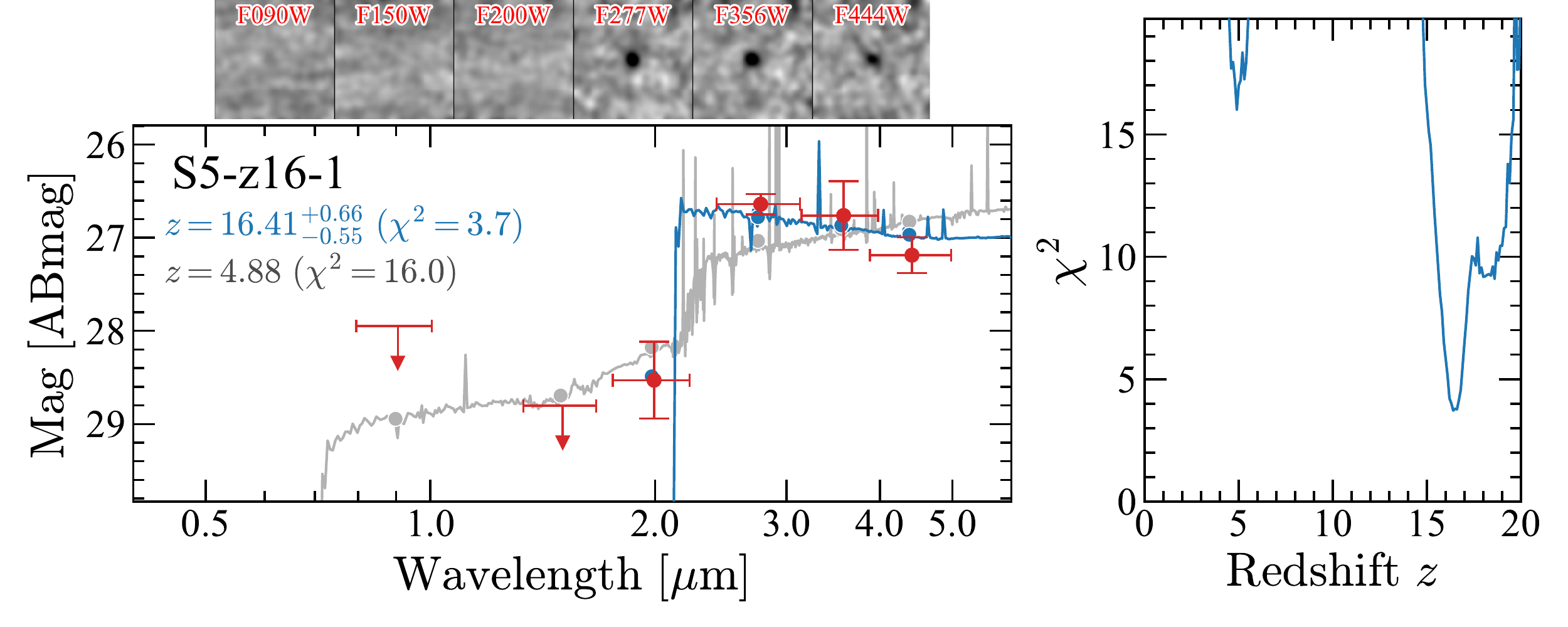}
\end{center}
\caption{Same as Figure \ref{fig_sed_z9}, but for $z\sim 12$ dropout galaxy candidate, CR2-z12-1 (top), and $z\sim \redc{16}$ dropout galaxy candidates, CR2-z16-1 (middle) and S5-z16-1 (bottom).
The open red symbols denote the measurements and the limits obtained with HST/ACS and WFC3.
The orange, pink, and brown open symbols in the middle panel are measurements in \citet{2022arXiv220712356D}, \citet{2022arXiv220802794N}, and \citet{2022arXiv221105792F}, respectively.
}
\label{fig_sed_z17}
\end{figure*}

Finally we select \redc{13}, 8, and 2 dropout galaxy candidates at $z\sim9$, 12, and \redc{16}, respectively (Table \ref{tab_numLBG}).
The photometric redshifts range from \redc{$z\sim8.7$ to $16.4$}, demonstrating the power of JWST exploring the early universe (Figure \ref{fig_z_Muv}).
Examples of the snapshots and SEDs of selected galaxies at $z\sim9$, 12, and \redc{16} are presented in Figures \ref{fig_sed_z9} and \ref{fig_sed_z17}.
These sources show a sharp discontinuity around the Lyman break band, a flat or blue continuum, and non-detection in the bluer bands than the break, all of which are consistent with a high redshift galaxy.
Photometric properties of our galaxy candidates are summarizes in Tables \ref{tab_photo_f115w}-\ref{tab_photo_f200w}.
\redc{Note that no objects appear in more than one final dropout sample.}

To investigate morphological properties of our galaxy candidates, we fit our galaxy candidates with the S\'{e}rsic profile using {\sc galfit} \citep{2010AJ....139.2097P}.
We find that some bright candidates, i.e., GL-z9-1, GL-z12-1, CR2-z12-1, CR-z16-1, and S5-z16-1, are clearly more extended than the PSF, although GL-z12-1 is compact compared to other candidates, implying potential AGN activity.
We stack images of other faint candidates at each redshift, and find that the stacked images also show extended profiles with respect to the PSF.
We thus conclude that the stellar contamination is negligible.
Details of the morphological properties of our candidates are presented in \citet{2022arXiv220813582O}.

One of the highest redshift source candidates in our catalogs is CR2-z16-1 at $z=16.25^{+0.24}_{-0.46}$ in the CEERS2 field.
As discussed later in Section \ref{ss_D22}, CR2-z16-1 is firstly identified as a $z=16.4$ source (ID 93116) in \citet{2022arXiv220712356D}.
As shown in the middle panel of Figure \ref{fig_sed_z17}, our measured fluxes are almost consistent with those presented in \citet{2022arXiv220802794N} and \citet{2022arXiv221105792F}, while fluxes in \citet{2022arXiv220712356D} are systematically fainter than our measurements, probably because \citet{2022arXiv220712356D} assume the PSF for the aperture correction.
The colors measured in these three studies (this study, \citealt{2022arXiv220712356D}, \citealt{2022arXiv220802794N}, and \citealt{2022arXiv221105792F}) consistently show a clear break around $F200W$-band, consistent with a $z=16.3$ galaxy, although there are some discussions about a possible solution of a dusty line emitter at $z\sim5$ \citep{2022arXiv220801816Z,2022arXiv220802794N}.
Although the NIRSpec spectroscopy is required to conclude the redshift of CR2-z16-1, we include this source as a $F200W$-dropout galaxy candidate.

The other candidate at $z\sim17$ is S5-z16-1 at $z=16.41^{+0.66}_{-0.55}$ identified in the Stephan's quintet field.
Although this source is located in a region whose exposure time is relatively short compared to the central region of the field, our position-dependent estimates of flux uncertainties indicate that the source detection, color, and non-detections are robust against the uncertainties. 
An emission line feature is detected with ALMA in S5-z16-1 \citep{2022arXiv221103896F}, which would be interpreted as either {\sc[Oiii]}52$\mu$m at $z=16.01$ or {\sc[Cii]1}58$\mu$m at $z=4.61$.
We include this possible candidate in the luminosity function calculation, although the luminosity is remarkably high compared to our expectations at this high redshift.

\begin{deluxetable}{cccc}
\setlength{\tabcolsep}{2pt}
\tablecaption{Number of Our Dropout Candidates}
\tablehead{\colhead{} & \colhead{$F115W$-drop} & \colhead{$F150W$-drop} & \colhead{$F200W$-drop}\\
\colhead{Field} & \colhead{$z\sim9$} & \colhead{$z\sim12$} & \colhead{$z\sim16$}}
\startdata
SMACS J0723 & \nodata & 1 & 0\\
GLASS & \redc{13} & 1 & 0\\
CEERS1 & \nodata & 0 & 0\\
CEERS2 & \nodata & \redc{4} & 1\\
CEERS3 & \nodata & 1 & 0\\
CEERS6 & \nodata & \redc{0} & 0\\
Stephan's Quintet & \nodata & 1 & 1\\
\hline
Total $(z)$ & $\redc{13}$ & $8$ & $2$\\
\hline
Total & \multicolumn{3}{c}{23}
\label{tab_numLBG}
\enddata
\end{deluxetable}

\begin{deluxetable*}{cccccccccccc}
\tablecaption{Summary of Our $F115W$-Dropoout Candidates at $z\sim 9$}
\setlength{\tabcolsep}{2pt}
\tablehead{\colhead{ID} & \colhead{R.A.} & \colhead{Decl.} & \colhead{$F356W$} & \colhead{$F115W$-$F150W$} & \colhead{$F150W$-$F277W$}& \colhead{$M_\m{UV}$}& \colhead{$z_\m{phot}$}& \colhead{$\Delta\chi^2$} & \colhead{Ref./Note}\\
\colhead{(1)}& \colhead{(2)}& \colhead{(3)}& \colhead{(4)} &  \colhead{(5)}& \colhead{(6)}& \colhead{(7)}& \colhead{(8)}& \colhead{(9)}& \colhead{(10)} }
\startdata
\multicolumn{10}{c}{GLASS}\\
GL-z9-1 & 00:14:02.85 & $-$30:22:18.6 & $26.6\pm0.1$ & $>4.1$ & $0.2\pm0.1$ & $-20.9\pm0.1$ & $10.49_{-0.72}^{+0.53}$ & 71.9 & This,N22,C22,D22\\
GL-z9-2 & 00:14:03.28 & $-$30:21:05.6 & $28.3\pm0.1$ & $>2.3$ & $-0.0\pm0.1$ & $-19.7\pm0.1$ & $10.46_{-0.99}^{+0.45}$ & 25.5 & This,C22\\
GL-z9-3 & 00:14:00.09 & $-$30:19:06.9 & $28.4\pm0.1$ & $1.7\pm0.5$ & $-0.4\pm0.2$ & $-18.8\pm0.1$ & $8.93_{-0.38}^{+0.39}$ & 22.6 & This\\
GL-z9-4 & 00:14:00.27 & $-$30:21:25.9 & $28.5\pm0.1$ & $>2.5$ & $0.2\pm0.1$ & $-19.4\pm0.1$ & $10.19_{-0.55}^{+0.63}$ & 27.2 & This,D22\\
GL-z9-5 & 00:14:03.10 & $-$30:22:26.3 & $28.7\pm0.1$ & $1.3\pm0.3$ & $-0.6\pm0.2$ & $-18.8\pm0.1$ & $8.69_{-0.15}^{+0.42}$ & 10.3 & This\\
GL-z9-6 & 00:14:04.37 & $-$30:20:39.6 & $28.7\pm0.1$ & $1.2\pm0.3$ & $-0.0\pm0.1$ & $-18.9\pm0.1$ & $8.97_{-0.36}^{+0.36}$ & 13.0 & This\\
GL-z9-7 & 00:14:02.52 & $-$30:21:57.0 & $28.9\pm0.1$ & $>2.0$ & $0.7\pm0.1$ & $-18.2\pm0.2$ & $10.32_{-0.82}^{+0.74}$ & 12.0 & This\\
GL-z9-8 & 00:14:00.83 & $-$30:21:29.8 & $29.1\pm0.1$ & $1.9\pm0.5$ & $-0.7\pm0.2$ & $-18.1\pm0.2$ & $9.08_{-0.32}^{+0.94}$ & 31.2 & This\\
GL-z9-9 & 00:14:03.71 & $-$30:21:03.6 & $29.1\pm0.1$ & $>2.1$ & $-0.1\pm0.2$ & $-18.1\pm0.2$ & $9.27_{-0.61}^{+1.28}$ & 20.5 & This\\
GL-z9-10 & 00:14:03.47 & $-$30:19:00.9 & $29.1\pm0.1$ & $1.2\pm0.4$ & $-0.2\pm0.2$ & $-18.2\pm0.2$ & $8.73_{-0.41}^{+0.68}$ & 22.0 & This\\
GL-z9-11 & 00:14:02.49 & $-$30:22:00.9 & $29.4\pm0.1$ & $1.9\pm0.5$ & $-0.8\pm0.2$ & $-18.6\pm0.1$ & $9.89_{-0.74}^{+0.21}$ & 41.0 & This\\
GL-z9-12 & 00:14:06.85 & $-$30:22:02.0 & $29.7\pm0.1$ & $1.3\pm0.5$ & $-0.7\pm0.3$ & $-18.2\pm0.2$ & $9.07_{-0.23}^{+1.02}$ & 50.9 & This\\
GL-z9-13 & 00:13:57.45 & $-$30:18:00.0 & $29.7\pm0.1$ & $1.1\pm0.3$ & $-0.8\pm0.2$ & $-18.1\pm0.3$ & $8.74_{-0.28}^{+0.57}$ & 110.3 & This\\
\hline
\multicolumn{10}{c}{Other Possible Candidates that did not Meet Our Selection Criteria}\\
\multicolumn{10}{c}{\citet{2022arXiv220709436C}}\\
GHZ3 & 00:14:06.94 & $-$30:21:49.7 & $27.0\pm0.1$ & $>2.8$ & $0.7\pm0.1$ & $-19.7\pm0.1$ & $11.02_{-0.47}^{+0.37}$ & 8.6 & 2\\
GHZ5 & 00:13:58.66 & $-$30:18:27.4 & $28.8\pm0.1$ & $1.1\pm0.4$ & $-0.2\pm0.2$ & $-18.8\pm0.2$ & $8.50_{-1.86}^{+0.86}$ & 22.3 & 1\\
GHZ6 & 00:13:54.97 & $-$30:18:53.7 & $27.8\pm0.1$ & $1.3\pm0.6$ & $0.8\pm0.1$ & $-18.8\pm0.1$ & $2.20_{-0.38}^{+8.99}$ & -2.8 & 2
\enddata
\tablecomments{(1) Name.
(2) Right ascension.
(3) Declination.
(4) Total magnitude in the $F356W$ band with $1\sigma$ errors.
(5) $F115W-F150W$ color with $1\sigma$ errors.
(6) $F150W-F277W$ color with $1\sigma$ errors.
(7) Absolute UV magnitude with $1\sigma$ errors. 
(8) Photometric redshift with $2\sigma$ errors.
(9) $\chi^2$ difference between the best high redshift solution and a lower redshift solution, $\Delta\chi^2=\chi^2(z_\m{low})-\chi^2(z_\m{high})$.
(10) Reference (This: this work, N22: \citet{2022arXiv220709434N}, C22: \citet{2022arXiv220709436C}, D22: \citet{2022arXiv220712356D}) and note for a reason why the source is not selected in this study (1: $>2\sigma$ detection in $F090W$, 2: $\Delta\chi^2<9.0$).
}
\label{tab_photo_f115w}
\end{deluxetable*}

\clearpage
\startlongtable
\centerwidetable
\begin{deluxetable*}{cccccccccccc}
\tablecaption{Summary of Our $F150W$-Dropoout Candidates at $z\sim 12$}
\setlength{\tabcolsep}{2pt}
\tablehead{\colhead{ID} & \colhead{R.A.} & \colhead{Decl.} & \colhead{$F356W$} & \colhead{$F150W$-$F200W$} & \colhead{$F200W$-$F356W$}& \colhead{$M_\m{UV}$}& \colhead{$z_\m{phot}$}& \colhead{$\Delta\chi^2$} & \colhead{Ref./Note}\\
\colhead{(1)}& \colhead{(2)}& \colhead{(3)}& \colhead{(4)} &  \colhead{(5)}& \colhead{(6)}& \colhead{(7)}& \colhead{(8)}& \colhead{(9)}& \colhead{(10)} }
\startdata
\multicolumn{10}{c}{SMACS J0723}\\
SM-z12-1 & 07:22:32.59 & $-$73:28:33.3 & $29.8\pm0.6$ & $1.9\pm0.9$ & $-1.2\pm0.6$ & $-18.5\pm0.1$ & $12.47_{-0.72}^{+1.19}$ & 14.5 & This,Y22\\
\multicolumn{10}{c}{GLASS}\\
GL-z12-1 & 00:13:59.74 & $-$30:19:29.1 & $27.2\pm0.1$ & $2.4\pm0.1$ & $-0.5\pm0.1$ & $-21.0\pm0.1$ & $12.28_{-0.07}^{+0.08}$ & 72.3 & This,N22,C22,D22\\
\multicolumn{10}{c}{CEERS2}\\
CR2-z12-1 & 14:19:46.36 & $+$52:56:32.8 & $28.2\pm0.1$ & $1.5\pm0.4$ & $-0.6\pm0.1$ & $-19.9\pm0.1$ & $11.63_{-0.53}^{+0.51}$ & 14.5 & This,F22,D22\\ 
CR2-z12-2 & 14:19:42.57 & $+$52:54:42.0 & $28.6\pm0.1$ & $1.3\pm0.6$ & $-0.1\pm0.1$ & $-19.0\pm0.2$ & $11.96_{-0.87}^{+1.44}$ & 14.0 & This\\
CR2-z12-3 & 14:19:41.61 & $+$52:55:07.6 & $28.8\pm0.1$ & $1.4\pm0.5$ & $-0.6\pm0.1$ & $-19.2\pm0.2$ & $11.66_{-0.71}^{+0.69}$ & 20.3 & This\\
CR2-z12-4 & 14:19:24.86 & $+$52:53:13.9 & $28.9\pm0.1$ & $1.6\pm1.1$ & $0.1\pm0.2$ & $-19.0\pm0.2$ & $12.08_{-1.25}^{+2.11}$ & 9.1 & This\\
\multicolumn{10}{c}{CEERS3}\\
CR3-z12-1 & 14:19:11.11 & $+$52:49:33.6 & $29.8\pm0.2$ & $1.0\pm0.7$ & $-0.5\pm0.2$ & $-18.4\pm0.4$ & $11.05_{-0.47}^{+2.24}$ & 47.1 & This\\
\multicolumn{10}{c}{Stephan's Quintet}\\
S5-z12-1 & 22:36:06.72 & $+$34:00:09.7 & $27.9\pm0.5$ & $>2.3$ & $-0.2\pm0.5$ & $-20.2\pm0.1$ & $12.58_{-0.46}^{+1.23}$ & 19.1 & This\\
\hline
\multicolumn{10}{c}{Other Possible Candidates that did not Meet Our Selection Criteria}\\
\multicolumn{10}{c}{\citet{2022arXiv220711217A}}\\
10234 & 07:22:39.60 & $-$73:30:06.2 & $29.8\pm0.6$ & $0.5\pm0.4$ & $-0.7\pm0.6$ & $-18.3\pm0.1$ & $5.34_{-0.04}^{+0.30}$ & -75.2 & 1,2,3\\
\multicolumn{10}{c}{\citet{2022arXiv220712338A}}\\
SMACS\_z12a & 07:22:47.38 & $-$73:30:01.7 & $28.0\pm0.3$ & $0.7\pm0.5$ & $0.2\pm0.3$ & $-18.5\pm0.1$ & $3.15_{-1.60}^{+1.15}$ & -36.4 & 2,3\\
SMACS\_z12b & 07:22:52.26 & $-$73:27:55.4 & $28.4\pm0.2$ & $1.5\pm0.6$ & $-0.1\pm0.2$ & $-18.4\pm0.1$ & $3.03_{-1.48}^{+0.29}$ & -5.4 & 3\\
\multicolumn{10}{c}{\citet{2022arXiv220712356D}}\\
10566 & 07:23:03.55 & $-$73:28:46.8 & $29.0\pm0.4$ & $1.3\pm0.2$ & $-1.6\pm0.4$ & $-19.0\pm0.1$ & $1.87_{-0.17}^{+9.73}$ & -3.2 & 3\\
1566 & 07:22:39.31 & $-$73:30:00.6 & $28.8\pm0.3$ & $1.9\pm1.1$ & $-0.4\pm0.4$ & $-18.1\pm0.1$ & $12.15_{-1.24}^{+1.54}$ & 4.8 & 3\\
27535\_4 & 14:19:27.31 & $+$52:51:29.2 & $27.6\pm0.1$ & $>1.8$ & $-0.1\pm0.1$ & $-19.0\pm0.2$ & $3.90_{-0.06}^{+0.29}$ & -7.1 & 3\\
\multicolumn{10}{c}{\citet{2022arXiv220711558Y}}\\
F150DB-007 & 07:23:23.97 & $-$73:27:58.7 & $28.0\pm0.3$ & $0.4\pm0.4$ & $0.4\pm0.4$ & $-18.5\pm0.1$ & $4.79_{-0.80}^{+0.45}$ & -101.7 & 2,3 \\
F150DB-011 & 07:23:27.39 & $-$73:27:58.0 & $29.1\pm0.7$ & $0.1\pm0.1$ & $-1.5\pm0.7$ & $-17.9\pm0.1$ & $2.57_{-0.02}^{+0.02}$ & -110.2 & 2,3 \\
F150DB-013 & 07:23:05.53 & $-$73:27:50.6 & $28.6\pm0.4$ & $1.2\pm0.4$ & $-1.0\pm0.4$ & $-18.7\pm0.1$ & $2.97_{-0.18}^{+0.17}$ & -69.6 & 3 \\
F150DB-021 & 07:23:12.64 & $-$73:27:45.2 & $27.1\pm0.1$ & $0.1\pm0.1$ & $0.2\pm0.1$ & $-18.8\pm0.1$ & $2.43_{-0.93}^{+4.22}$ & -37.7 & 2,3 \\
F150DB-026 & 07:23:23.74 & $-$73:27:40.6 & $29.6\pm0.6$ & $0.1\pm0.3$ & $-0.4\pm0.6$ & $-16.5\pm0.3$ & $10.38_{-1.26}^{+1.08}$ & 18.1 & 2 \\
F150DB-031 & 07:23:21.43 & $-$73:27:36.3 & $28.8\pm0.4$ & $1.1\pm0.5$ & $-0.4\pm0.4$ & $-16.6\pm0.3$ & $11.59_{-9.48}^{+0.90}$ & 3.9 & 3 \\
F150DB-033 & 07:23:30.55 & $-$73:27:33.1 & $27.2\pm0.1$ & $0.8\pm0.2$ & $0.5\pm0.1$ & $-18.3\pm0.1$ & $11.38_{-0.58}^{+0.42}$ & 4.4 & 2,3 \\
F150DB-040 & 07:23:11.94 & $-$73:27:24.9 & $29.1\pm0.3$ & $-0.2\pm0.3$ & $0.1\pm0.4$ & $-15.7\pm0.6$ & $4.74_{-0.62}^{+0.15}$ & -87.3 & 2,3 \\
F150DB-041 & 07:23:06.63 & $-$73:27:25.4 & $27.3\pm0.1$ & $0.2\pm0.2$ & $0.4\pm0.2$ & $-17.9\pm0.1$ & $1.69_{-1.01}^{+0.19}$ & -49.9 & 2,3 \\
F150DB-044 & 07:23:39.31 & $-$73:27:22.3 & $28.8\pm0.4$ & $0.7\pm0.3$ & $-0.7\pm0.4$ & $-18.5\pm0.1$ & $3.45_{-0.27}^{+0.36}$ & -99.8 & 2,3 \\
F150DB-048 & 07:23:01.57 & $-$73:27:18.0 & $27.6\pm0.1$ & $1.3\pm0.7$ & $0.8\pm0.2$ & $-17.8\pm0.1$ & $12.39_{-10.33}^{+2.69}$ & 0.8 & 3 \\
F150DB-050 & 07:23:24.58 & $-$73:27:15.0 & $29.0\pm0.4$ & $0.1\pm0.2$ & $-0.6\pm0.4$ & $-16.6\pm0.3$ & $3.00_{-0.79}^{+0.07}$ & -109.3 & 2,3 \\
F150DB-052 & 07:23:28.14 & $-$73:27:13.8 & $29.2\pm0.5$ & $0.5\pm0.5$ & $-0.2\pm0.5$ & $-15.7\pm0.7$ & $3.75_{-0.32}^{+1.09}$ & -68.0 & 2,3 \\
F150DB-054 & 07:23:12.51 & $-$73:27:10.7 & $29.6\pm0.6$ & $0.0\pm0.5$ & $-0.1\pm0.7$ & $-16.0\pm0.4$ & $4.60_{-0.75}^{+0.63}$ & -55.9 & 2,3 \\
F150DB-069 & 07:23:04.26 & $-$73:26:54.2 & $29.4\pm0.5$ & $1.1\pm0.5$ & $-1.0\pm0.5$ & $-17.5\pm0.1$ & $11.99_{-0.81}^{+1.47}$ & 33.8 & 4\\
F150DB-075 & 07:23:02.23 & $-$73:26:41.5 & $27.1\pm0.1$ & $0.8\pm0.1$ & $-0.3\pm0.1$ & $-19.8\pm0.1$ & $2.87_{-1.55}^{+0.10}$ & -45.8 & 2,3 \\
F150DB-076 & 07:23:29.41 & $-$73:26:39.7 & $28.6\pm0.4$ & $0.4\pm0.2$ & $-0.6\pm0.5$ & $-18.2\pm0.1$ & $3.78_{-0.03}^{+0.04}$ & -100.7 & 2,3 \\
F150DB-079 & 07:23:13.15 & $-$73:26:29.6 & $28.9\pm0.4$ & $0.9\pm0.4$ & $-0.5\pm0.4$ & $-18.1\pm0.1$ & $2.75_{-0.08}^{+0.40}$ & -32.2 & 2,3 \\
F150DB-082 & 07:23:22.75 & $-$73:26:25.6 & $27.9\pm0.2$ & $1.2\pm0.6$ & $0.1\pm0.3$ & $-18.5\pm0.1$ & $3.05_{-1.86}^{+0.36}$ & -13.5 & 3 \\
F150DB-084 & 07:23:07.54 & $-$73:26:23.8 & $29.2\pm0.4$ & $0.2\pm0.3$ & $-0.6\pm0.5$ & $-18.8\pm0.1$ & $3.00_{-0.27}^{+0.00}$ & -14.3 & 2,3 \\
F150DB-088 & 07:23:14.04 & $-$73:26:17.3 & $26.3\pm0.1$ & $1.2\pm0.3$ & $0.2\pm0.1$ & $-19.6\pm0.1$ & $3.06_{-0.32}^{+0.48}$ & -41.2 & 3 \\
F150DB-090 & 07:23:26.23 & $-$73:26:13.8 & $25.7\pm0.1$ & $0.8\pm0.1$ & $0.6\pm0.1$ & $-20.6\pm0.1$ & $3.19_{-0.20}^{+0.87}$ & -93.5 & 2,3 \\
F150DB-095 & 07:23:24.76 & $-$73:26:01.2 & $28.3\pm0.3$ & $0.2\pm0.2$ & $-0.5\pm0.3$ & $-19.1\pm0.1$ & $2.85_{-0.14}^{+0.43}$ & -113.6 & 2,3 \\
F150DB-C\_4 & 07:23:25.96 & $-$73:26:39.9 & $23.2\pm0.1$ & $2.0\pm0.3$ & $2.7\pm0.1$ & $-21.4\pm0.1$ & $5.28_{-0.12}^{+0.16}$ & -94.0 & 3 \\
F150DA-005 & 07:22:41.01 & $-$73:29:54.9 & $27.9\pm0.2$ & $0.4\pm0.2$ & $-0.3\pm0.2$ & $-19.8\pm0.1$ & $3.01_{-0.24}^{+1.09}$ & -113.8 & 2,3 \\
F150DA-007 & 07:22:44.88 & $-$73:29:53.6 & $28.5\pm0.2$ & $0.3\pm0.3$ & $0.1\pm0.2$ & $-19.3\pm0.1$ & $10.99_{-0.56}^{+0.74}$ & 7.9 & 2,3 \\
F150DA-008 & 07:22:52.75 & $-$73:29:51.6 & $28.1\pm0.3$ & $0.7\pm0.3$ & $-0.2\pm0.3$ & $-19.6\pm0.1$ & $2.99_{-1.91}^{+0.30}$ & -49.2 & 2,3 \\
F150DA-010 & 07:22:40.09 & $-$73:29:46.1 & $28.5\pm0.4$ & $0.3\pm0.3$ & $-0.4\pm0.4$ & $-19.1\pm0.1$ & $3.54_{-1.92}^{+0.98}$ & -79.9 & 2,3 \\
F150DA-015 & 07:22:44.74 & $-$73:29:26.8 & $28.0\pm0.2$ & $0.7\pm0.3$ & $0.1\pm0.2$ & $-19.6\pm0.1$ & $3.01_{-0.24}^{+1.63}$ & -106.8 & 2,3 \\
F150DA-018 & 07:22:56.02 & $-$73:29:21.9 & $29.0\pm0.3$ & $1.6\pm0.8$ & $-0.5\pm0.4$ & $-18.7\pm0.1$ & $12.84_{-1.27}^{+0.97}$ & 19.1 & 4\\
F150DA-019 & 07:22:39.40 & $-$73:29:20.5 & $28.7\pm0.3$ & $0.5\pm0.3$ & $-0.3\pm0.3$ & $-19.5\pm0.1$ & $2.23_{-0.22}^{+9.39}$ & -2.8 & 2,3 \\
F150DA-020 & 07:22:55.87 & $-$73:29:17.4 & $28.3\pm0.2$ & $0.9\pm0.3$ & $-0.1\pm0.2$ & $-19.5\pm0.1$ & $3.50_{-0.22}^{+0.68}$ & -98.4 & 2,3 \\
F150DA-024 & 07:22:33.46 & $-$73:29:09.5 & $28.8\pm0.3$ & $0.5\pm0.2$ & $-0.7\pm0.3$ & $-18.8\pm0.1$ & $2.62_{-0.06}^{+0.06}$ & -77.3 & 2,3 \\
F150DA-026 & 07:22:46.02 & $-$73:29:08.1 & $30.6\pm1.2$ & $0.0\pm0.6$ & $<-0.7$ & $-18.9\pm0.1$ & $4.58_{-0.35}^{+0.64}$ & -60.9 & 2,3 \\
F150DA-027 & 07:23:01.03 & $-$73:29:07.1 & $20.0\pm0.1$ & $0.1\pm0.1$ & $-0.9\pm0.1$ & $-27.1\pm0.1$ & $3.47_{-3.21}^{+0.07}$ & -109.0 & 2,3 \\
F150DA-031 & 07:22:40.65 & $-$73:29:00.5 & $28.4\pm0.2$ & $1.1\pm0.6$ & $-0.0\pm0.3$ & $-19.3\pm0.1$ & $11.95_{-9.28}^{+2.69}$ & 3.9 & 3 \\
F150DA-038 & 07:23:02.95 & $-$73:28:46.1 & $29.0\pm0.4$ & $0.7\pm0.4$ & $-0.6\pm0.5$ & $-18.8\pm0.1$ & $2.74_{-0.89}^{+9.09}$ & -2.7 & 2,3 \\
F150DA-039 & 07:23:00.58 & $-$73:28:47.0 & $28.3\pm0.3$ & $0.5\pm0.3$ & $-0.3\pm0.3$ & $-18.8\pm0.1$ & $3.02_{-1.29}^{+0.28}$ & -45.2 & 2,3 \\
F150DA-050 & 07:22:45.00 & $-$73:28:36.9 & $28.1\pm0.2$ & $-0.2\pm0.2$ & $-0.2\pm0.2$ & $-19.1\pm0.1$ & $1.94_{-0.19}^{+0.40}$ & -78.1 & 2,3 \\
F150DA-052 & 07:22:26.94 & $-$73:28:33.8 & $28.2\pm0.2$ & $0.7\pm0.5$ & $0.2\pm0.3$ & $-19.2\pm0.1$ & $1.58_{-0.25}^{+1.95}$ & -13.2 & 2,3 \\
F150DA-054 & 07:22:38.89 & $-$73:28:30.8 & $28.8\pm0.3$ & $0.7\pm0.3$ & $-0.3\pm0.3$ & $-19.2\pm0.1$ & $3.04_{-0.33}^{+0.31}$ & -105.0 & 2,3 \\
F150DA-058 & 07:22:48.28 & $-$73:28:27.3 & $27.8\pm0.2$ & $0.6\pm0.2$ & $-0.1\pm0.2$ & $-19.5\pm0.1$ & $2.96_{-1.88}^{+0.32}$ & -71.1 & 2,3 \\
F150DA-060 & 07:22:40.75 & $-$73:28:23.7 & $28.6\pm0.3$ & $0.5\pm0.4$ & $0.0\pm0.3$ & $-18.9\pm0.1$ & $3.74_{-1.33}^{+0.69}$ & -70.3 & 2,3 \\
F150DA-062 & 07:22:54.22 & $-$73:28:23.5 & $28.6\pm0.2$ & $0.4\pm0.2$ & $-0.2\pm0.2$ & $-19.3\pm0.1$ & $1.80_{-0.14}^{+9.47}$ & -1.1 & 2,3 \\
F150DA-063 & 07:22:53.83 & $-$73:28:23.2 & $27.1\pm0.1$ & $0.0\pm0.1$ & $-0.0\pm0.1$ & $-19.9\pm0.1$ & $6.99_{-5.26}^{+0.02}$ & 1.6 & 2,3 \\
F150DA-066 & 07:22:39.61 & $-$73:28:12.1 & $29.1\pm0.3$ & $0.8\pm0.2$ & $-1.3\pm0.4$ & $-19.3\pm0.1$ & $2.99_{-0.19}^{+0.06}$ & -118.6 & 2,3 \\
F150DA-075 & 07:22:38.35 & $-$73:27:57.1 & $28.1\pm0.2$ & $0.5\pm0.2$ & $-0.1\pm0.2$ & $-19.1\pm0.1$ & $3.27_{-0.91}^{+0.58}$ & -59.3 & 2,3 \\
F150DA-077 & 07:22:52.23 & $-$73:27:55.4 & $28.4\pm0.2$ & $1.5\pm0.6$ & $-0.1\pm0.2$ & $-19.0\pm0.1$ & $1.45_{-0.06}^{+1.37}$ & -4.6 & 3 \\
F150DA-078 & 07:22:49.24 & $-$73:27:49.8 & $28.1\pm0.2$ & $0.1\pm0.2$ & $0.1\pm0.2$ & $-19.3\pm0.1$ & $1.94_{-0.78}^{+0.32}$ & -16.8 & 2,3 \\
F150DA-081 & 07:22:49.24 & $-$73:27:44.5 & $27.9\pm0.2$ & $0.3\pm0.3$ & $0.3\pm0.2$ & $-19.1\pm0.1$ & $1.43_{-0.09}^{+1.47}$ & -16.9 & 2,3 \\
F150DA-082 & 07:22:52.78 & $-$73:27:41.9 & $29.2\pm0.4$ & $1.1\pm0.4$ & $-0.9\pm0.4$ & $-18.3\pm0.1$ & $2.58_{-0.02}^{+0.03}$ & -78.6 & 3 \\
F150DA-083 & 07:22:42.72 & $-$73:27:32.3 & $28.8\pm0.3$ & $0.2\pm0.3$ & $-0.3\pm0.3$ & $-19.3\pm0.1$ & $4.38_{-0.09}^{+0.34}$ & -72.1 & 2,3 \\
\enddata
\tablecomments{(1) Name.
(2) Right ascension.
(3) Declination.
(4) Total magnitude in the $F356W$ band with $1\sigma$ errors.
(5) $F150W-F200W$ color with $1\sigma$ errors.
(6) $F200W-F356W$ color with $1\sigma$ errors.
(7) Absolute UV magnitude with $1\sigma$ errors. Values of galaxies in the SMACS J0723 field are after the lensing magnification correction with {\sc glafic}.
(8) Photometric redshift with $2\sigma$ errors.
(9) $\chi^2$ difference between the best high redshift solution and a lower redshift solution, $\Delta\chi^2=\chi^2(z_\m{low})-\chi^2(z_\m{high})$.
(10) Reference (This: this work, N22: \citet{2022arXiv220709434N}, C22: \citet{2022arXiv220709436C},
D22: \citet{2022arXiv220712356D}, F22: \citet{2022arXiv220712474F}, Y22: \citet{2022arXiv220711558Y}) and note for a reason not selected in this study (1: $>2\sigma$ detection in $F090W$, 2: $F150W-F200W<1.0$, 3: $\Delta\chi^2<9.0$, 4: $<5\sigma$ in the detection image).
}
\label{tab_photo_f150w}
\end{deluxetable*}

\startlongtable
\centerwidetable
\begin{deluxetable*}{cccccccccccc}
\tablecaption{Summary of Our $F200W$-Dropoout Candidates at $z\sim 16$}
\setlength{\tabcolsep}{2pt}
\tablehead{\colhead{ID} & \colhead{R.A.} & \colhead{Decl.} & \colhead{$F356W$} & \colhead{$F200W$-$F277W$} & \colhead{$F277W$-$F444W$}& \colhead{$M_\m{UV}$}& \colhead{$z_\m{phot}$}& \colhead{$\Delta\chi^2$} & \colhead{Ref./Note}\\
\colhead{(1)}& \colhead{(2)}& \colhead{(3)}& \colhead{(4)} &  \colhead{(5)}& \colhead{(6)}& \colhead{(7)}& \colhead{(8)}& \colhead{(9)}& \colhead{(10)} }
\startdata
\multicolumn{10}{c}{CEERS2}\\
CR2-z16-1 & 14:19:39.48 & $+$52:56:34.9 & $26.4\pm0.1$ & $1.5\pm0.1$ & $-0.2\pm0.1$ & $-21.9\pm0.1$ & $16.25_{-0.46}^{+0.24}$ & 15.5 & This, D22\\
\multicolumn{10}{c}{Stephan's Quintet}\\
S5-z16-1 & 22:36:03.81 & $+$33:54:16.7 & $26.8\pm0.4$ & $1.9\pm0.5$ & $-0.5\pm0.2$ & $-21.6\pm0.3$ & $16.41_{-0.55}^{+0.66}$ & 12.3 & This\\
\hline
\multicolumn{10}{c}{Other Possible Candidates that did not Meet Our Selection Criteria}\\
\multicolumn{10}{c}{\citet{2022arXiv220712338A}}\\
SMACS\_z16a & 07:23:26.39 & $-$73:28:04.5 & $28.0\pm0.3$ & $0.5\pm0.1$ & $-0.2\pm0.1$ & $-18.4\pm0.8$ & $10.61_{-8.55}^{+0.51}$ & -20.2 & 1,2\\
SMACS\_z16b & 07:22:39.57 & $-$73:30:08.2 & $28.3\pm0.2$ & $0.7\pm0.1$ & $-0.3\pm0.1$ & $-19.3\pm0.3$ & $15.15_{-1.22}^{+0.50}$ & 6.9 & 1,2\\
\multicolumn{10}{c}{\citet{2022arXiv220711558Y}}\\
F200DB-045 & 07:23:22.77 & $-$73:27:39.7 & $29.6\pm0.6$ & $0.1\pm0.3$ & $-0.3\pm0.4$ & $-16.9\pm2.6$ & $4.39_{-0.82}^{+1.73}$ & -9.8 & 1,2 \\
F200DB-086 & 07:23:06.42 & $-$73:27:19.8 & $29.9\pm0.5$ & $>1.4$ & $0.4\pm0.2$ & $-16.8\pm2.1$ & $17.13_{-13.12}^{+2.87}$ & 1.6 & 2 \\
F200DB-159 & 07:23:25.35 & $-$73:26:46.0 & $29.4\pm0.6$ & $-0.6\pm0.2$ & $-0.2\pm0.3$ & $-17.7\pm1.1$ & $3.42_{-0.23}^{+0.02}$ & -6.1 & 1,2 \\
F200DA-006 & 07:22:40.35 & $-$73:30:10.3 & $28.5\pm0.3$ & $0.2\pm0.2$ & $-0.7\pm0.3$ & $-19.7\pm0.2$ & $10.50_{-1.32}^{+0.73}$ & -10.8 & 1,2 \\
F200DA-033 & 07:22:43.92 & $-$73:29:15.7 & $25.8\pm0.1$ & $2.8\pm0.3$ & $1.4\pm0.1$ & $-22.5\pm0.1$ & $5.75_{-0.72}^{+0.40}$ & -7.0 & 2 \\
F200DA-034 & 07:23:05.20 & $-$73:29:13.4 & $28.7\pm0.3$ & $>1.5$ & $-0.2\pm0.3$ & $-19.6\pm0.3$ & $5.41_{-0.01}^{+0.16}$ & -72.0 & 2 \\
F200DA-040 & 07:23:03.93 & $-$73:29:06.1 & $28.7\pm0.3$ & $0.3\pm0.2$ & $-0.3\pm0.1$ & $-19.2\pm0.2$ & $3.94_{-0.46}^{+0.97}$ & -102.5 & 1,2 \\
F200DA-056 & 07:22:37.03 & $-$73:28:41.5 & $29.2\pm0.3$ & $1.0\pm0.3$ & $-0.2\pm0.1$ & $-19.2\pm0.2$ & $5.19_{-0.35}^{+0.33}$ & -95.3 & 2 \\
F200DA-061 & 07:22:31.69 & $-$73:28:38.6 & $28.1\pm0.2$ & $0.5\pm0.2$ & $-0.2\pm0.1$ & $-19.6\pm0.2$ & $4.91_{-1.06}^{+0.70}$ & -97.1 & 1,2 \\
F200DA-089 & 07:22:32.43 & $-$73:28:06.8 & $28.1\pm0.2$ & $1.6\pm0.9$ & $0.7\pm0.1$ & $-19.8\pm0.2$ & $15.73_{-10.22}^{+4.21}$ & 0.3 & 2 \\
F200DA-098 & 07:22:34.80 & $-$73:28:00.2 & $29.7\pm0.9$ & $>1.2$ & $0.3\pm0.3$ & $-18.6\pm0.7$ & $5.58_{-0.70}^{+0.50}$ & -17.0 & 2 \\
\enddata
\tablecomments{(1) Name.
(2) Right ascension.
(3) Declination.
(4) Total magnitude in the $F356W$ band with $1\sigma$ errors.
(5) $F200W-F277W$ color with $1\sigma$ errors.
(6) $F277W-F444W$ color with $1\sigma$ errors.
(7) Absolute UV magnitude with $1\sigma$ errors. Values of galaxies in the SMACS J0723 field are after the lensing magnification correction with {\sc glafic}.
(8) Photometric redshift with $2\sigma$ errors.
(9) $\chi^2$ difference between the best high redshift solution and a lower redshift solution, $\Delta\chi^2=\chi^2(z_\m{low})-\chi^2(z_\m{high})$.
(10) Reference (This: this work, D22: \citet{2022arXiv220712356D}) and note for a reason not selected in this study (1: $F200W-F277W<1.0$, 2: $\Delta\chi^2<9.0$).
}
\label{tab_photo_f200w}
\end{deluxetable*}

\subsection{Comparison with Previous Studies}\label{ss_comparison}

Some other studies identified galaxy candidates at $z>9$ using the JWST NIRCam ERO and/or ERS datasets.
Here we review these studies and compare their samples with our galaxy samples.
Tables \ref{tab_photo_f115w}-\ref{tab_photo_f200w} summarize properties of other possible candidates that were selected in other studies but did not meet our selection criteria.
\redc{These comparisons were conducted on November 20, 2022, and we clarify the version of the paper we compared in the following sections.}

\subsubsection{\citet{2022arXiv220709434N}}

Using the ERS CEERS and GLASS datasets, \citet{2022arXiv220709434N} found two bright galaxy candidates at $z\sim10$ and $12$, GLASS-z10 and GLASS-z12, which correspond to GL-z9-1 and GL-z12-1 in our sample, respectively.
Their estimates of the photometric redshifts ($z=10.35^{+0.38}_{-0.51}$ and $z=12.38^{+0.13}_{-0.27}$ for GL-z10-1 and GL-z12-1, respectively, with {\sc prospector}, \redc{from the ApJL published version}) are consistent with our estimates ($z=10.49^{+0.53}_{-0.72}$ and $z=12.28^{+0.08}_{-0.07}$).

\subsubsection{\citet{2022arXiv220709436C}}

\citet{2022arXiv220709436C} identified seven galaxy candidates at $z\sim9-12$ with the color selection using the ERS GLASS dataset.
\redc{Among the six candidates from the ApJL published version,} three candidates, GHZ1, GHZ2, and GHZ4, are selected in our selection.
GHZ1 (GHZ2) is GL-z9-1 (GL-z13-1) in our sample, and their photometric redshift, $z=10.53-10.63$ ($z=12.11-12.30$) is comparable with our estimates.
GHZ4 was identified in our selection as GL-z9-2, and their photometric redshift ($z=9.93-10.08$) agrees with our estimate ($z=10.46^{+0.45}_{-0.99}$).
The other three candidates, GHZ3, GHZ5, and GHZ6, did not meet our selection criteria, due to a possible detection in the $F090W$-band or $\Delta\chi^2<9$, although GHZ3 and GHZ5 were also reported to have low redshift solutions in \citet{2022arXiv220709436C}.

\subsubsection{\citet{2022arXiv220515388L}} 

\citet{2022arXiv220515388L} studied photometric properties of galaxies at $7<z<9$ using the ERS GLASS dataset.
\redc{We refer to the manuscript version 2 that was submitted to arXiv on October 4.}
Since their galaxies are identified from the $F090W$-dropout selection, we do not expect significant overlap between their galaxy sample and ours.

\subsubsection{\citet{2022arXiv220711217A}}

\citet{2022arXiv220711217A} identified four galaxy candidates at $9<z<12$ in the ERO dataset taken in the SMACS J0723 field.
\redc{We refer to the manuscript version 2 that was submitted to arXiv on August 9.}
Since three of them are expected to be at $z<10$, it is reasonable that they are not identified in our $F150W$-dropout selection ($z\sim12$).
The other source (ID 10234) is estimated to be at $z=11.42$, around the edge of our redshift selection window (see Figure \ref{fig_completeness}), and not selected in our study due to its insufficient $F150W-F200W$ color and a possible detection in the $F090W$-band.

\subsubsection{\citet{2022arXiv220712338A}}

Using the ERO dataset in the SMACS J0723 field, \citet{2022arXiv220712338A} selected 10 galaxy candidates at $10<z<16$.
\redc{We refer to the manuscript version 2 that was submitted to arXiv on October 31}.
Among them, four candidates, SMACS\_z12a, SMAC\_z12b, SMACS\_z16a, and SMACS\_z16b, have photometric redshifts of $z>12$, and are expected to be overlapped in our galaxy catalogs.
However, none of them are selected as high redshift galaxy candidates in our study, due to their insufficient colors or $\Delta\chi^2<9$.

\subsubsection{\citet{2022arXiv220712356D}}\label{ss_D22}

\citet{2022arXiv220712356D} selected 45 galaxies at $z>8.5$ using ERO SMACS J0723 and ERS GLASS and CEERS datasets.
\redc{We refer to the manuscript version 2 that was submitted to arXiv on October 22}.
Among the 45 galaxies, three galaxies (IDs 1698, 6415, and 17487) are identified in the GLASS dataset, and are also selected in this study as GL-z9-1, GL-z9-4, and GL-z12-1, respectively.
Their photometric redshifts ($z=10.45^{+0.26}_{-0.16}$, $z=10.79^{+0.45}_{-0.66}$, and $z=12.42^{+0.27}_{-0.21}$ for IDs 1698, 6415, and 17487, respectively) are consistent with our estimates ($z=10.49^{+0.53}_{-0.72}$, $z=10.19^{+0.63}_{-0.55}$, and $z=12.28^{+0.08}_{-0.07}$).
The brightest candidate in \citet{2022arXiv220712356D} is ID 93316 at $z=16.39^{+0.32}_{-0.22}$, which is CR2-z16-1 at $z=16.25^{+0.24}_{-0.46}$ in our catalog.
In the version 2, \citet{2022arXiv220712356D} newly selected ID 32395\_2 at $z=12.29^{+0.91}_{-0.32}$, which is also selected in this study as CR2-z12-1 at $z=11.63^{+0.51}_{-0.53}$, which was firstly identified in \citet{2022arXiv220712474F}
\citet{2022arXiv220712356D} presented other three candidates at $z>12$, but these candidates are not selected in this study due to $\Delta\chi^2<9$.

\subsubsection{\citet{2022arXiv220712474F}}

One galaxy candidate at $z\sim14$, dubbed Maisie's Galaxy in \citet{2022arXiv220712474F}, is also selected in this study as CR2-z12-1.
\redc{We refer to the manuscript version 2 that was submitted to arXiv on September 7}.
The photometric redshift presented in \citet{2022arXiv220712474F} is $z=11.8^{+0.3}_{-1.2}$, consistent with our estimate ($z=11.63^{+0.51}_{-0.53}$).

\begin{figure*}
\centering
\begin{minipage}{0.32\hsize}
\begin{center}
\includegraphics[width=0.99\hsize, bb=16 5 343 356]{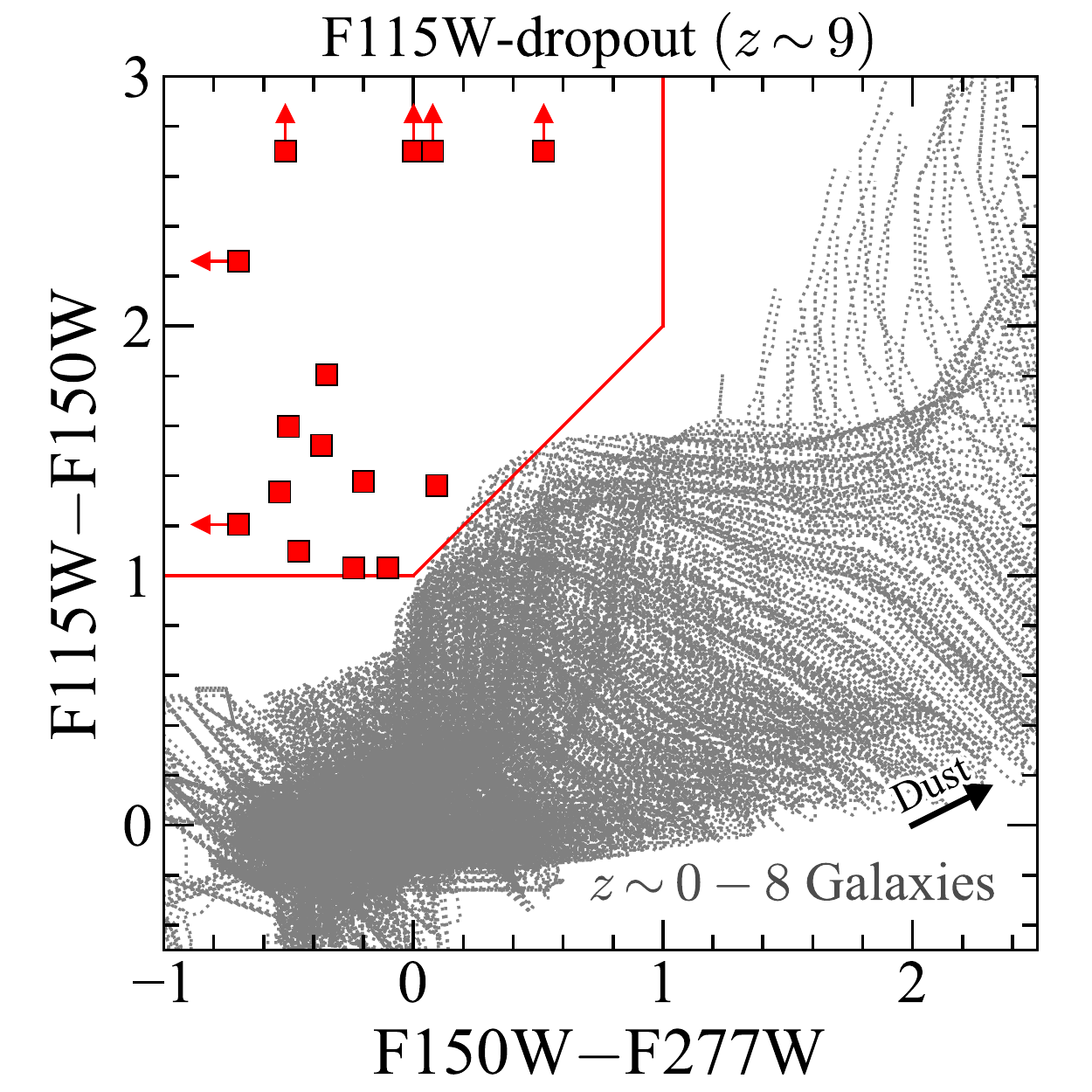}
\end{center}
\end{minipage}
\begin{minipage}{0.32\hsize}
\begin{center}
\includegraphics[width=0.99\hsize, bb=16 5 343 356]{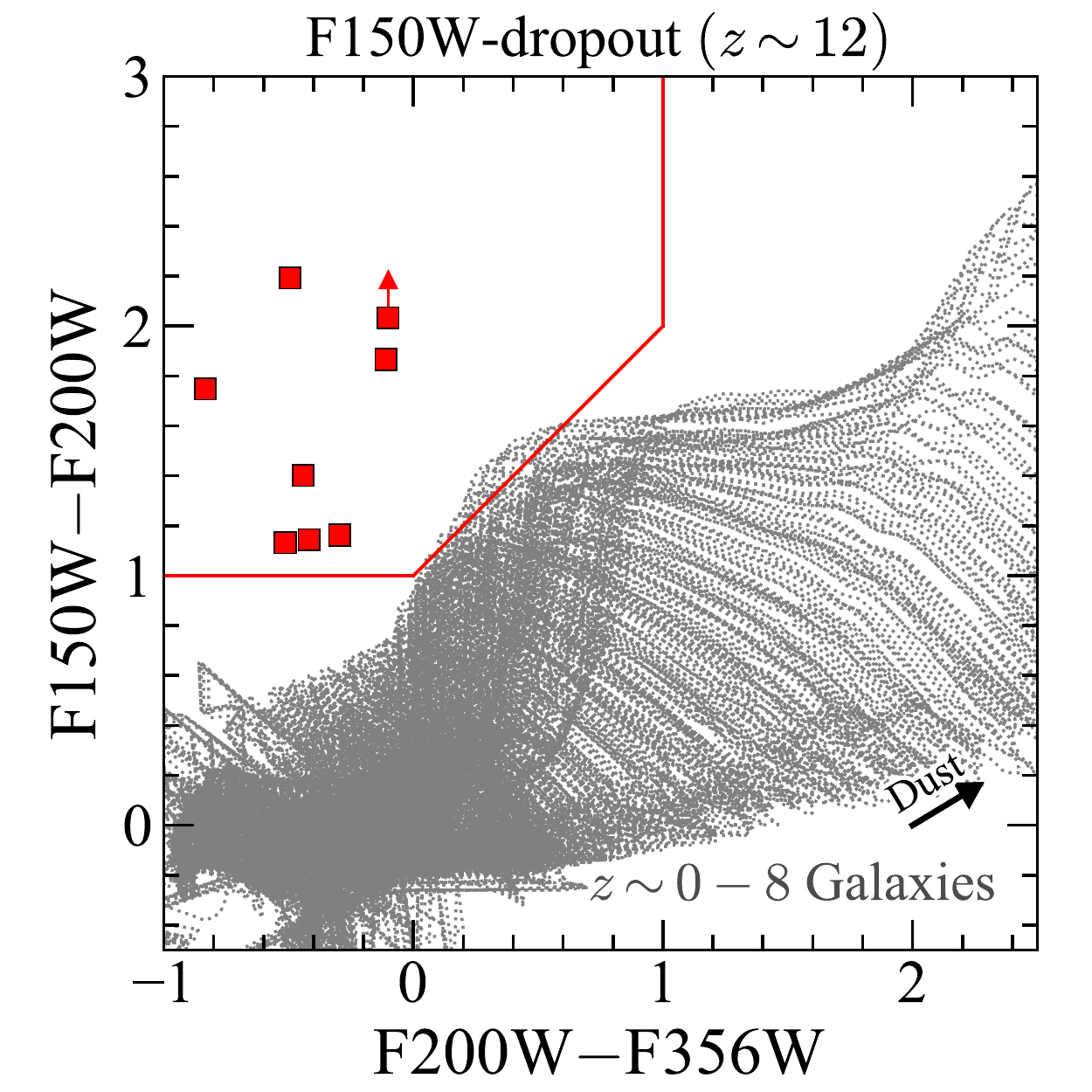}
\end{center}
\end{minipage}
\begin{minipage}{0.32\hsize}
\begin{center}
\includegraphics[width=0.99\hsize, bb=16 5 343 356]{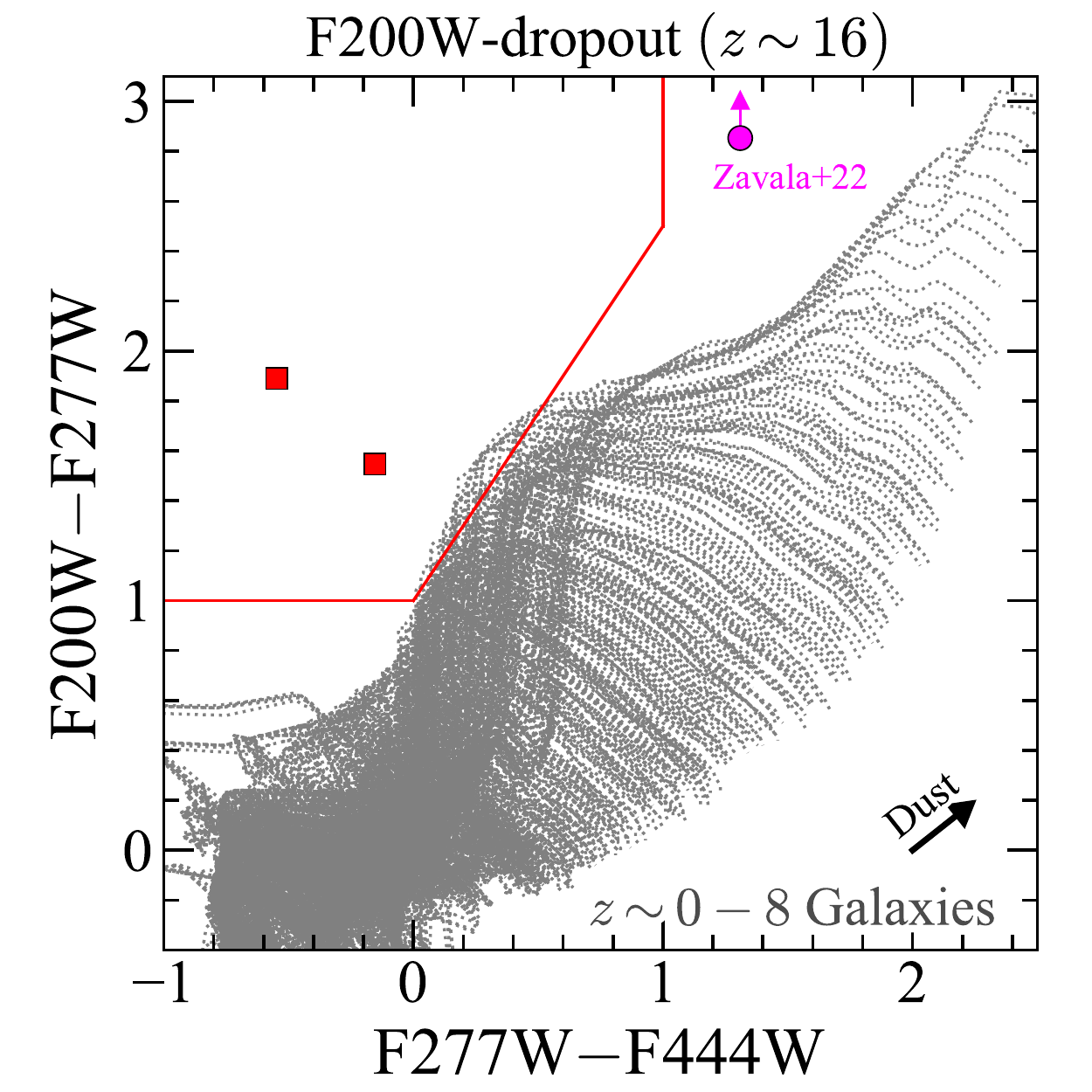}
\end{center}
\end{minipage}
\caption{Same as Figure \ref{fig_2color}, but for evaluating the interlopers of foreground galaxies. The gray curves indicate colors of model galaxies at $z=0-8$ that are produced with {\sc PANHIT} \citep{2020IAUS..341..285M}. See texts for details of the models.
The black arrow indicates a shift of the colors with dust extinction of $\Delta E(B-V)=+0.1$.
\redc{The magenta circle in the right panel is a dusty starburst galaxy at $z\sim5$ that may appear as a $z>15$ galaxy discussed in \citet{2022arXiv220801816Z}.}
Our color selection criteria avoid these low-redshift interlopers at $z\sim0-8$.
}
\label{fig_2color_wlowz}
\end{figure*}

\subsubsection{\citet{2022arXiv220711558Y}}

\citet{2022arXiv220711558Y} identified a total of 88 galaxy candidates at $z\sim11-20$ in the ERO SMACS J0723 field, including 63 $F150W$-dropouts and 15 $F200W$-dropouts, possibly overlapping with our $F150W$ and $F200W$-dropout candidates, respectively.
\redc{We refer to the manuscript version 1 that was submitted to arXiv on July 23}.
Out of 61 and 15 sources in their $F150W$-dropouts and $F200W$-dropouts, we identify 54 and 11 objects in our original photometric catalogs, respectively.
However, we cannot identify counterparts of the remaining 11 sources, F150DA-013, F150DA-047, F150DA-057, F150DB-004, F150DB-023, F150DB-056, F150DB-058, F200DB-015, F200DB-109, F200DB-175, and F200DB-181, probably because their SNRs are not sufficient to be identified in this study, the source is severely affected by nearby bright objects, or a WCS offset between \citet{2022arXiv220711558Y} and this study is too large to identify the counterparts.
Among the 54 objects identified as $F150W$-dropouts, F150DA-053 at $z=11.71^{+1.56}_{-0.54}$ is SM-z12-1 at $z=12.47^{+1.19}_{-0.72}$ in this study.
We have checked photometry of the other 53 and 11 objects identified in our original photometric catalogs, but none of them are selected as high redshift candidates in this study, due to their insufficient colors of the break, $\Delta\chi^2<9$, and/or an insufficient S/N in the detection image.

\subsubsection{Summary of the Comparisons}

In summary, we have found that bright candidates reported in previous studies are reproduced in this study, such as GL-z9-1, GL-z12-1, CR2-z12-1, and CR2-z16-1.
However, some of faint candidates reported in other studies are not selected in our selection criteria, because most of these faint candidates are selected by photometric redshifts but with a weak criterion (e.g., $\Delta\chi^2>4$) or by relatively weak color selection criteria (e.g., $F150W-F200W>0.5$).
It is expected that the contamination fraction in such faint candidates is high, given the small $\Delta\chi^2$ values (see discussions in Section \ref{ss_SEDfit}).
These comparisons indicate that our selection criteria are conservative enough to remove foreground interlopers while keeping bright and reliable candidates.

\begin{figure*}
\centering
\begin{center}
\includegraphics[width=0.85\hsize, bb=0 0 1008 174]{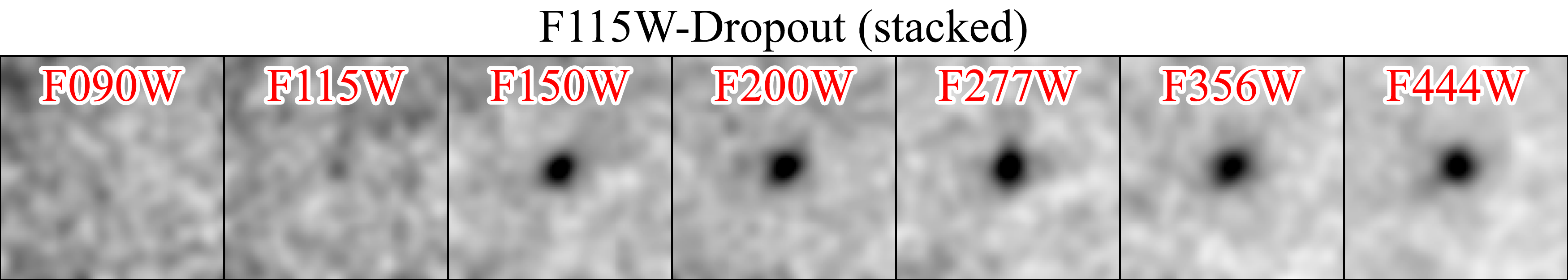}
\includegraphics[width=0.99\hsize, bb=0 0 1152 174]{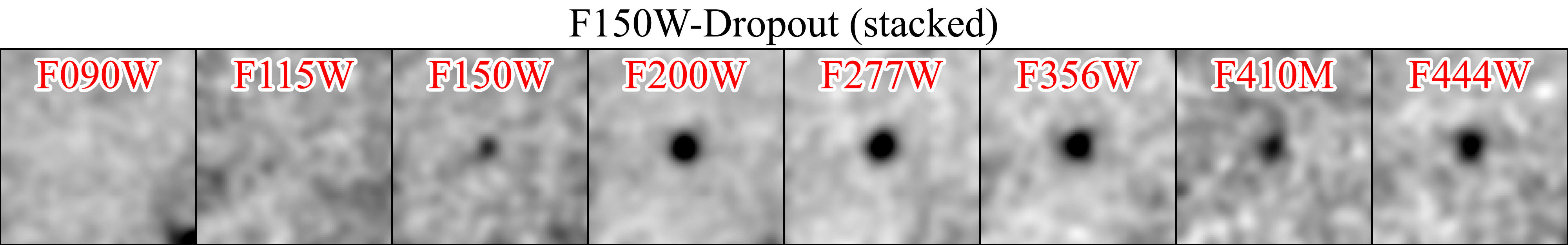}
\end{center}
\caption{Stacked images of our $F115W$-dropouts (top) and $F150W$-dropouts (bottom). The size of the images is $1.\carcsec5\times1.\carcsec5$.
There are no positive signals found at the positions of the dropouts in the $F090W$ image (top) 
and the $F090W$ and $F115W$ images (bottom) whose wavelength ranges
(rest-frame $<1216$\AA) do not include emission from $z\sim 9$ and $12$ sources, indicating that our samples are not significantly contaminated by foreground interlopers.
}
\label{fig_snapshot_stack}
\end{figure*}

\subsection{Contamination}

We check whether our sample is largely contaminated by foreground interlopers or not.
One of the major sources for contamination is low redshift galaxies whose Balmer breaks are redshifted to the wavelength of the Lyman break of our dropout galaxies.
To test the effect of such contamination, we make a mock catalog of galaxies with Balmer breaks at $z=0-8$.
We first generate model spectra of galaxies by using PANHIT \citep{2020IAUS..341..285M} assuming a delayed-$\tau$ star formation history, $\tau=0.01$, $0.03$, $0.1$, $0.3$, and $1$ Gyr; stellar age of $0.01-1.3$ Gyr; and metallicity of $Z=0.0001$, 0.0004, 0.004, 0.008, 0.02, and 0.05.
In Figure \ref{fig_2color_wlowz} we plot model tracks of the $z=0-8$ galaxies with colors of our selected galaxy candidates and our color selection criteria.
We find that the $z=0-8$ galaxies with Balmer breaks have a relatively small break color (e.g., $F150W-F200W<1.0$ in the $F150W$-dropout selection) or larger break color and red continuum color ($F150W-F200W>1.0$ and $F200W-F356W>0$).
Our color selection criteria avoid model tracks of these $z=0-8$ galaxies, and most of our candidates are located far from these model tracks.
\redc{In the right panel of Figure \ref{fig_2color_wlowz}, we also plot a dusty starburst galaxy at $z\sim5$ that may appear as a $z>15$ galaxy discussed in \citet{2022arXiv220801816Z}.
Such a dusty interloper is also removed from our sample due to the red continuum color ($F277W-F444W>1.0$).}

\redc{To evaluate the effect of contamination from low redshift objects scattering into our selection criteria due to the photometric noise at the depth of the observations, we conduct Monte Carlo simulations using the real datasets in the same manner as previous studies \citep[e.g.,][]{2015ApJ...803...34B,2018PASJ...70S..10O,2022ApJS..259...20H}.
We start from multi-band catalogs constructed in the GLASS field whose images are sufficiently deep.
We create 100 mock catalogs by perturbing the measured fluxes by adding photometric scatters based on the flux uncertainties in each band in the CEERS and Stephan's Quintet fields whose depths are shallower than the GLASS field.
We select high redshift galaxies from the mock catalogs with the same selection criteria as our real selection.
In the same manner as \citet{2015ApJ...803...34B}, we classify sources that are selected but that show detections in the band bluer than the break in the original catalogs as contaminants.
Based on these simulations, we find that the contamination rate due to the scatter is $<6\%$ for the $F150W$ and $F200W$-dropout selections.
The contamination rates for the $F115W$-dropout selection cannot be evaluated with this procedure, because the galaxy selection is conducted only in the GLASS field where we start the simulation.
However, the good agreement in $z\sim9$ number densities between our results and previous studies (Section \ref{ss_result}) indicates that the contamination is not significant in our $F115W$-dropout sample.
}

To further test the contamination, we stack images of the 13 $F115W$-dropout candidates and 8 $F150W$-dropout candidates.
If the sample is significantly contaminated by low redshift interlopers, the stacked images should show signals in a band whose wavelength is bluer than the Lyman break.
Figure \ref{fig_snapshot_stack} presents stacked images of our $F115W$- and $F150W$-dropout candidates.
There are no significant positive signals found in the $F090W$-band and in the $F090W$- and $F115W$-bands for the $F115W$- and $F150W$-dropout candidates, respectively, suggesting that our samples are not significantly contaminated by low redshift interlopers.
These tests and comparisons in Section \ref{ss_comparison} indicate that our conservative selection criteria with careful screening of low redshift interlopers provide a reliable sample of $z\sim9-\redc{16}$ galaxy candidates, suitable for statistical studies such as luminosity function measurements.

\section{Mass Model} \label{ss_model}

Various mass models for SMACS J0723 are produced
by parametric mass modeling algorithms.
The RELICS survey \citep{2019ApJ...884...85C} team
provides the mass models\footnote{https://archive.stsci.edu/prepds/relics/}
developed with the
{\sc Lenstool} \citep{2007NJPh....9..447J,2022ApJ...928...87F} and 
{\sc glafic} \citep{2010PASJ...62.1017O} codes, both of which are constructed using the HST data. 
A new model with the Lenstool code is constructed with the JWST ERO data in \citet{2022arXiv220707101M}. 
\citet{2022arXiv220705007G} recently present 
a mass model of SMACS J0723 developed
by the light-traces-mass ({\sc LTM}; \citealt{2005ApJ...621...53B,2009MNRAS.396.1985Z,2015ApJ...801...44Z}) approach before the JWST ERO data release, and subsequently \citet{2022arXiv220707102P} present 
the {\sc LTM} mass modeling with the JWST ERO data.
\citet{2022arXiv220707567C} also develop the mass model of SMACS J0723 using {\sc Lenstool}.

%
In this paper, we construct an updated {\sc glafic} \citep{2010PASJ...62.1017O,2021PASP..133g4504O} strong lens mass model of SMACS0723 using the new JWST ERO data.
The magnification factors predicted by the updated {\sc glafic} mass model are compared with those from the other existing mass models to evaluate the lens-model uncertainty.
The {\sc glafic} code performs the so-called parametric lens mass modeling, where shapes of the mass distributions of the cluster are described by a superposition of a small number of lens mass components with known profile shapes, and parameters characterizing the lens mass components are determined so as to reproduce observed positions of multiple images.

As a specific procedure, we largely follow the methodology described in \citet{2016ApJ...819..114K}.
We model the dark matter halo by an elliptical Navarro-Frenk-White \citep[NFW;][]{1997ApJ...490..493N} density profile with an approximation to speed up the calculation of its lensing property (see \citealt{2021PASP..133g4504O}).
Model parameters associated with the NFW component are the mass, the center, the ellipticity and its position angle, and the concentration parameter.
In addition to the main NFW halo, we place an additional NFW component whose center is fixed to a bright cluster member galaxy located at North-West, (R.A., Decl.)=($110.7928634$, $-73.4476417$).
We also fix the concentration parameter of the additional NFW component to $c=10$, and fit its mass, ellipticity and position angle only.
Cluster member galaxies selected by photometric redshifts from the RELICS HST data \citep{2019ApJ...884...85C} are modeled by an elliptical pseudo Jaffe profile.
In order to reduce the number of parameters, the velocity dispersion $\sigma$ and the truncation radius $r_{\mathrm{trunc}}$ of each cluster member galaxy are assumed to scale with its luminosity (in HST F814W band) as $\sigma \propto L^{1/4}$ and $r_{\mathrm{trunc}} \propto L^\eta$ with their normalization and $\eta$ being treated as free parameters.
In addition we include an external shear to improve the fitting.
For multiple image sets without spectroscopic redshifts, we simultaneously fit their redshifts.

We search for the best-fitting model by the standard $\chi^2$ minimization, where $\chi^2$ is computed from the differences between observed and model-predicted positions.
We assume the positional error of $0\farcs 4$, which is a typical positional accuracy achieved by the parametric strong lens mass modeling \citep{2016ApJ...819..114K}. 
The $\chi^2$ is evaluated in the source plane, taking account of the full magnification tensor at each multiple image position (see Appendix 2 of \citealt{2010PASJ...62.1017O}).
Errors on model parameters are derived using the standard Markov Chain Monte Carlo technique.
Multiple images are identified iteratively, starting with secure sets of multiple images that are obvious from their colors, morphologies, and redshifts, constructing a preliminary mass model with those sets of multiple images, and searching for new multiple sets with help of the preliminary mass model.
In this work, we use conservative 12 sets of multiple images for our strong lens mass modeling, with the total number of multiple images of 38.
These multiple image sets are mostly consistent with other work using different lens modeling codes \citep{2022arXiv220707102P,2022arXiv220707101M}.
We adopt spectroscopic redshifts for five sets of multiple images given in the literature \citep{2022arXiv220705007G,2022arXiv220707102P,2022arXiv220707101M}.
Our best-fitting model has $\chi^2=28.3$ for degree of freedom of 32, representing a good fit.
The root-mean-square (rms) of differences between observed and predicted multiple image positions is $0\farcs 35$.

With the updated {\sc glafic} mass model, we calculate
the magnification factors $\mu$ of our dropout galaxy candidates and the effective survey volume.
Table \ref{tab_mu} summarizes the magnification factors of our dropout galaxy candidate and spectroscopically confirmed galaxies at $z>7$ calculated by {\sc glafic}, \citet{2022arXiv220707101M}, \citet{2022arXiv220707102P}, and \citet{2022arXiv220707567C}.
We find that the magnification factor calculated by each model agree well typically within $\sim20\%$.

\begin{deluxetable}{ccccccc}
\tablecaption{Summary of Magnification Factors Estimated by Various Mass Models}
\label{tab_mu}
\tablehead{\colhead{ID} & \colhead{$z$} & \colhead{$\mu_\m{glafic}$} & \colhead{$\mu_\m{M22}$} & \colhead{$\mu_\m{P22}$} & \colhead{$\mu_\m{C22}$} \\
\colhead{(1)}& \colhead{(2)}& \colhead{(3)} & \colhead{(4)}  & \colhead{(5)}  & \colhead{(6)} }
\startdata
SM-z12-1 & $12.47_{-0.72}^{+1.19}$ &1.22 & 1.11 & 1.00 &1.00 \\
s04590 & 8.495 & 8.69 & 5.81 & 6.90 & 7.42\\
s06355 & 7.664 & 1.78 & 1.68 & 1.43 & 1.68\\
s10612 & 7.659 & 1.86 & 1.68 & 1.61 & 1.66
\enddata
\tablecomments{
(1) Name.
(2) Spectroscopic or photometric redshift.
(3)-(6) Magnification factors estimated by {\sc grafic}, \citet{2022arXiv220707101M}, \citet{2022arXiv220707102P}, and \citet{2022arXiv220707567C}.}
\end{deluxetable}

\section{Luminosity Function} \label{ss_LF}

\subsection{Sample Completeness} \label{ss_completeness}

\begin{figure}
\centering
\begin{center}
\includegraphics[width=0.95\hsize, bb=7 13 349 279]{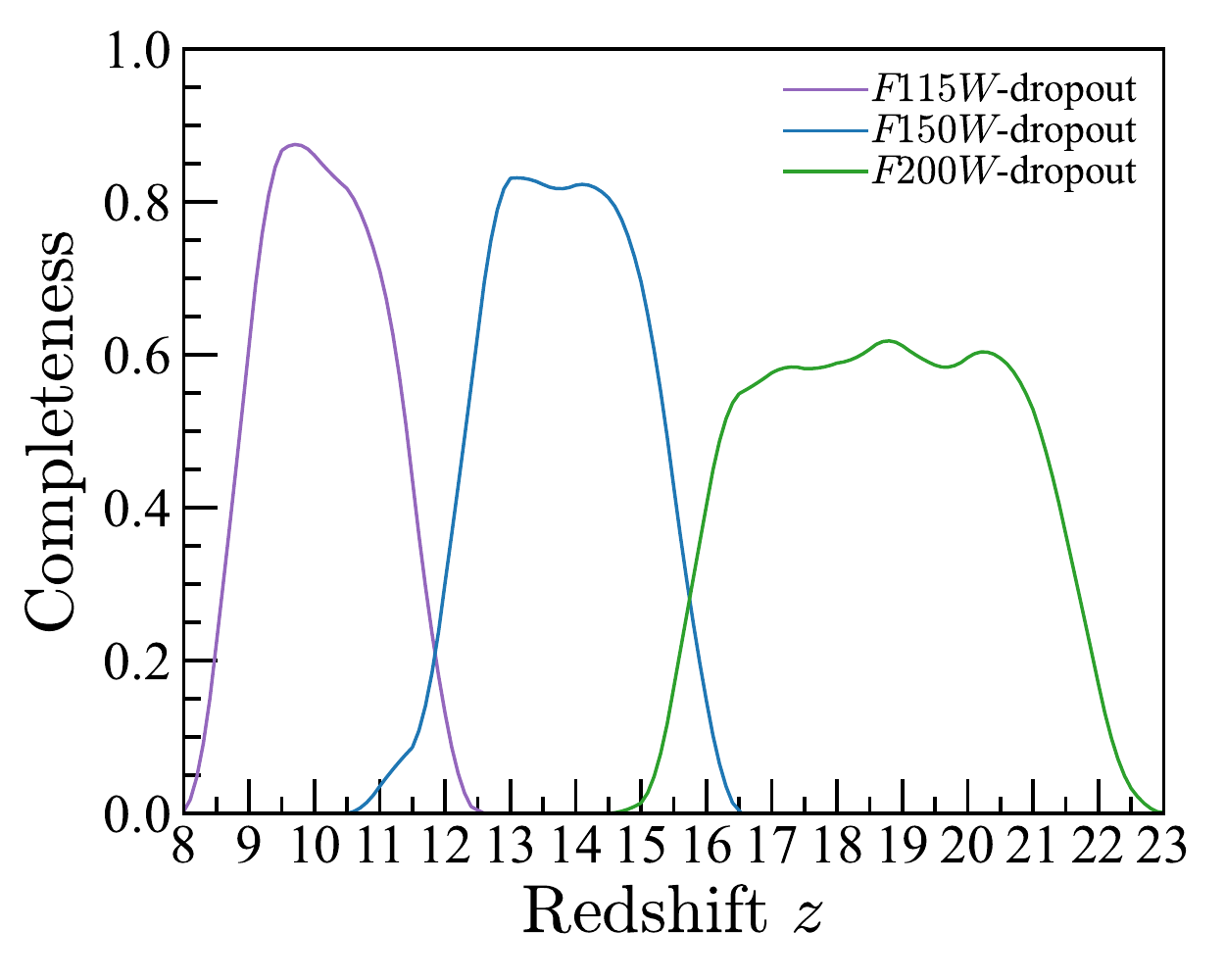}
\end{center}
\caption{
Selection completeness for our dropout galaxies.
The purple, blue, and green curves show selection completeness for the $F115W$-, $F150W$-, and $F200W$-dropout galaxies whose rest-frame UV ($\sim1500\ \m{\AA}$) magnitudes are $F200W=27.0$ mag, $F277W=27.0$ mag, and $F356W=27.0$ mag, respectively.
Each selection window is smoothed by $\Delta z=1.0$.
}
\label{fig_completeness}
\end{figure}

To derive the rest-frame UV luminosity function, we estimate the completeness of our dropout galaxy selection in the same manner as previous studies \citep[e.g.,][]{2009ApJ...706.1136O,2018PASJ...70S..10O,2022ApJS..259...20H}.
We conduct Monte Carlo simulations with real NIRCam images and artificial galaxies mocking high redshift galaxies.
The mock high redshift galaxies follow the size-$M_{\rm UV}$ redshift distribution revealed with the HST legacy data sets for galaxies  at $z\sim 0-10$ \citep{2015ApJS..219...15S} that is extrapolated to our redshift ranges, where the size-$M_{\rm UV}$ distribution is the log-normal distribution.
Our initial measurements of sizes for our $z>9$ galaxy candidates are consistent with this assumption within the uncertainties.
We adopt the S\'{e}rsic index $n=1$ found in typical galaxies  at $z\sim 5-10$  \citep{2013ApJ...777..155O,2015ApJS..219...15S} and the flat distribution of the intrinsic ellipticity in the range of 0.0-0.8.
Recent studies indicate that morphologies of $z\sim9-16$ galaxies identified in the JWST datasets are consistent with these assumptions \citep[e.g,][]{2022arXiv220813582O}.
The SEDs of the mock high redshift galaxies uniformly distribute over magnitude and redshift, and  have a color distribution agreeing with the $M_{\rm UV}$-$\beta_\m{UV}$ relation observationally determined at $z\sim8$ \citep{2014ApJ...793..115B}, where 
$\beta_\m{UV}$ is the UV spectral slope index. 
The IGM absorption of \citet{2014MNRAS.442.1805I} is applied to the SEDs, which produces absorption features in the wavelengths shorter than the Ly$\alpha$ line.
We produce 100 artificial objects of the mock high redshift galaxies with {\sc IRAF} \texttt{mkobject} in each redshift and magnitude bin, and place the artificial objects on  the real JWST NIRCam images. With the images, we perform the object
detections, photometry, the color selection, \redc{and the SED fitting} in the same manner as Section \ref{ss_catalog_selection}.
In the SMACS J0723 field, we consider the source magnification and multiply lensed images by using the mass model made with {\sc glafic} described in Section \ref{ss_model}.
Finally we calculate the selection completeness 
as a function of magnitude and redshift,
$C(m,z)$, with the photometric catalogs of the artificial high redshift galaxies. 
Figure \ref{fig_completeness} presents examples of the selection completeness thus obtained.
Although the average redshifts are $z=10.1$ for $F115W$-dropouts, $z=13.8$ for $F150W$-dropouts, and $z=18.7$ for $F200W$-dropouts, we use the median of photometric redshifts of our selected candidates, $z=9.1$ for $F115W$-dropouts, $z=12.0$ for $F150W$-dropouts, and $z=16.3$ for $F200W$-dropouts as the representative redshifts of our each dropout sample.

Based on the results of these selection completeness simulations, we estimate the survey volume per unit area as a function of apparent magnitude \citep{1999ApJ...519....1S},
\begin{equation}
V_\m{eff}(m)=\int C(m,z)\frac{dV(z)}{dz}dz,
\end{equation}
where $dV(z)/dz$ is the differential comoving volume as a function of redshift.
The space number density of our galaxy candidates that is corrected for incompleteness is calculated with the following equation:
\begin{equation}
\psi(m)=\frac{n(m)}{V_\m{eff}(m)},
\end{equation}
where $n(m)$ is the surface number density of selected galaxies in an apparent magnitude bin of $m$.
We convert the number density as a function of apparent magnitude, $\psi(m)$, into the UV luminosity functions, $\Phi[M_\m{UV}(m)]$, which are the number densities of galaxies as a function of rest-frame UV absolute magnitude.
Assuming a flat rest-frame UV continuum, we calculate the absolute UV magnitudes of galaxies from their apparent magnitudes in the bluest band not affected by the Lyman break, i.e., $F200W$, $F277W$, and $F356W$-bands for $F115W$, $F150W$, and $F200W$-dropout galaxy candidates, respectively.
The 1$\sigma$ uncertainty is calculated by taking into account the Poisson confidence limit \citep{1986ApJ...303..336G} and the cosmic variance.
We estimate the cosmic variance in the number densities using the bias values of $z\sim7$ galaxies obtained in \citet{2016ApJ...821..123H}, following the procedures in \citet{2004ApJ...600L.171S}.

\begin{figure*}
\centering
\begin{minipage}{0.48\hsize}
\begin{center}
\includegraphics[width=0.99\hsize, bb=7 9 430 358]{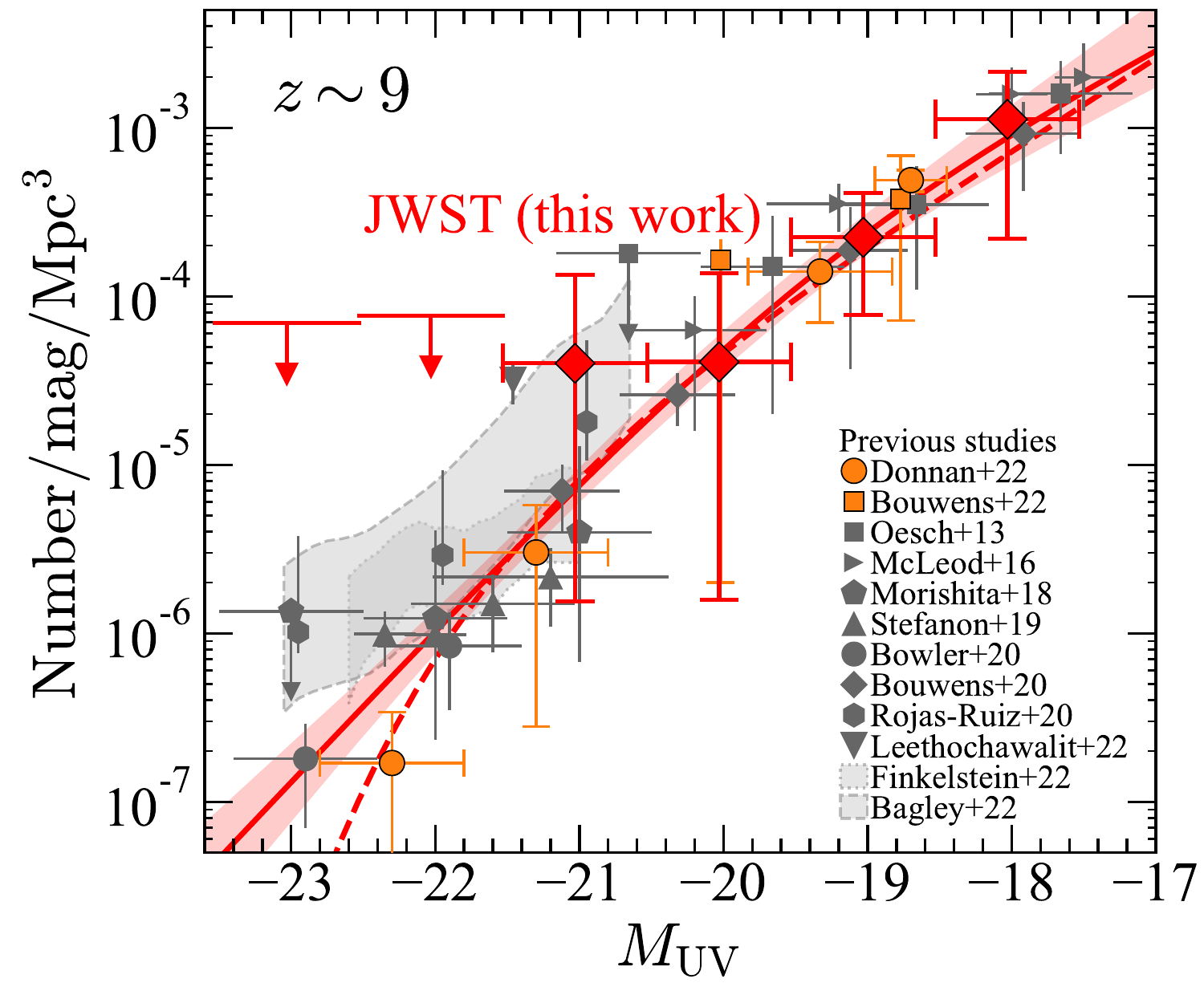}
\end{center}
\end{minipage}
\begin{minipage}{0.48\hsize}
\begin{center}
\includegraphics[width=0.99\hsize, bb=7 9 430 358]{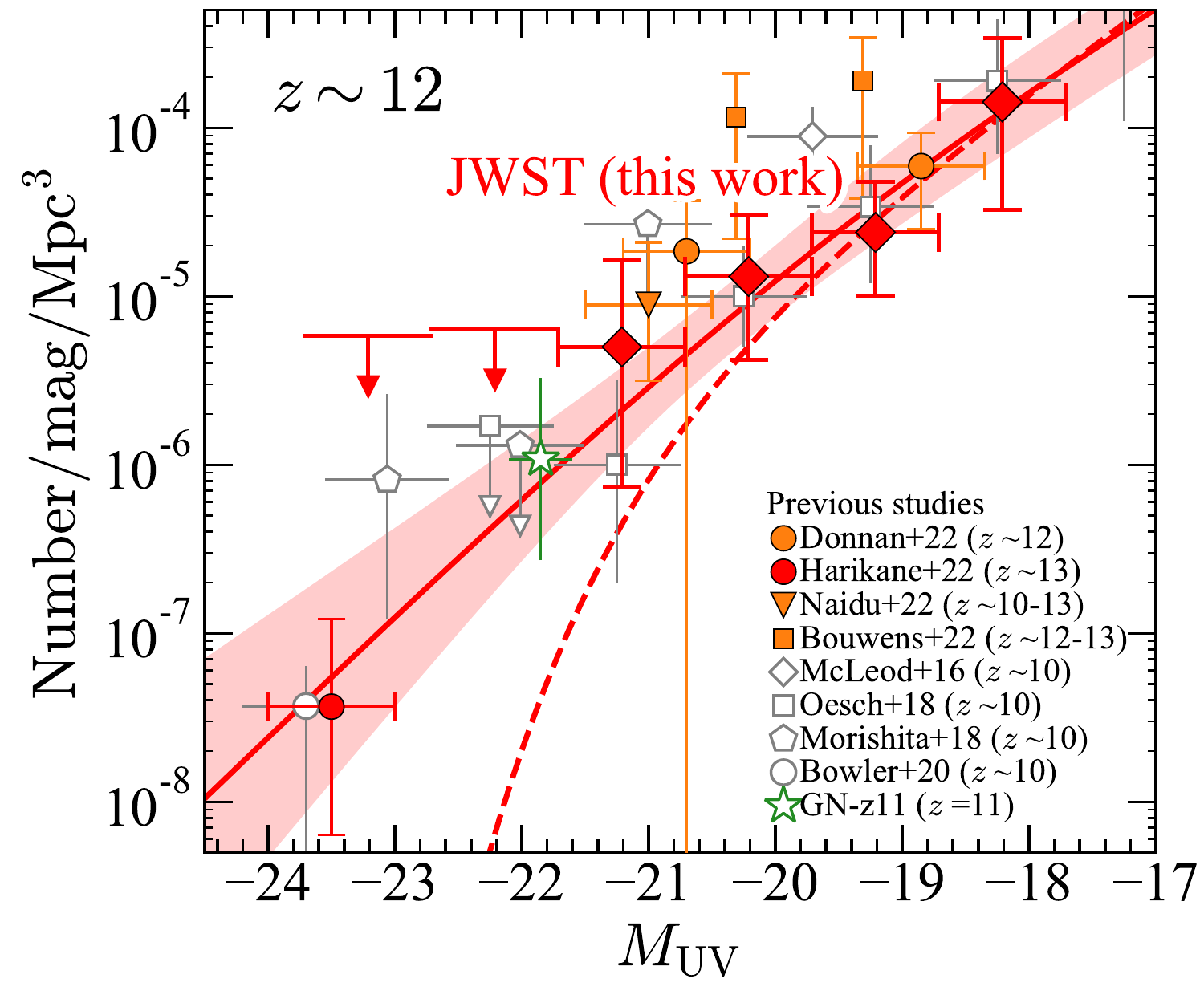}
\end{center}
\end{minipage}
\caption{UV luminosity functions at $z\sim 9$ (left) and $12$ (right).
The red diamonds represent the number densities of our galaxy candidates, while
the red arrows indicate the $1\sigma$ upper limits.
The errors include the cosmic variance (see text).
The red solid and dashed lines are our best-fit double power-law
and Schechter functions, respectively.
In the left panel, the orange circles indicates the luminosity functions at $z\sim9$ obtained in \citet{2022arXiv220712356D} using JWST data, and the gray symbols and shades 
denote the results at $z\sim 9$ derived
by the previous studies using HST or ground-based telescope data, 
\citet[squares]{2013ApJ...773...75O},
\citet[right-pointing triangles]{2016MNRAS.459.3812M},
\citet[pentagons]{2018ApJ...867..150M},
\citet[triangles]{2019ApJ...883...99S},
\citet[circles]{2020MNRAS.493.2059B},
\citet[diamonds]{2021AJ....162...47B},
\citet[hexagons]{2020ApJ...891..146R},
\citet[down-pointing triangle]{2022arXiv220515388L},
\citet[shade with dotted lines]{2022ApJ...928...52F},
and
\citet[shade with dashed lines]{2022arXiv220512980B}.
In the right panel, the orange circles, the red circle, the orange down-pointing triangle, and the orange squares indicate the number density of galaxies at $z\sim12$, $z\sim13$, $z\sim 10-13$, and $z\sim12-13$
reported by \citet{2022arXiv220712356D}, \citet{2022ApJ...929....1H}, \citet{2022arXiv220709434N}, and \citet{2022arXiv221102607B}, respectively.
The gray open symbols indicate
the luminosity functions at $z\sim 10$ obtained by
\citet[diamonds]{2016MNRAS.459.3812M},
\citet[squares]{2018ApJ...855..105O},
\citet[pentagons]{2018ApJ...867..150M}, and
\citet[circle]{2020MNRAS.493.2059B}.
The green open star mark represents the number density of GN-z11 \citep{2016ApJ...819..129O}.
See \citet{2022ApJ...929....1H} for the estimate of the number density and the UV magnitude of GN-z11.
Our estimated luminosity functions at $z\sim9$ and 12 agree well with previous HST and JWST results.
}
\label{fig_uvlf_1}
\end{figure*}

\begin{figure}
\centering
\begin{center}
\includegraphics[width=0.95\hsize, bb=7 9 430 358]{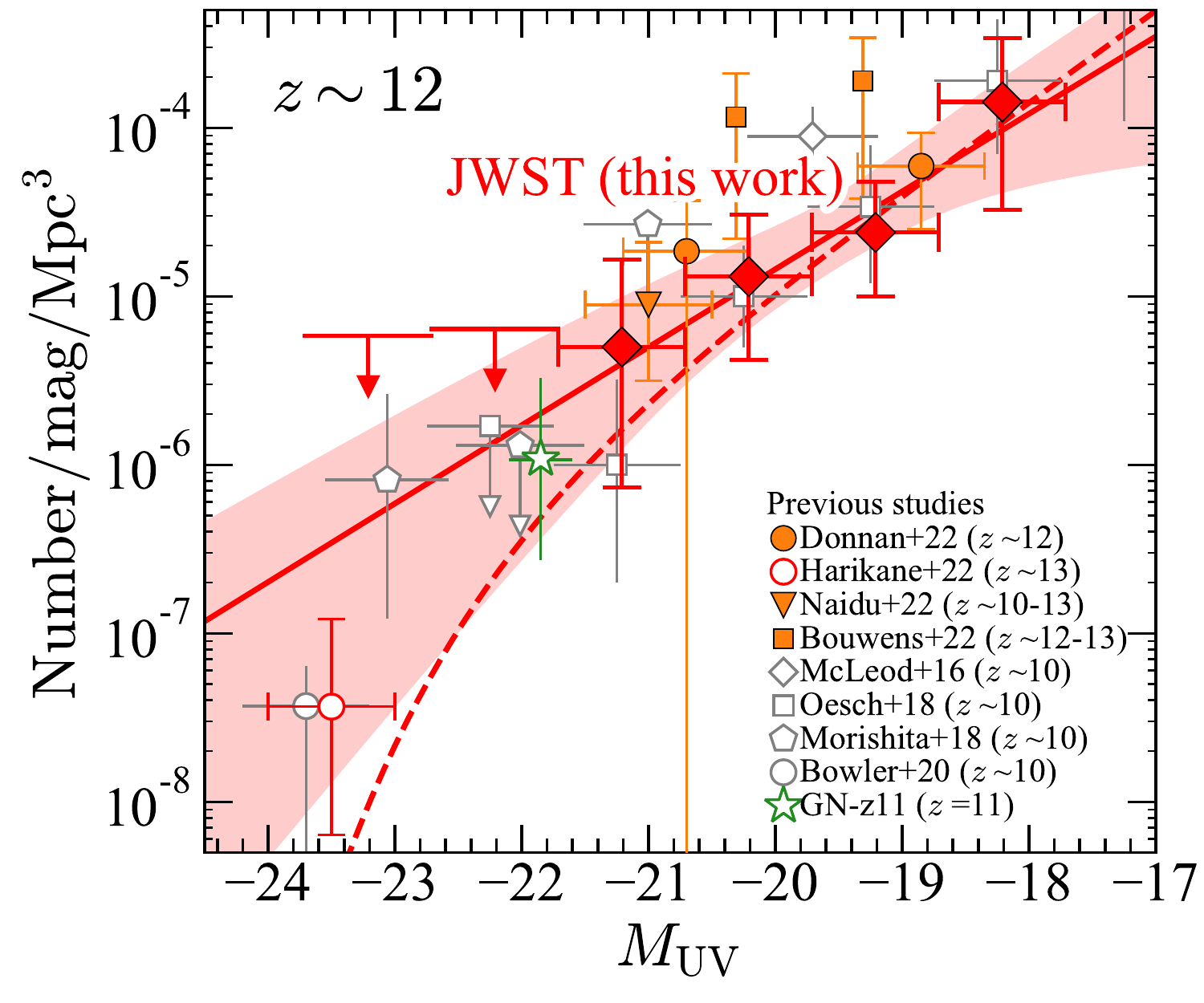}
\end{center}
\caption{
\redc{Same as the right panel of Figure \ref{fig_uvlf_1}, but with the fitting results without the brightest datapoint in \citet{2022ApJ...929....1H}.}
}
\label{fig_uvlf_1_woHD1}
\end{figure}

\begin{figure}
\centering
\begin{center}
\includegraphics[width=0.95\hsize, bb=7 9 430 358]{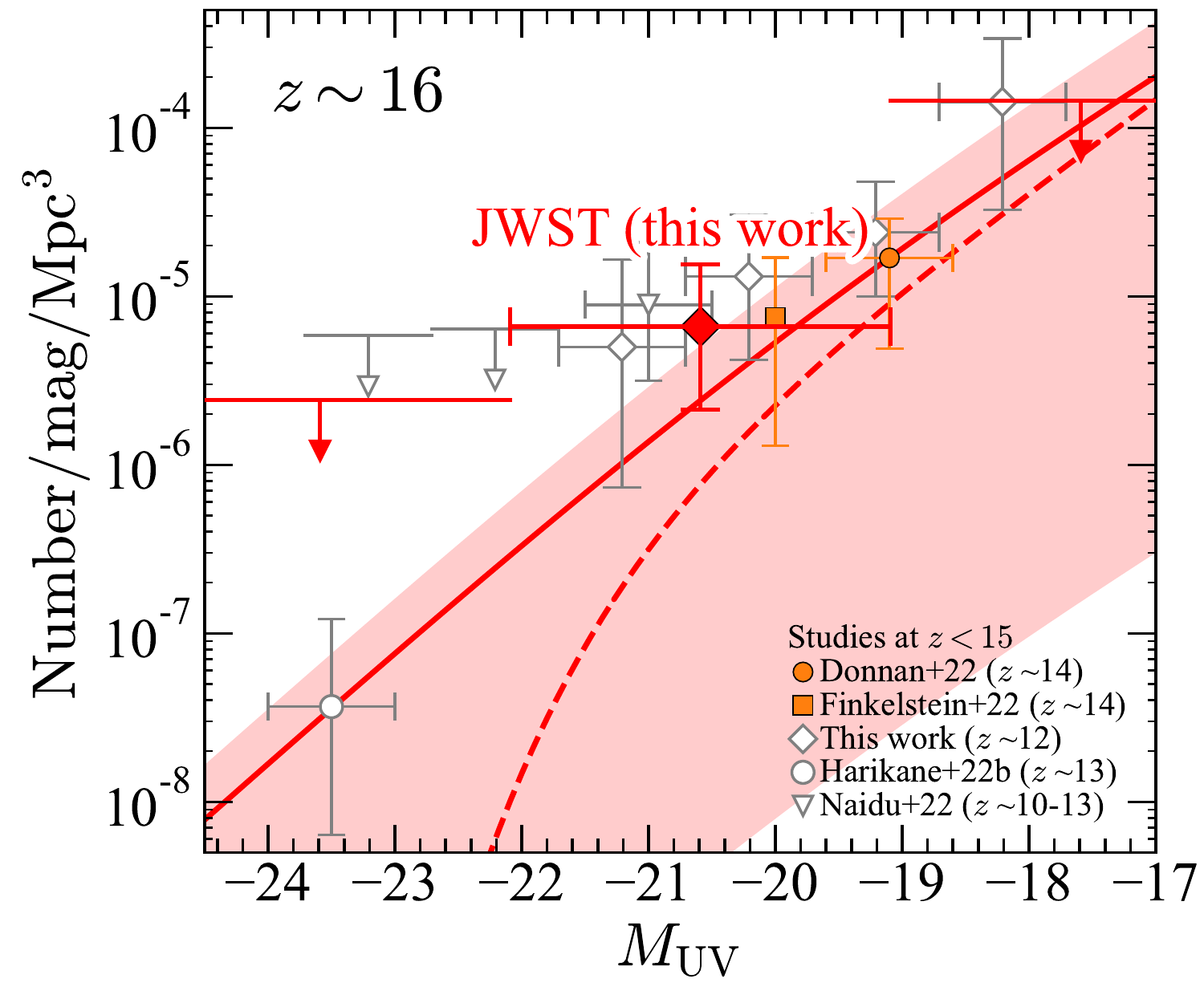}
\end{center}
\caption{
UV luminosity function at $z\sim \redc{16}$. The red diamond and the arrows
represent the number density of our galaxy candidates and the $1\sigma$ upper limits, respectively. For reference, we show
the UV luminosity functions at the lower redshifts,
$z\sim14$ (\citealt{2022arXiv220712356D}; orange filled circle),
$z\sim14$ (\citealt{2022arXiv220712474F}; orange filled square)
$z\sim 12$ (this study; gray open diamonds),
$z\sim 13$ (\citealt{2022ApJ...929....1H}; gray open circle), and
$z\sim 10-13$ (\citealt{2022arXiv220709434N}; gray open down-pointing triangle).
}
\label{fig_uvlf_2}
\end{figure}

\begin{figure}
\centering
\begin{center}
\includegraphics[width=0.95\hsize, bb=7 9 430 358]{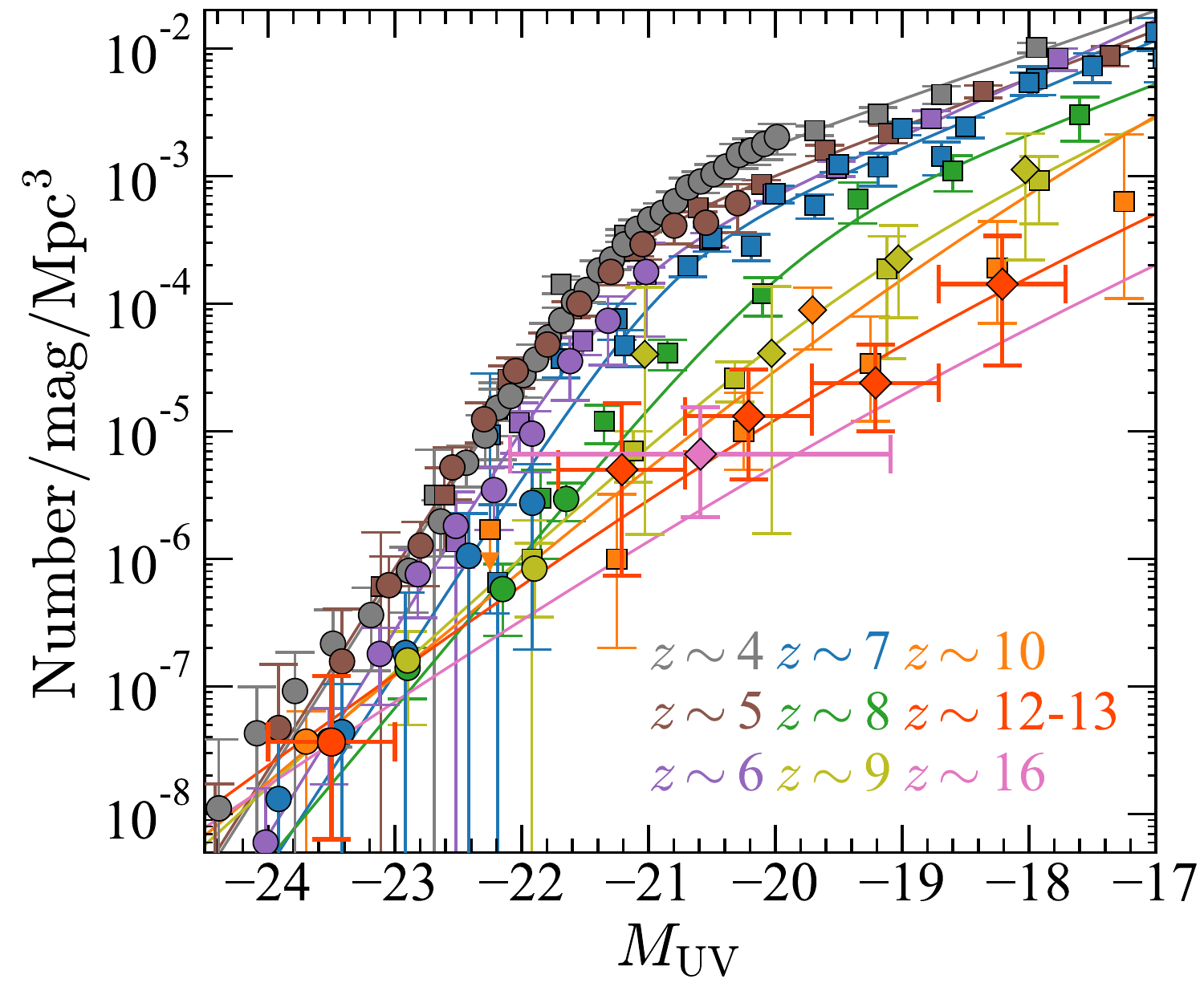}
\end{center}
\caption{
Evolution of the UV luminosity functions from $z\sim 4$ to $z\sim \redc{16}$.
The yellow, red, and pink diamonds represent our measurements of 
the luminosity functions at $z\sim9$, $12$, and $\redc{16}$, respectively,
whereas the red circle is the one obtained by \citet{2022ApJ...929....1H} at $z\sim13$.
The orange, yellow, green, blue, purple, brown, and gray symbols
indicate the luminosity functions at $z\sim 10$, $9$, $8$, $7$, $6$, $5$, 
and $4$, respectively. The circles at $z\sim 4-7$ and $8-10$ are the data
taken from \citet{2022ApJS..259...20H} and \citet{2020MNRAS.493.2059B},
respectively. The squares at $z\sim 4-9$ and $z\sim 10$ are the data
of \citet{2021AJ....162...47B} and \citet{2018ApJ...855..105O}, respectively.
The diamond at $z\sim 10$ represents the result of \citet{2016MNRAS.459.3812M}.
The lines denote the double power-law functions derived by
the previous studies for $z\sim 4-7$ \citep{2022ApJS..259...20H}
and $z\sim 8-13$ \citep{2020MNRAS.493.2059B}.
For clarity, we shift the data point of \citet{2020MNRAS.493.2059B} at $z\sim 10$
by $-0.2$ mag.
%
%
}
\label{fig_uvlf_evol}
\end{figure}

\begin{figure*}
\centering
\begin{minipage}{0.48\hsize}
\begin{center}
\includegraphics[width=0.99\hsize, bb=7 9 430 358]{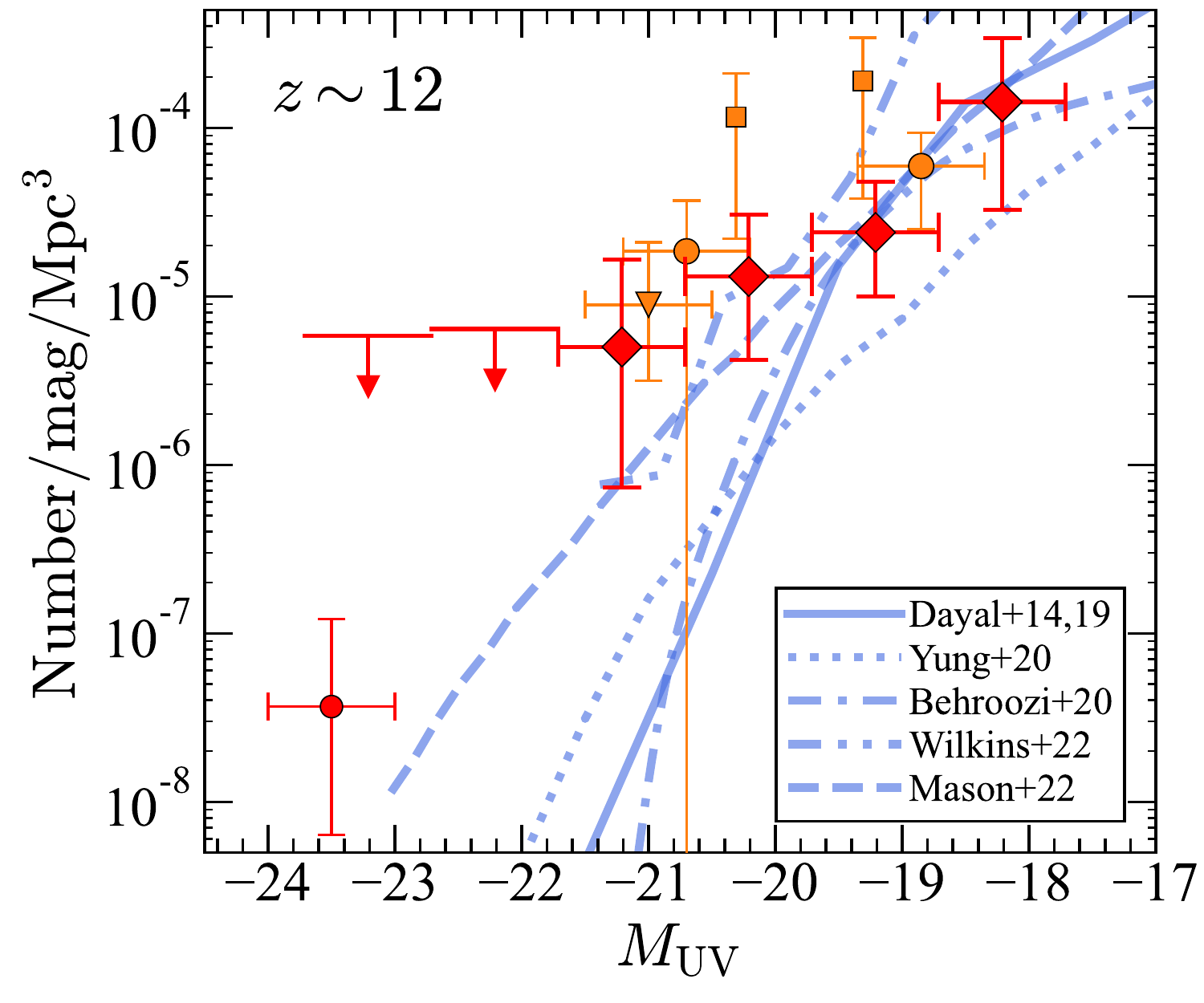}
\end{center}
\end{minipage}
\begin{minipage}{0.48\hsize}
\begin{center}
\includegraphics[width=0.99\hsize, bb=7 9 430 358]{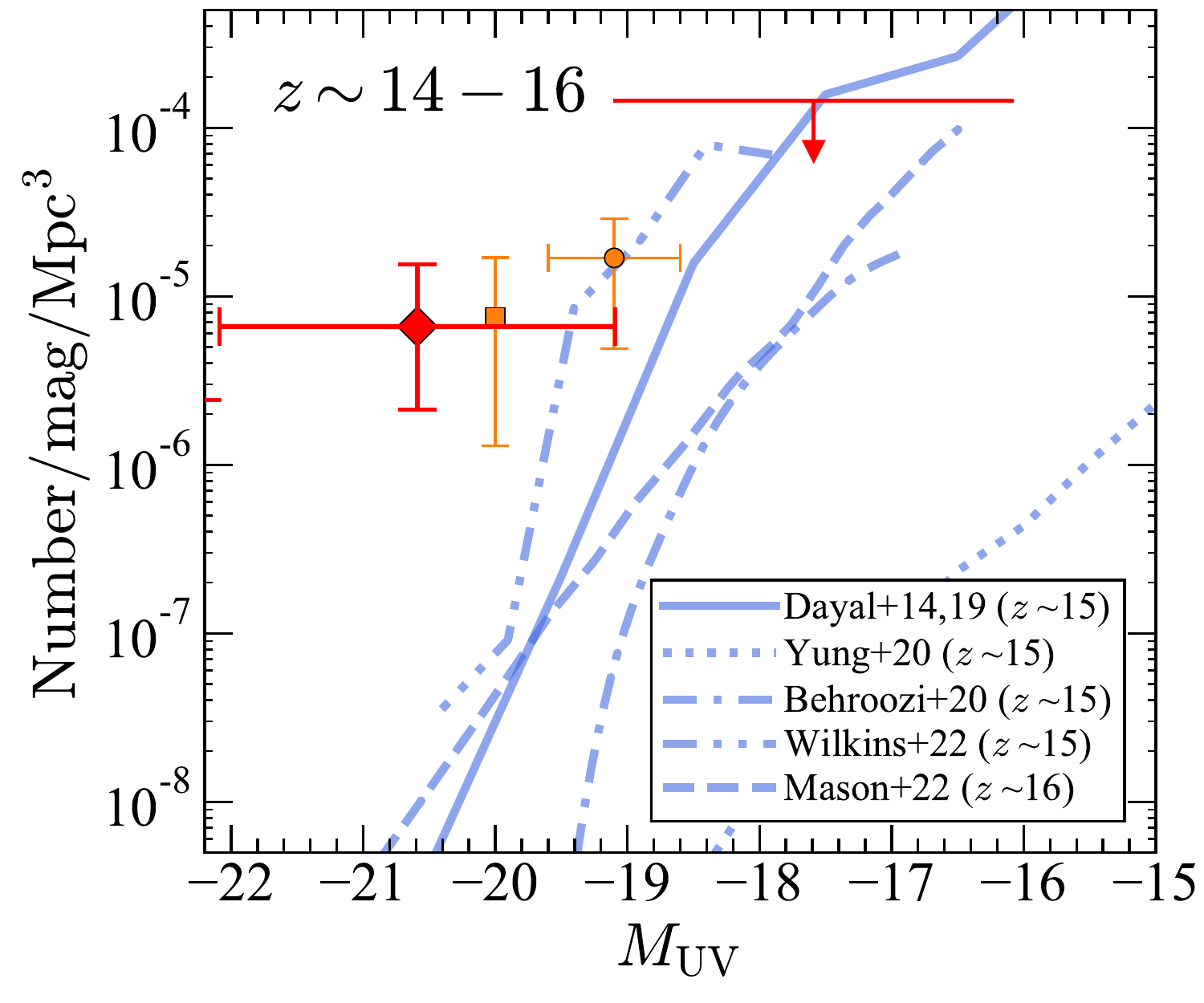}
\end{center}
\end{minipage}
\caption{
Comparison of the luminosity-function measurements 
with theoretical predictions and the empirical models at $z\sim 12$ (left) and $z\sim\redc{16}$ (right).
The blue lines show the theoretical and empirical models
obtained by 
\citet[solid line]{2014MNRAS.445.2545D,2019MNRAS.486.2336D},
\citet[dotted line]{2020MNRAS.496.4574Y},
\citet[dotted-dashed line]{2020MNRAS.499.5702B},
\citet[double-dotted dashed line]{2022arXiv220409431W}, and 
\citet[dashed line; their no dust model]{2022arXiv220714808M}.
The red and orange symbols show observational results in the same manner as Figures \ref{fig_uvlf_1} and \ref{fig_uvlf_2}.
The red diamonds and arrows represent the measurements and upper limits
obtained by this study. 
The orange circles, the red circle, the down-pointing orange triangle, and the orange square in the left (right) panel indicate the number densities reported by \citet{2022arXiv220712356D}, \citet{2022ApJ...929....1H}, \citet{2022arXiv220709434N}, and \citet{2022arXiv221102607B} (\citet{2022arXiv220712474F}), respectively.
}
\label{fig_uvlf_z12model}
\end{figure*}

\begin{figure}
\centering
\begin{center}
\includegraphics[width=0.95\hsize, bb=11 1 381 358]{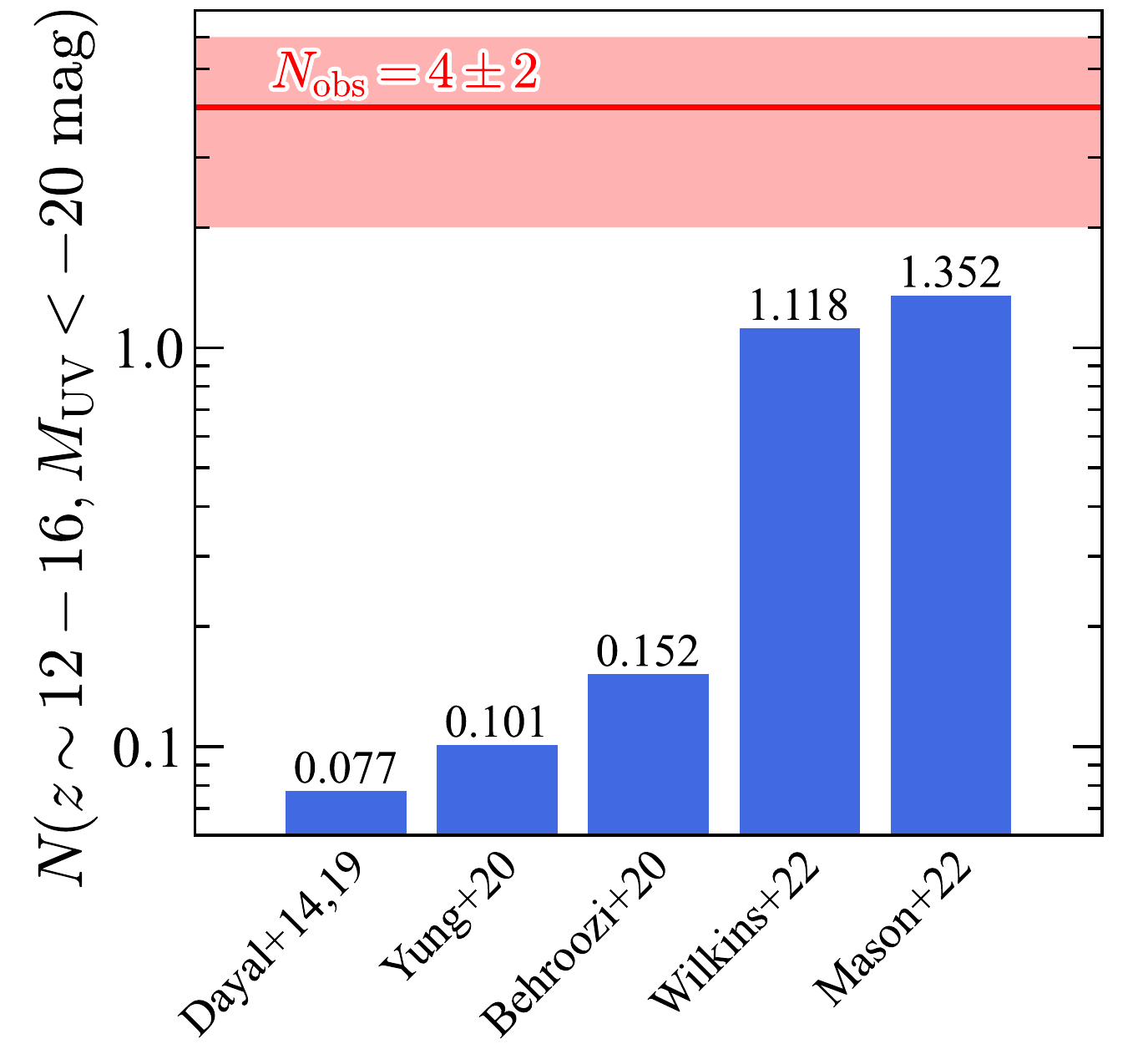}
\end{center}
\caption{
\redc{Theoretical predictions for the number of bright galaxies at $z\sim12-16$ with $M_\m{UV}<-20$ mag detected in our survey area of $\sim90\ \m{arcmin^2}$.
These numbers are based on the theoretical models of \citet{2014MNRAS.445.2545D,2019MNRAS.486.2336D}, \citet{2020MNRAS.496.4574Y}, \citet{2020MNRAS.499.5702B}, \citet{2022arXiv220409431W}, and \citet{2022arXiv220714808M}.
The red horizontal line with the shaded region indicates the number of observed galaxies at $z\sim12-16$ with $M_\m{UV}<-20$ mag ($N_\m{obs}=4\pm2$), which is higher than these model predictions.
}}
\label{fig_num_hist}
\end{figure}

\begin{figure}
\centering
\begin{center}
\includegraphics[width=0.95\hsize, bb=4 7 354 319]{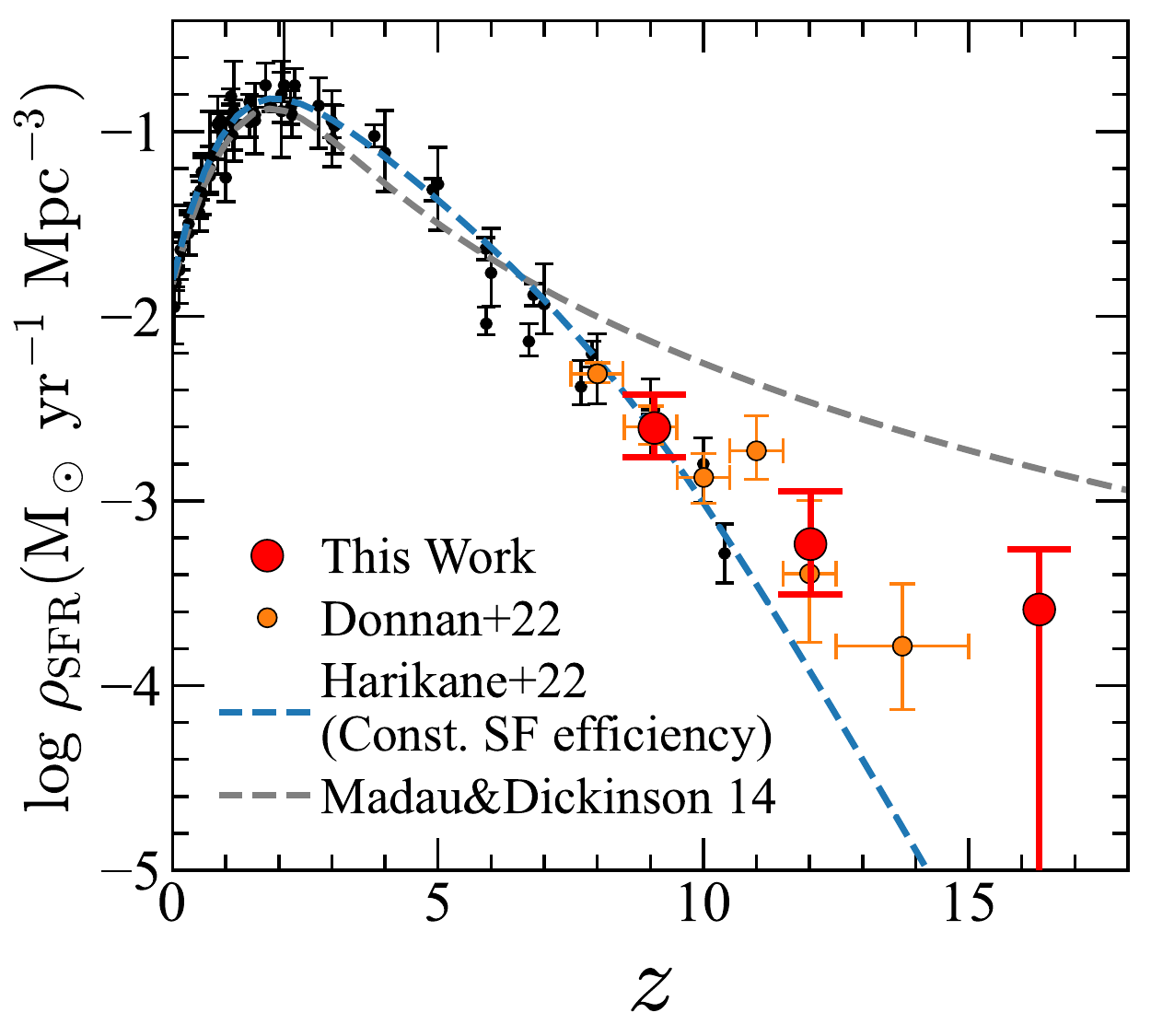}
\end{center}
\caption{
Cosmic SFR density evolution. The red circles represent the cosmic SFR densities obtained by our study, with the double power-law luminosity functions integrated down to $M_\m{UV}=-17$ mag. The black circles indicate the cosmic SFR densities derived by \citet{2014ARA&A..52..415M}, \citet{2015ApJ...810...71F}, \citet{2016MNRAS.459.3812M}, \citet{2019MNRAS.486.3805B}, and \citet{2020ApJ...902..112B}. 
The orange circles are results in \citet{2022arXiv220712356D}.
The blue dashed curve is the best-fit function of the cosmic SFR densities in \citet[their Equation 60]{2022ApJS..259...20H}.
In \citet{2022ApJS..259...20H}, they assume the constant star formation efficiency at $z>10$, resulting in the power-law decline with $\rho_\m{SFR}\propto10^{-0.5(1+z)}$.
The gray dashed curve shows the best-fit function at $z\lesssim 8$ determined by \citet{2014ARA&A..52..415M} extrapolated to $z>8$.
All results are converted to those of the \citet{1955ApJ...121..161S} IMF.
%
}
\label{fig_cSFR}
\end{figure}

\subsection{Results}\label{ss_result}
%
%
%

Figures \ref{fig_uvlf_1} and \ref{fig_uvlf_2} present our luminosity functions at $z\sim 9$, $12$, and $\redc{16}$ together with luminosity functions
obtained by previous work including the latest JWST studies \citep{2022arXiv220709434N,2022arXiv220712356D,2022arXiv220712474F,2022arXiv221102607B}.
Our measurements of the luminosity functions are summarized
in Table \ref{tab_LFdata}.
%
%
Comparing with the previous measurements of the luminosity functions,
we find that our luminosity functions at $z\sim 9$ and $12$ 
agree well with those of the previous HST and JWST studies 
within the uncertainties, as shown in Figure \ref{fig_uvlf_1}. 
%
%
In Figure \ref{fig_uvlf_2}, we compare 
the luminosity function of our possible candidates at $z\sim \redc{16}$
newly determined by this study 
with those available at lower redshifts at $z\sim 14$ constrained by JWST.
We confirm that these luminosity functions are comparable.

We conduct $\chi^2$ minimization fitting of 
the double power-law and Schechter functions
to the luminosity functions that include
the measurements at the bright end in the literature.
In the fitting, we use the results of this study, \citet{2018ApJ...867..150M}, \citet{2020MNRAS.493.2059B}, and \citet{2021AJ....162...47B} for the $z\sim9$ luminosity function, the results of this study and \citet{2022ApJ...929....1H} for the $z\sim12$ luminosity function assuming that the UV luminosity function does not rapidly change at $z\sim12-13$, and the result of this study for the $z\sim\redc{16}$ luminosity function. 
We show the best-fit functions in Figures \ref{fig_uvlf_1} and \ref{fig_uvlf_2}, and present the best-fit parameters in Table \ref{tab_LFpar}.
At $z\sim 9$, the $\chi^2$ values of the fitting suggest that
the double power-law function explains the luminosity functions
($\chi^2/{\rm dof}=2.3/9$) better than the Schechter functions
($\chi^2/{\rm dof}=3.6/10$), albeit with the moderately small
difference of $\chi^2$ ($\sim1\sigma$).
At $z\sim 12$ and $16$, we find no significant differences between
the double power-law and Schechter functions in
the $\chi^2$ values, probably due to the large uncertainties
of the measurements.
\redc{At $z\sim12$, we also fit only the measurements of this study, excluding the brightest datapoint in \citet{2022ApJ...929....1H}, as shown in Figure \ref{fig_uvlf_1_woHD1}.
The best-fit DPL and Schechter functions are slightly flatter than the fitting results with the datapoint in \cite{2022ApJ...929....1H} at the bright end.}

Figure \ref{fig_uvlf_evol} presents
the redshift evolution of the luminosity function.
We find the continuous decrease of luminosity functions
from $z\sim 5$ to $z\sim 12$. We do not find a significant
decrease from $z\sim 12$ to $\redc{16}$ beyond the uncertainty.
There is a hint of a small evolution from $z\sim 12$ to $\redc{16}$,
while the small number statistics do not allow us to conclude 
whether the evolutionary trend changes from $z\sim 5-12$ to $12-\redc{16}$.

Figure \ref{fig_uvlf_z12model} compares 
the observed luminosity functions at $z\sim 12$ and $\redc{16}$ with those predicted by theoretical models \citep{2014MNRAS.445.2545D,2019MNRAS.486.2336D,2020MNRAS.496.4574Y,2020MNRAS.499.5702B,2022arXiv220409431W,2022arXiv220714808M}.
At $z\sim12$, most of the models in Figure \ref{fig_uvlf_z12model}
explain the observational measurements in
the faint magnitude range from $-20$ to $-18$ mag,
while some models do not reproduce the moderately high
number densities of the observational measurements
at the bright magnitude of $M_\m{UV}<-20$ mag.
At $z\sim\redc{16}$, most of the models cannot reproduce the observed number density of bright galaxies at $M_\m{UV}<-20$ mag, except for the FLARES \citep{2021MNRAS.500.2127L,2021MNRAS.501.3289V,2022arXiv220409431W} whose prediction at $z\sim15$ agrees with our number density estimate within uncertainties. 
\redc{Similarly, Figure \ref{fig_num_hist} shows the predicted number of bright galaxies at $z\sim12-16$ with $M_\m{UV}<-20$ mag.
Figure \ref{fig_num_hist} indicates that the models underpredict the number of galaxies compared to the observation, although the significance is small and more data are needed to obtain the conclusion.}
This difference of the observations and models
would suggest that the feedback effects in the models
may be too strong to produce abundant bright galaxies, lower dust obscuration in these bright galaxies than the model assumptions, 
and/or that there exist hidden AGNs 
that produce radiation comparable with or more than 
stellar components of the galaxies
\citep[e.g.,][]{2014MNRAS.440.2810B,2020MNRAS.493.2059B,2018PASJ...70S..10O,2018ApJ...863...63S,2022PASJ...74...73S,2022ApJS..259...20H,2022MNRAS.514L...6P,2022arXiv220714808M}, although there is also a possibility that this difference
may be caused by other physical processes, as discussed in Section \ref{ss_dis}.

\begin{deluxetable}{cc}
\tablecaption{Obtained Luminosity Function at Each Redshift}
\label{tab_LFdata}
\tablehead{\colhead{$M_\m{UV}$} & \colhead{$\Phi$} \\
\colhead{(ABmag)}& \colhead{$\m{Mpc^{-3}}\ \m{mag^{-1}}$}}
\startdata
\multicolumn{2}{c}{F115W-Drop $(z\sim9)$}\\
$-23.03$ & $<6.95\times10^{-5}$\\
$-22.03$ & $<7.67\times10^{-5}$\\
$-21.03$ & $(4.00^{+9.42}_{-3.85})\times10^{-5}$\\
$-20.03$ & $(4.08^{+9.60}_{-3.92})\times10^{-5}$\\
$-19.03$ & $(2.24^{+1.87}_{-1.46})\times10^{-4}$\\
$-18.03$ & $(1.12^{+1.03}_{-0.90})\times10^{-3}$\\
\hline
\multicolumn{2}{c}{F150W-Drop $(z\sim12)$}\\
$-23.21$ & $<5.85\times10^{-6}$\\
$-22.21$ & $<6.40\times10^{-6}$\\
$-21.21$ & $(5.00^{+11.56}_{-4.27})\times10^{-6}$\\
$-20.21$ & $(1.31^{+1.75}_{-0.89})\times10^{-5}$\\
$-19.21$ & $(2.40^{+2.38}_{-1.40})\times10^{-5}$\\
$-18.21$ & $(1.42^{+1.97}_{-1.10})\times10^{-4}$\\
\hline
\multicolumn{2}{c}{F200W-Drop $(z\sim16)$}\\
$-23.59$ & $<2.42\times10^{-6}$\\
$-20.59$ & $(6.62^{+8.84}_{-4.49})\times10^{-6}$
\enddata
\tablecomments{Errors and upper limits are $1\sigma$.}
\end{deluxetable}

\begin{deluxetable*}{ccccccc}
\tablecaption{Fit Parameters for Luminosity Functions}
\label{tab_LFpar}
\tablehead{\colhead{Redshift} & \colhead{Fitted Function} & \colhead{$M_\m{UV}^*$} & \colhead{$\m{log}\phi^*$} & \colhead{$\alpha$} & \colhead{$\beta$} & \colhead{$\chi^2/\m{dof}$}\\
\colhead{}& \colhead{}& \colhead{(ABmag)}& \colhead{($\m{Mpc^{-3}}$)} &  \colhead{}& \colhead{}& \colhead{} }
\startdata
$z\sim9$ & DPL & $-19.33^{+2.24}_{-0.96}$& $-3.50^{+1.53}_{-0.65}$& $-2.10$(fixed)& $-3.27^{+0.34}_{-0.37}$& $2.1/9$\\
& Schechter & $-21.24^{+0.45}_{-0.59}$& $-4.83^{+0.37}_{-0.49}$& $-2.35$(fixed)& \nodata & $3.4/10$\\
$z\sim12$ & DPL & $-19.60$(fixed)& $-4.33^{+0.22}_{-0.22}$& $-2.10$(fixed)& $-2.83^{+0.50}_{-0.44}$& $0.8/3$\\
& Schechter & $-20.47^{+1.94}_{-0.15}$& $-5.06^{+1.51}_{-0.17}$& $-2.35$(fixed)& \nodata & $1.2/3$\\
$z\sim16$ & DPL & $-19.60$(fixed)& $-4.71^{+0.33}_{-2.83}$& $-2.10$(fixed)& $-2.70^{+0.00}_{-0.00}$& $1.4/1$\\
& Schechter & $-20.80$(fixed)& $-5.84^{+0.47}_{-4.03}$& $-2.35$(fixed)& \nodata & $1.9/2$\\
\hline
$z\sim12^\dagger$ & DPL & $-19.60$(fixed)& $-4.32^{+0.22}_{-0.22}$& $-2.10$(fixed)& $-2.21^{+1.07}_{-1.06}$& $0.3/2$\\
& Schechter & $-21.97^{+2.88}_{-0.11}$& $-5.95^{+1.84}_{-0.18}$& $-2.35$(fixed)& \nodata & $0.5/2$
\enddata
\tablecomments{Errors are $1\sigma$.\\
\redc{$^\dagger$ Fit parameters without the brightest datapoint in \citet{2022ApJ...929....1H}, which are shown in Figure \ref{fig_uvlf_1_woHD1}.}}
\end{deluxetable*}

\begin{deluxetable*}{cccc}
\tablecaption{Obtained Cosmic UV Luminosity Density and SFR Density}
\label{tab_cSFR}
\tablehead{\colhead{Redshift} & \colhead{$\m{log}\rho_\m{UV}$} & \colhead{$\m{log}\rho_\m{SFR,UV}$}  & \colhead{$\m{log}\rho_\m{SFR}$} \\
\colhead{}& \colhead{($\m{erg\ s^{-1}\ Hz^{-1}\ Mpc^{-3}}$)}& \colhead{($\m{M_\odot\ yr^{-1}\ Mpc^{-3}}$)} & \colhead{($\m{M_\odot\ yr^{-1}\ Mpc^{-3}}$)} }
\startdata
$z\sim9$ & $25.28_{-0.16}^{+0.19}$ & $-2.65_{-0.16}^{+0.19}$ & $-2.61_{-0.16}^{+0.18}$\\
$z\sim12$ & $24.61_{-0.26}^{+0.26}$ & $-3.33_{-0.26}^{+0.26}$  & $-3.23_{-0.27}^{+0.29}$ \\
$z\sim16$ & $24.24_{-2.83}^{+0.33}$ & $-3.70_{-2.83}^{+0.33}$ & $-3.59_{-2.83}^{+0.33}$\\
\enddata
\tablecomments{Errors are $1\sigma$. $\rho_\m{SFR,UV}$ and $\rho_\m{SFR}$ are SFR densities without and with dust extinction correction, respectively.}
\end{deluxetable*}

\subsection{Cosmic SFR Density}
\label{ss_cSFR}

%
%

We derive the cosmic SFR densities
at $z\sim 9$, $12$, and $\redc{16}$.
We integrate the best-fit double power-law functions
(Table \ref{tab_LFpar}) down to $-17$ mag, the same limit as previous studies \citep[e.g.,][]{2015ApJ...803...34B,2018ApJ...855..105O,2022ApJ...929....1H},
and obtain the UV luminosity densities, $\rho_{\rm UV}$.
%
We correct $\rho_{\rm UV}$ for the dust extinction,
following the attenuation-UV slope ($\beta_{\rm UV}$) 
relation \citep{1999ApJ...521...64M} 
and $\beta_{\rm UV}$-$M_{\rm UV}$ relation at $z=8$ in \citet{2014ApJ...793..115B}.
\redc{The choice of these assumptions (e.g., using the attenuation-UV slope law in \citealt{2014A&A...563A..81D} instead) does not affect our conclusions because the correction factor is very small ($\lesssim0.1$ dex).}
We calculate SFRs from UV luminosities, $L_{\rm UV}$, 
corrected for dust extinction
by the relation,
\begin{equation}
SFR (M_\odot\ {\rm yr^{-1}})= \mathcal{K}_{\rm UV} L_{\rm UV} ({\rm erg\ s^{-1}\ Hz^{-1}}),
\label{eq_SFR}
\end{equation}
where $\mathcal{K}_{\rm UV}$ is the conversion factor
that depends on the recent star-formation history, metal
enrichment history, and the choice of the IMF.
Here we apply $\mathcal{K}_{\rm UV}=1.15\times10^{-28}\ M_\odot\ \m{yr^{-1}/(erg\ s^{-1}\ Hz^{-1})}$
that is used in \citet{2014ARA&A..52..415M}.
This value of $\mathcal{K}_{\rm UV}$ is valid for
the \citet{1955ApJ...121..161S} IMF,
and consistent with the cosmic star-formation history
and the evolved stellar metallicity ($10^{-0.15z}Z_\odot$;
\citealt{2014ARA&A..52..415M}) up to $z\sim 10$.
Table \ref{tab_cSFR} summarizes our measurements of the cosmic UV luminosity density, SFR densities without and with dust extinction correction at each redshift.

Figure \ref{fig_cSFR} presents the cosmic SFR density evolution.
In this figure, we show the cosmic SFR density measurements at $z\sim 0-10$ obtained by previous studies, all of which are converted to the calibration of \citet{2014ARA&A..52..415M} 
with the \citet{1955ApJ...121..161S} IMF 
(Equation (\ref{eq_SFR})).
We confirm that our SFR density at $z\sim 9$
is consistent with the previous measurements.
We compare the observational measurements of 
the SFR densities with the constant star-formation efficiency ($SFR/\dot{M}_\m{h}(z)=\m{const.}$)
model \citep{2022ApJS..259...20H} 
together with the extrapolation
of the \citet{2014ARA&A..52..415M} estimates at $z=0-8$.
We find that the cosmic SFR densities significantly decrease from $z\sim 9$ to $12$. A decrease of the cosmic SFR densities 
may exist from $z\sim 12$ to $\redc{16}$, while the decrease is
not larger than the errors.
Interestingly, The constant star-formation efficiency model 
explains the evolution of the cosmic SFR densities 
up to $z\sim 10$ \citep{2022ApJS..259...20H},
while our measurement at $z\sim 12$ is higher than the model prediction beyond the uncertainty. Moreover, 
there is a hint of a high cosmic SFR density 
at $z\sim \redc{16}$ above the model, although it is not
statistically significant due to the large error.
Such higher SFR densities than the constant efficiency model at $z\sim15$ is actually consistent with observations of Balmer break galaxy candidates at $z\sim6$ \citep{2020ApJ...889..137M}.

\begin{figure*}
\centering
\begin{minipage}{0.48\hsize}
\begin{center}
\includegraphics[width=0.9\hsize, bb=4 7 354 319]{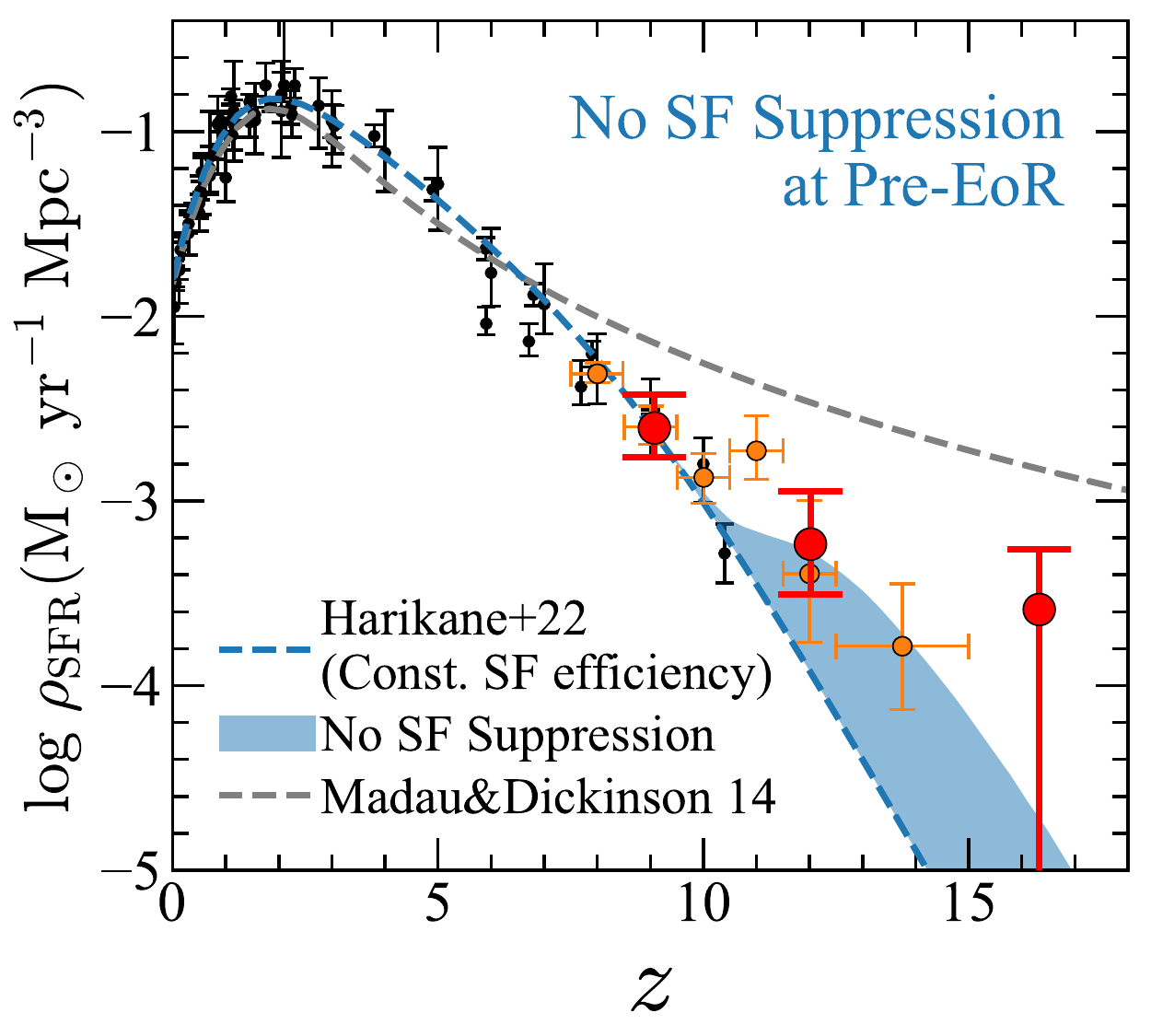}
\end{center}
\end{minipage}
\begin{minipage}{0.48\hsize}
\begin{center}
\includegraphics[width=0.9\hsize, bb=4 7 354 319]{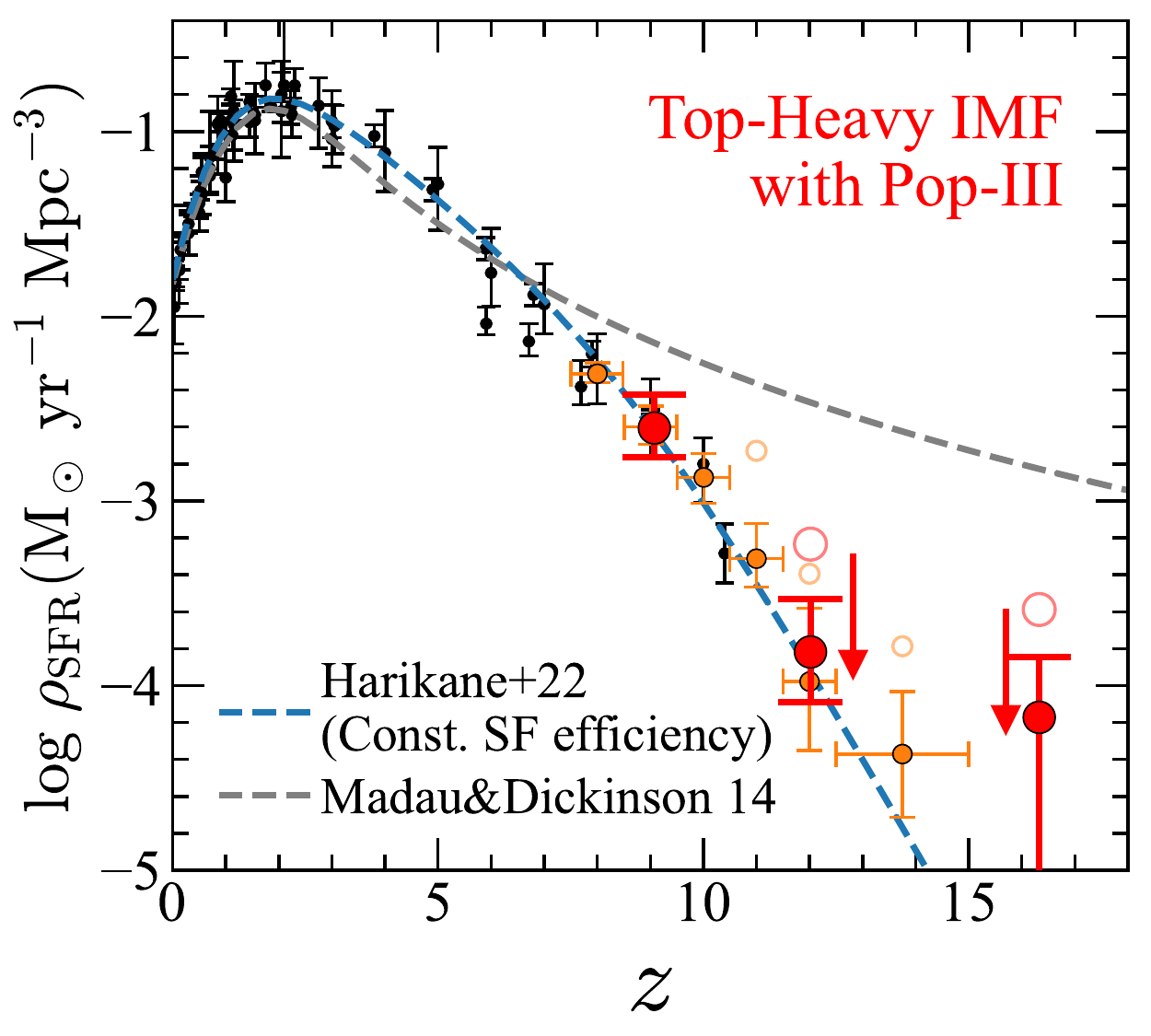}
\end{center}
\end{minipage}
\caption{
Possible scenarios to explain the observed SFR densities at $z>10$.
(Left:) Scenario of no star formation suppression at pre-reionization epoch.
At the reionization epoch and after that, star formation in low-mass halos is suppressed by strong UV background radiation, while before the reionization epoch such a suppression of star formation activity does not occur. 
The upper edge of the blue shaded region indicates the enhancement of the star formation by this effect \citep{2000ApJ...539...20B}, which explains the observed SFR densities (see texts for details).
(Right:) Scenario of Pop III star formation.
As shown in Figure \ref{fig_kuv}, Pop III stellar populations with a top-heavy IMF produces a significant amount of UV photons at a given SFR, resulting in the overestimates of the SFR densities if we use the canonical UV-SFR conversion factor.
The red and orange filled circles at $z>10.5$ are SFR densities calculated based on the conversion factor for a Pop III stellar population with a top-heavy IMF (the PopIII.1 model in Figure \ref{fig_kuv}), which agree well with the constant star formation efficiency model in \citet{2022ApJS..259...20H}.
The open circles are SFR densities based on the canonical conversion factor.
}
\label{fig_cSFR_2}
\end{figure*}

\begin{figure}
\centering
\begin{center}
\includegraphics[width=0.95\hsize, bb=6 10 359 350]{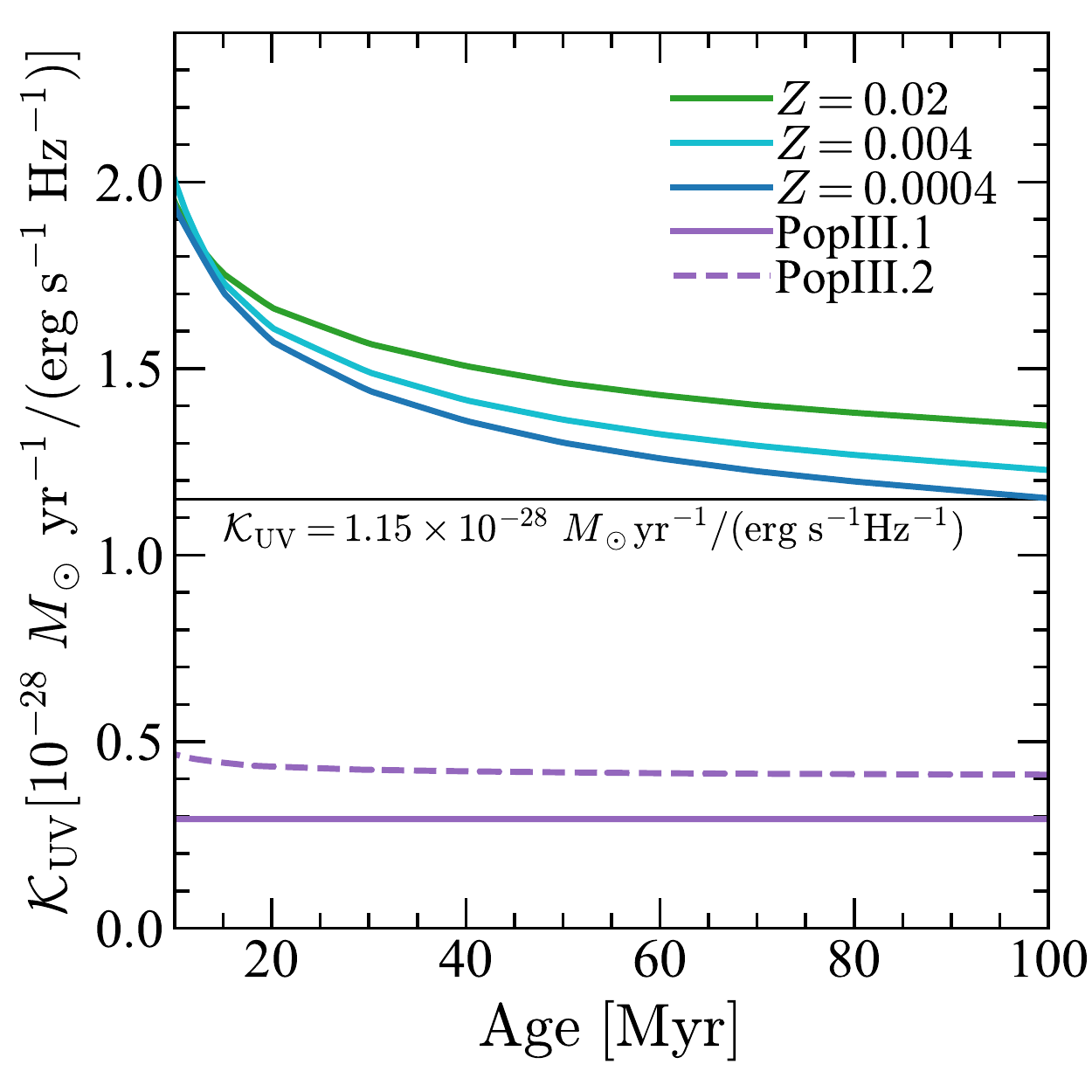}
\end{center}
\caption{
UV luminosity-SFR conversion factor, $\mathcal{K}_{\rm UV}$, for various metallicities as a function of stellar age.
The green, cyan, and blue curves show the conversion factor for metallicities of $Z=0.02$, $0.004$, $0.0004$, respectively, calculated with {\sc Yggdrasil} \citep{2011ApJ...740...13Z} assuming a constant star formation history with a unity gas covering fraction.
These factors are values for a UV luminosity at 1500 $\m{\AA}$ in the \citet{1955ApJ...121..161S} IMF in the interval of $0.1-100$ $M_\odot$.
Note that the original outputs from {\sc Yggdrasil} are for the \citet{2001MNRAS.322..231K} IMF, and we correct for the IMF difference by multiplying the outputs by 1.49.
The solid (PopIII.1) and dashed (PopIII.2) purple curves show the conversion factors for Pop III stellar populations with an extremely top-heavy IMF (50-500 $M_\odot$, the \citealt{1955ApJ...121..161S} slope) and a moderately top-heavy IMF (log-normal with characteristic mass of $M_\m{c}=10\ M_\odot$, dispersion $\sigma=1\ M_\odot$, and wings extending from $1-500$ $M_\odot$).
If galaxies at $z>10$ are dominated by Pop III stellar populations, the conversion factor is significantly lower than the typically assumed value ($\mathcal{K}_{\rm UV}=1.15\times10^{-28}\ M_\odot\m{yr^{-1}/(erg\ s^{-1}\ Hz^{-1})}$; the black line), resulting in the overestimate of the SFR.
}
\label{fig_kuv}
\end{figure}

\section{Discussion}\label{ss_dis}

\subsection{Possible High Cosmic SFR Density at $z>10$}\label{ss_dis_cSFR}

Our observational measurements suggest that
the SFR densities at $z\sim 12-\redc{16}$ are higher than 
the constant star-formation efficiency model of \citet{2022ApJS..259...20H}.
Although the constant star-formation efficiency model 
well explains the cosmic SFR densities at $z\sim 0-10$,
this model underpredicts those at $z\sim 12-\redc{16}$.
Here we discuss the following three possibilities that explain the observed high SFR densities at $z\sim12-\redc{16}$.

\begin{itemize}
    \item[(A)] {\it No star formation suppression at the pre-reionization epoch.}
    The universe at $z\sim12-\redc{16}$ is at the pre-reionization epoch when the IGM is highly neutral \citep{2020ARA&A..58..617O,2021arXiv211013160R}.
    At the epoch of reionizaton (EoR; $z\sim6-12$) and the epoch of post-reionization (post-EoR; $z\lesssim6$), galaxies and AGN produce UV radiation by their star-formation and nuclear activity, and produce strong UV background radiation.
    The UV background radiation heats up {\sc Hi} gas in low-mass halos of $M_{\rm h} \lesssim 10^{8-9} M_\odot$ with negligible {\sc Hi} self-shielding, suppressing star-formation at the EoR and post-EoR \citep{2000ApJ...539...20B,2004ApJ...600....1S,2006MNRAS.371..401H,2009MNRAS.396L..46P,2009MNRAS.399.1650M,2010MNRAS.402.1599S,2015ApJ...807..154B}.
    Although the halo masses of our galaxies at $z\sim 12-\redc{16}$ are unknown, the maximum halo mass existing at $z\sim 15$ given the survey volume of this study is $M_{\rm h} \simeq 3\times 10^9 M_\odot$ in the structure formation model with the Planck cosmology \citep{2020MNRAS.499.5702B}.
    In other words, most of halos (with $\lesssim 10^9 M_\odot$) at $z\sim 15$ are not affected by the UV background at the pre-reionization epoch, while the similar halos at $z\lesssim 10$ experience the suppression of star formation by the UV background at the EoR and post-EoR. 
    To test whether this effect can quantitatively explains the observed SFR densities, we construct a model of the SFR density evolution including the enhancement of the star formation that is free from the suppression by the UV background at the pre-EoR.
    We use a model in \citet{2000ApJ...539...20B} with a reionization redshift of $z_\m{reion}=13$, and multiply the prediction of the constant star formation efficiency model \citep{2022ApJS..259...20H} by a factor of the star formation rate enhancement due to no suppression by the UV background in \citet{2000ApJ...539...20B}.
    The left panel of Figure \ref{fig_cSFR_2} presents this hybrid model including the effect of star formation enhancement at pre-EoR, which reproduces the observed SFR densities at $z\sim12-\redc{16}$ within uncertainties.
    This agreement indicates a possibility that the star formation efficiency at $z\sim12-\redc{16}$ is higher than those at $z\lesssim10$ due to no suppression of the star formation activity at the pre-EoR.
    
    \item[(B)] {\it Presence of AGN activity.}
    Another possibility is that a large fraction of the observed UV luminosity densities at $z\sim12-\redc{16}$ is produced by AGN, and there is no excessive SFR densities at $z\sim 12-\redc{16}$ beyond the constant star-formation efficiency model.
    This is an interesting scenario that mitigates the existence of supermassive black holes (SMBH) at $z\sim 7$ (\citealt{2011Natur.474..616M}, \citealt{2018Natur.553..473B}, \citealt{2021ApJ...907L...1W}) by efficient gas accretion on SMBHs creating AGNs, while a standard gas accretion limited by the Eddington accretion rate does not explain the existence of the SMBHs at $z\sim 7$.
    However, our $z\sim 12-\redc{16}$ candidates except for GL-z12-1 show extended morphologies (Section \ref{ss_sample}).
    Thus the fraction of AGN radiation dominated galaxies is as small as $\sim 10$\% $(=1/10)$ at $z\sim 12-\redc{16}$. 
    Although the excessive SFR density estimate at $z\sim \redc{16}$ is unclear due to the small statistics, the one at $z\sim 12$ cannot be explained by AGN activity.
    
    \item[(C)] {\it A top-heavy IMF.}
    The third possibility is an overestimate of the SFR density due to a top-heavy IMF possibly with the Population III (Pop III) stellar population.
    In our estimate of the SFR density, we use the canonical UV luminosity-to-SFR conversion factor of $\mathcal{K}_{\rm UV}=1.15\times10^{-28}\ M_\odot\ \m{yr^{-1}/(erg\ s^{-1}\ Hz^{-1})}$, which is for the \citet{1955ApJ...121..161S} IMF, while $\mathcal{K}_{\rm UV}$ depends on star-formation history, metallicities, and IMFs \cite[e.g.,][]{2014ARA&A..52..415M,2018ApJ...868...92T}.
    Indeed in the early universe, the IMF is expected to be more top-heavy because of a lower metallicity or a higher CMB temperature \citep[e.g.,][]{2005ApJ...626..627O,2022MNRAS.514.4639C}, resulting in a higher Jeans mass, especially for Pop III stellar populations \citep[e.g.,][]{2014ApJ...781...60H,2015MNRAS.448..568H}.
    To test whether this effect can explain the observed densities, we calculate the UV-to-SFR conversion factor, $\mathcal{K}_{\rm UV}$, for different metallicity and IMF assumptions using {\sc Yggdrasil} \citep{2011ApJ...740...13Z}.
    Figure \ref{fig_kuv} presents $\mathcal{K}_{\rm UV}$ for different metallicities and IMFs as a function of stellar age.
    We find that Pop III stellar populations with top-heavy IMFs (PopIII.1 and PopIII.2 in {\sc Yggdrasil}) produce $\sim3-4$ times more UV photons than the canonical assumption given the SFR, because nebular continuum emission boosts the UV luminosity as discussed in previous studies \citep[e.g.,][]{2008ApJ...676L...9Z,2009A&A...502..423S,2010A&A...515A..73S}.
    This low conversion factor reduces the SFR density estimates at $z\sim12-\redc{16}$ as shown in the right panel of Figure \ref{fig_cSFR_2}, resulting in the SFR densities consistent with the constant star formation efficiency model.

\end{itemize}

Based on these discussions, we conclude that (A) {\it no star formation suppression at pre-reionization epoch} or (C) {\it a top-heavy IMF with a Pop III-like star formation} can explain the observed high SFR densities at $z\sim12-\redc{16}$.
These possibilities can be further investigated by follow-up observations with JWST/MIRI covering a longer wavelength than the Balmer break to obtain the robust stellar mass measurements and star formation history, or with JWST/NIRSpec and MIRI spectroscopy to search for signatures of Pop III-like stellar populations and AGN activity.

\subsection{Properties of Luminous Galaxy Candidates}

In this study, we have found several luminous galaxy candidates 
at the early epoch of $z\sim10-\redc{16}$, when the age of the universe is only $\sim200-500$ Myrs after the Big Bang.
Here we discuss physical properties of these luminous galaxy candidates.

Table \ref{tab_MsSFR} summarizes SFRs and stellar masses of six galaxy candidates whose UV magnitudes are brighter than $M_\m{UV}=-19.5$ mag, constrained by the SED fitting in Section \ref{ss_SEDfit} assuming the \citet{2003PASP..115..763C} IMF.
Our estimates of SFRs and stellar masses agree with previous estimates by \citet{2022arXiv220709434N}, \citet{2022arXiv220712356D}, and \citet{2022arXiv220712474F}, indicating that these luminous galaxies are very massive with stellar masses as high as $M_*\sim(1-10)\times10^8\ M_\odot$ at $z\sim10-\redc{16}$.
While the contributions from AGN radiation to the SEDs may be suspected in
one of the objects, GL-z12-1 (see Sections \ref{ss_sample} and \ref{ss_dis_cSFR}),
at least the rest of the objects (i.e. $\sim 80$\% of the bright $z\sim 10-\redc{16}$ galaxies)
would be truly stellar massive.
Although the NIRCam photometry is limited to $<5\ \mu$m and does not trace the SEDs beyond the Balmer break (4000\ \AA) corresponding to $5-7\ \mu$m in the observed frame at $z\sim10-\redc{16}$, these stellar mass estimates provide rough lower limits that miss the contribution from old stellar populations beyond the Balmer break, given high specific SFRs of these galaxy candidates, $SFR/M_*\sim10^{-8}\ \m{yr^{-1}}$.

\begin{figure}
\centering
\begin{center}
\includegraphics[width=0.95\hsize, bb=8 7 354 281]{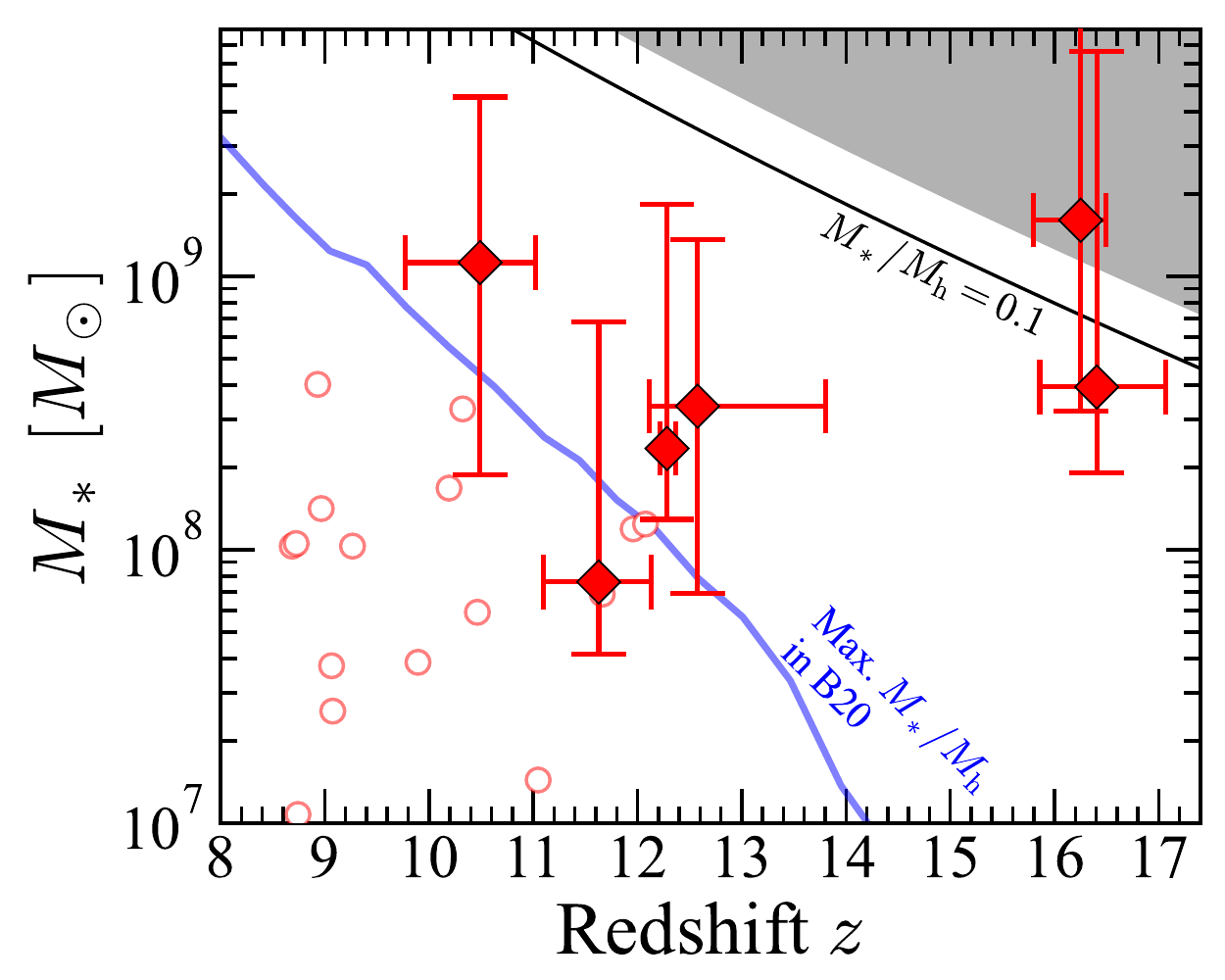}
\end{center}
\caption{
Stellar masses of our galaxy candidates as a function of redshift.
The red filled diamonds show the stellar mass estimates for six luminous galaxy candidates with $M_\m{UV}<-19.5$ mag at $z\sim10-16$ (Table \ref{tab_MsSFR}), and open red circles are results for other candidates.
The gray shaded region indicates the stellar mass whose number density is below the observational limit, calculated from the cosmic baryon fraction ($\Omega_\m{b}/\Omega_\m{m}=0.16$) and the maximum halo mass that can be observed with the survey volume of this study.
The black and blue curves indicate the stellar masses calculated from the maximum halo mass with $M_*/M_\m{h}=0.1$ and the maximum $M_*/M_\m{h}$ value at each redshift in \citet{2020MNRAS.499.5702B}, respectively.
The massive stellar masses ($M_*\sim10^9\ M_\odot$) of the two $z\sim16$ candidates can be explained by a very high SHMR of $M_*/M_\m{h}=0.1$, indicating a star formation efficiency as high as $\sim60\%$.
The other four luminous candidates at $z\sim10-13$ also show higher stellar masses compared to the predictions from \citet{2020MNRAS.499.5702B}.
}
\label{fig_Ms_z}
\end{figure}

\begin{deluxetable}{cccc}
\tablecaption{SFRs and Stellar Masses of Luminous Galaxy Candidates with $M_\m{UV}<-19.5$ mag}
\label{tab_MsSFR}
\tablehead{\colhead{ID} & \colhead{$z_\m{phot}$} & \colhead{$SFR$}  & \colhead{$M_*$} \\
\colhead{}& \colhead{}& \colhead{($\m{M_\odot\ yr^{-1}}$)} & \colhead{($\m{M_\odot}$)} }
\startdata
GL-z9-1 & $10.49_{-0.72}^{+0.53}$ & $14.2_{-11.2}^{+25.0}$ & $(1.1_{-0.9}^{+3.4})\times10^{9}$\\
CR2-z12-1 & $11.63_{-0.53}^{+0.51}$ & $0.9_{-0.1}^{+4.9}$ & $(7.6_{-3.5}^{+60.4})\times10^{7}$\\
GL-z12-1$^\dagger$ & $12.28_{-0.07}^{+0.08}$ & $2.9_{-1.0}^{+10.9}$ & $(2.3_{-1.1}^{+16.0})\times10^{8}$\\
S5-z12-1 & $12.58_{-0.46}^{+1.23}$ & $5.5_{-4.4}^{+4.7}$ & $(3.4_{-2.7}^{+10.3})\times10^{8}$\\
CR2-z16-1 & $16.25_{-0.46}^{+0.24}$ & $31.2_{-30.8}^{+25.8}$ & $(1.6_{-1.3}^{+16.8})\times10^{9}$\\
S5-z16-1 & $16.41_{-0.55}^{+0.66}$ & $5.1_{-1.8}^{+21.7}$ & $(3.9_{-2.0}^{+62.4})\times10^{8}$
\enddata
\tablecomments{Assuming the \citet{2003PASP..115..763C} IMF and metallicity of $Z=0.2\ Z_\odot$. The SFR is averaged over the past 50 Myr in the same manner as \citet{2022ApJ...927..170T}. See Section \ref{ss_SEDfit} for the details of the SED fitting.\\
$^\dagger$ This candidate shows a compact morphology indicative of AGN activity, while the profile is spatially extended more than the PSF (Section \ref{ss_sample}).
}
\end{deluxetable}

Here is a question how these galaxies with 
the large stellar masses form at this early epoch of $z\sim 10-\redc{16}$.
To discuss the formation scenario of these massive galaxy candidates, we estimate the stellar-to-halo mass ratio (SHMR) of these galaxies.
Using the abundance matching technique in the same manner as \citet[][their Equation (66)]{2016ApJ...821..123H},  we estimate the halo mass of the most massive halo that can be observed with the survey volume in this study, resulting in $5\times10^{10}\ M_\odot$ and $5\times10^9\ M_\odot$ at $z\sim12$ and $z\sim16$, respectively.
From the stellar mass estimates discussed above, the SHMRs of $z\sim12$ and $z\sim16$ galaxies are $\sim0.01$ and $\sim0.1$, respectively.
Because the cosmic baryon to dark matter density ratio is
$\Omega_\m{b}/\Omega_\m{m}=0.16$ \citep{2020A&A...641A...6P}, 
the SHMRs of the $z\sim16$ galaxies
reach $\sim60\%$ of the cosmic baryon fraction, as shown in Figure \ref{fig_Ms_z}.
In other words, more than a half of baryon gas in the halos is
converted to stars, which is unlikely found 
in lower-redshift and present-day galaxies whose SHMRs are
$\sim0.02-0.03$ at maximum \citep[e.g.,][]{2016ApJ...821..123H,2019MNRAS.488.3143B}.
A similar conclusion is obtained from the comparison of the UV luminosity functions \citep{2022ApJ...938L..10I}.
However, theoretical models predict such 
efficient star formation at the pre-reionization epoch
where $70-80$\% of baryon are converted stars \citep{2004ApJ...600....1S} in halos with $10^{8-9}M_\odot$ masses
in a few hundred Myr,
when the UV background radiation is too weak to suppress
star formation (see discussion (A) in Section \ref{ss_dis_cSFR}).
The other four galaxies at $z\sim11-14$ also show higher stellar masses compared to the predictions from the maximum SHMR in \citet{2020MNRAS.499.5702B}, indicating elevated star formation efficiencies, probably due to no suppression of star formation activity at the pre-reionization epoch.
Another possibility is that the SFR and the stellar mass of these bright galaxies are overestimated due to the assumption of the IMF and metallicity in the SED fitting, as discussed in Section \ref{ss_dis_cSFR} (discussion (C)).
Indeed, if we assume that the stellar population of these galaxies are dominated by Pop III with a top-heavy IMF, the SFR and stellar mass are reduced by a factor of $\sim3-4$, more comparable to the observed SHMRs at lower redshifts.
These comparisons, together with the discussions in Section \ref{ss_dis_cSFR}, indicate that the observed properties of $z\sim10-\redc{16}$ galaxies (i.e., high cosmic SFR densities and massive stellar masses) can be explained by either no star formation suppression by UV background radiation at the pre-reionization epoch or a top heavy IMF possibly with a Pop III-like stellar population.


\section{Summary}\label{ss_summary}
In this paper, we have conducted comprehensive analyses for 
the JWST/NIRCam images taken by the JWST ERO SMACS J0723, Stephan's Quintet,
ERS GLASS, and CEERS projects, covering a total of $\sim88.7\ \m{arcmin^2}$, 
in conjunction with the supports of the ERO SMACS J0723 NIRSpec spectra. 
\redc{We reduced the NIRCam datasets using the new calibration parameters released in October, based on calibration observations of three different standard stars placed in all of the 10 NIRCam detectors.}
Our major findings are summarized below:

\begin{enumerate}
\item
We have selected dropout galaxy candidates at $z\sim 9$,
$z\sim 12$, and $z\sim \redc{16}$ showing significant 
continuum breaks in the NIRCam $F115W$, $F150W$, and $F200W$-bands,
respectively, by the color criteria,
confirming clear non-detections in the band(s) whose wavelength is
shorter than the continuum breaks including the $F090W$ band (Section \ref{ss_selection}, Figure \ref{fig_2color}).
Because we have found that a weak photo-$z$ criterion of $\Delta \chi^2>4$ 
cannot remove a number of foreground interlopers 
on the bases of the JWST simulation data produced by the CEERS project team (Figure \ref{fig_ceers}),
we apply a stringent photo-$z$ determination criterion
of $\Delta \chi^2>9$ with the {\sc prospector} code for our 
galaxy selection.
We thus identify \redc{13}, 8, and 2 dropout galaxy candidates at $z\sim 9$,
$z\sim 12$, and $z\sim \redc{16}$, respectively (Table \ref{tab_numLBG}). 
We confirm that our photometric redshifts agree well with 
the spectroscopic redshifts, by applying our photometric redshift technique 
to galaxies at $z_\m{spec}\sim 8-9$ found by the ERO NIRSpec observations (Figures \ref{fig_specz_1} and \ref{fig_specz_2}).

\item
We have thoroughly compared our dropout galaxy candidates with other high redshift galaxies reported in a number of recent studies 
in the ERO SMACS J0723 and 
the ERS GLASS+CEERS NIRCam fields. 
We have summarized the candidates so far claimed
in the literature together with our dropouts in Tables \ref{tab_photo_f115w}-\ref{tab_photo_f200w}.
For bright galaxy candidates, we find that a reasonable fraction of 
galaxies are commonly selected in our and previous studies.
We confirm that, among all of the candidates,
our dropout galaxies show the significant Ly$\alpha$ 
continuum breaks and flat UV continua with non-detections of continua 
whose wavelengths are shorter than the break (Figures \ref{fig_sed_z9}-\ref{fig_sed_z17}), and conclude that 
we do not miss many reliable candidates
in the redshift range of $z\sim 9-\redc{16}$ in our selection.

\item 
We have derived the UV luminosity functions at $z\sim 9$, $12$, and $\redc{16}$ (Figures \ref{fig_uvlf_1} and \ref{fig_uvlf_2}).
The UV luminosity functions 
at $z\sim 9$ and $12$ agree with those of previous HST and JWST studies 
within uncertainties including the cosmic variance, and the UV luminosity function at $z\sim \redc{16}$
is newly constrained. 
The double power-law function is preferred to 
the Schechter function at $z\sim 9$, albeit with the moderately small
difference of $\chi^2$.

\item
The cosmic SFR densities at $z\sim 9$, $12$, and $\redc{16}$
are derived by the integration of 
the best-fit UV luminosity functions (Figure \ref{fig_cSFR}).
By the comparisons with the previous low-redshift 
determinations of cosmic SFR densities, we find that 
the cosmic SFR densities significantly decrease from $z\sim 9$ to $12$. 
A decrease of the cosmic SFR densities 
may exist from $z\sim 12$ to $\redc{16}$, while the decrease is
not larger than the errors.
Our measurements of the cosmic SFR density at $z\sim12$ is 
higher than predictions from the constant star-formation efficiency model \citep{2022ApJS..259...20H},
while the model explains the cosmic star-formation history at $z\lesssim 10$. 
Moreover, there is a hint of a high cosmic SFR density 
at $z\sim \redc{16}$ above the model, although it is not
statistically significant due to the large error.
%

\item
There are several luminous and massive galaxy candidates with $M_{\rm UV}<-19.5$ mag 
at the early epoch of $z\sim 10-\redc{16}$, when
the age of the universe is only $\sim 200-500$ Myrs after the Big Bang (Figure \ref{fig_Ms_z}).
We confirm that our stellar mass estimates
are comparable with those of the previous studies. 
Although one of the objects may have contributions of UV radiation 
from an AGN suggested by their morphologies, 
a majority ($\sim 80$\%) of the galaxies may be truly stellar massive.
By the comparisons with the structure formation models 
that provide the upper limits of the dark-matter halo masses observed in this study, the SHMR of the luminous galaxy candidates at $z\sim16$ is $M_*/M_{\rm h}\sim0.1$, corresponding to $\sim60\%$ of the baryon to dark matter density ratio in the Planck cosmology, indicating that most of baryon may be converted to stars, unlike lower-redshift and present-day galaxies
with a reasonably small SHMR up to $M_*/M_{\rm h}\simeq 0.02-0.03$
\citep[e.g.,][]{2016ApJ...821..123H,2019MNRAS.488.3143B}.
The other candidates at $z\sim10-13$ also have stellar masses more massive than predictions from the maximum SHMR in \citet{2020MNRAS.499.5702B}.

\item
This study identifies two interesting observational properties of
galaxies at $z\sim 10-\redc{16}$, 
the cosmic SFR densities higher than the constant star-formation efficiency model 
and the existence of the UV-luminous galaxies with high stellar masses.
The possibility of the AGN contribution can be ruled out, because
the small fraction of galaxies have compact morphologies 
suggesting no dominant radiation from the AGN activity.
Instead, there are two scenarios that explain 
the observational properties (Figure \ref{fig_cSFR_2}). One scenario is that 
the UV background radiation does not suppress the star formation
at the pre-reionization epoch unlike at the EoR and post-EoR.
Efficient star formation may take place at $z\sim 10-\redc{16}$, producing
the high cosmic SFR densities and the stellar massive galaxies.
The other scenario is that a top-heavy IMF possibly with Pop III (or similarly metal poor) stellar populations produces strong UV radiation.
The strong UV radiation may result in the overestimates of SFR densities above the
constant star-formation efficiency model and of the
stellar mass of the luminous galaxies.
%
Further observational and theoretical studies are needed 
to test these two scenarios.
%



\end{enumerate}

\acknowledgments
\redcc{We thank the anonymous referee for a careful reading and valuable comments that improved the clarity of the paper.}
We thank the CEERS team, especially Micaela Bagley and Steven Finkelstein, for providing many useful scripts for the NIRCam data reduction and datasets of the CEERS simulated images, and \redcc{L. Y. Aaron Yung for providing SEDs of the mock galaxies}.
We are grateful to Rychard Bouwens, Seiji Fujimoto, Kohei Inayoshi, Akio Inoue, Tadayuki Kodama, Sandro Tacchella, Ken-ichi Tadaki, and Masayuki Umemura for useful comments and discussions.
We thank Pratika Dayal and L. Y. Aaron Yung for sharing their data of the luminosity function.
This work is based on observations made with the NASA/ESA/CSA James Webb Space Telescope. The data were obtained from the Mikulski Archive for Space Telescopes at the Space Telescope Science Institute, which is operated by the Association of Universities for Research in Astronomy, Inc., under NASA contract NAS 5-03127 for JWST. These observations are associated with programs 2732, 2736, 1324, and 1345.
The authors acknowledge the
ERO, GLASS, and CEERS teams
led by
Klaus M. Pontoppidan,
Tommaso Treu, and
Steven L. Finkelstein, respectively,
for developing their observing programs with a zero-exclusive-access period.
This publication is based upon work supported by World Premier International Research Center Initiative (WPI Initiative), MEXT, Japan, and KAKENHI (20H00180, 20H00181, 20H05856, 21K13953, 21H04467, 22H01260) through Japan Society for the Promotion of Science.
This work was supported by the joint research program of the Institute for Cosmic Ray Research (ICRR), University of Tokyo.

\software{PANHIT \citep{2020IAUS..341..285M}, Prospector \citep{2021ApJS..254...22J}, SExtractor \citep{1996A&AS..117..393B}, SWarp \citep{2002ASPC..281..228B}, Yggdrasil \citep{2011ApJ...740...13Z}}

\clearpage
\bibliographystyle{apj}
\bibliography{apj-jour,reference}

\end{document}